\documentclass[12pt]{article}
\newenvironment{ISItext1}
{\normalsize\rm\setlength{\parindent}{1cm}\setlength{\parskip}{0pt}}
{\vskip 12pt}

\usepackage{graphicx}  
\usepackage{graphics}  
\usepackage{subfig}
\usepackage{amssymb}
\usepackage{amsmath}
\usepackage{epsfig}
\begin{document}
\newcommand{\ISItitle}[1]{\vskip 0pt\setlength{\parindent}{0cm}\Large\textbf{#1}\vskip 12pt}

\newcommand{\ISIsubtitleA}[1]{\normalsize\rm\setlength{\parindent}{0cm}\textbf{#1}\vskip 12pt}
\newcommand{\ISIsubtitleB}[1]{\normalsize\rm\setlength{\parindent}{0cm}\textbf{#1}\vskip 12pt}
\newcommand{\ISIsubtitleFig}[1]{
\normalsize\rm\setlength{\parindent}{0cm}
\textbf{\textit{#1}}\vskip 12pt}

\newcommand{\ISIauthname}[1]{\normalsize\rm\setlength{\parindent}{0cm}#1 \\}
\newcommand{\ISIauthaddr}[1]{\normalsize\rm\setlength{\parindent}{0cm}\it #1 \vskip 12pt}

\ISItitle{Magnetic Nanostructures} \ISIauthname{K. Bennemann}
\ISIauthaddr{Institute of Theoretical Physics
FU-Berlin\\Arnimallee 14, 14195 Berlin\\keywords: Magnetism,
Nanostructures}

\tableofcontents
\begin{ISItext1}
\section*{Abstract}
Characteristic results of magnetism in small particles
and thin films are presented. As a consequence of the reduced
atomic coordination in small clusters and thin films the
electronic states and density of states modify. Thus magnetic
moments and magnetization are affected. In tunnel junctions interplay of magnetism,
spin currents and superconductivity are of particular interest. Results are given for
single transition metal clusters, cluster ensembles, thin films
and tunnel systems.
\end{ISItext1}
\setcounter{page}{1}
\setcounter{section}{0}
\section{Introduction}
\begin{ISItext1}
Due to advances in preparing small particles, thin films, film
multilayers and tunnel junctions the area of nanostructures in
physics has received increased interest. Clearly engineering on an
atomic scale condensed matter offers many possibilities regarding
new physics and technical applications.

Of particular interest is the occurrence of magnetism in
nanostructures like single transition metal clusters (Ni, Co, Fe
etc.), ensembles of such clusters, for example in lattice like
arrangements\cite{B1}, ferromagnetic thin films, multilayers of such films \cite{B2}
and tunnel junctions\cite{B3}. The latter are of interest, for example,
regarding switching of electric, charge and spin currents. Size
and orientation of magnetic moments and magnetization depend in
general sensitively on atomic environment, atomic coordination
(surfaces, interfaces etc.).

Of course, one expects that magnetic fluctuations are significant
due to the reduced dimension of nanostructures. Regarding magnetic
anisotropy, this will play an important role in general and also
with respect to the role played effectively by the magnetic
fluctuations. Due to magnetic anisotropy phase transitions occur
in reduced dimensions and ultrathin films.

\begin{figure}[htbp]
\centerline{\includegraphics[width=0.6\textwidth]{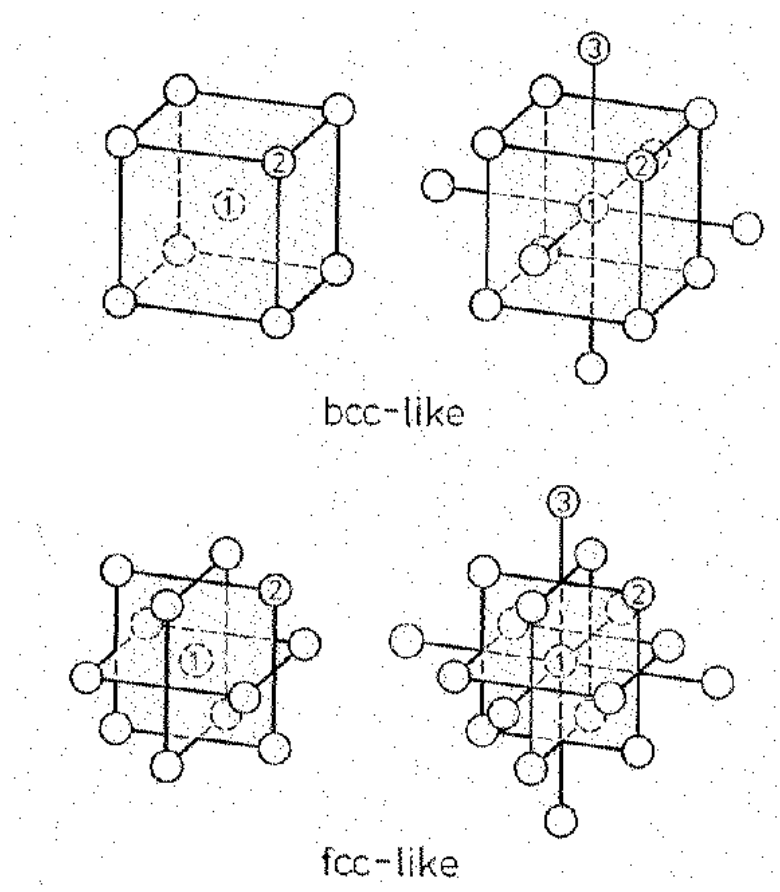}}
\caption{Illustration of b.c.c.-- and f.c.c. like clusters.
Different atomic shells surrounding the center 1 are labelled by 2, 3 etc.}
\end{figure}
In general the topology of the nanostructure affects strongly the
electronic structure and the orientation of the magnetization, its
domain structure, and magnetic relaxation. The electron energy
spectrum gets discrete and quantum well states (QWS) occur in thin
films. Note, in nanostructured thin films magnetic domain
structure is frequently present. The size of the domains and their relaxation
depends on atomic structure of the film, anisotropy and temperature.
The reversal of a domain magnetization may be relatively slow, but may speed up at
nonequilibrium for example due to hot electrons.

In growing thin films interesting nonequilibrium behavior may
result from the interdependence of atomic morphology of the
nanostructure and its magnetic domain formation. This may be seen
observing time resolved properties of growing nanostructures.
Thus, it is of interest to study, time resolved, the occurrence of
uniform ferromagnetism in thin films as a function of film
thickness and domain density and size, in general on growth
conditions for producing nanostructures.

For tunneling interplay of magnetism and superconductivity is of particular interest. Between two
magnets one expects as a result of the spin continuity equation and Landau--Lifshitz equation Josephson
like spin currents.

In the following magnetic nanostructures having different geometry and tunneling
are discussed:

\subsection{Magnetic Clusters}

While small clusters may have a complicated atomic structure, this
will tend to become bulk like for larger ones. Hence,
approximately one may assume first liquid like structures as the
number of atoms $N$ in the cluster increases and then bulk like
structures for larger clusters. The cluster volume $V$ is given by
$V\sim N^3$ and the surface area by $S\sim N^{2/3}$ \cite{B1}.

In Fig.1 f.c.c.- and b.c.c. like cluster structures are shown.
These may approximate magnetic transition metal clusters. In such clusters DOS and
magnetic moments are site dependent.

In Fig.2 the discretization of the electron energy spectrum of
such clusters is illustrated. Levels are occupied up to the
Fermi--level $\varepsilon_F$ and interesting odd, even effects
occur as a function of cluster size (number $N$ of atoms).

The magnetism of the cluster is characterized by its atomic shell
$(i)$ dependent magnetic moments $\mu_i(N)$, by the magnetization
$M(T,N)$ and Curie--temperature $T_c(N)$. These magnetic properties
may be calculated using an electronic theory and for example a
tight--binding Hubbard like hamiltonian, or on a more
phenomenological level a Heisenberg hamiltonian including magnetic
anisotropy. Thus, one may estimate the Curie--temperature from
\begin{equation}\label{eq1}
    T_c \sim aJq_{eff}(N)\quad ,
\end{equation}
where a is a fitting parameter, $J$ the interatomic exchange
coupling integral and $q_{eff}$ the effective coordination number \cite{B2}.

In Fig.2 size effects for the electronic structure are sketched.
Note, screening of Coulombic interactions and width of d--electron
states varies with cluster size and affects thus magnetic
activity. Also spacing of the n electronic states is
\begin{equation}\label{eq2}
\delta\approx\varepsilon_F/n\approx(\hbar
v_F/R)(k_{F}R)^{-2}(N(0)V)^{-1} \quad .
\end{equation}
Here, $\varepsilon_F$ is the Fermi energy and $v_F$ and $k_F$ the
corresponding Fermi velocity and wave--vector. $V$ refers to the
volume and $N(0)$ to the density of states (D.O.S.) at
$\varepsilon_F$. One expects for $T>\delta$ that the
discretization plays no great role. However, discretization might affect many properties in grains and quantum dots,
in particular magnetism and superconductivity (formation of magnetic moments, magnetization, size of Cooper pairs, superfluid density etc.).

The local electron density of states of atom $i$ is
\begin{figure}
\centerline{\includegraphics[width=0.5\textwidth]{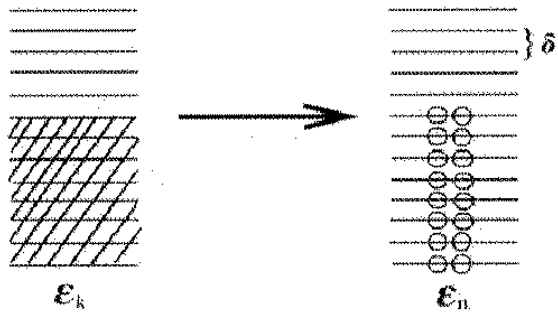}}
\caption{Sketch of size effects in small clusters having a
diameter $R$ and $n$ electrons with energy $\varepsilon_n$. The
spacing of the electronic energy levels is $\delta$. ($\epsilon_k\rightarrow$ discrete spectrum as particle size
decreases). The electronic spectrum exhibits shell structure which reflects characteristically the atomic structure. Many properties should reflect this, in particular magnetism and superconductivity.}
\end{figure}
given by $N_i(\varepsilon, N)$ and generally spin dependent ($N_{i\sigma}(\varepsilon$, $N$)). This determines occurrence and
size of local magnetic moments. Their direction is determined by
magnetic anisotropy. Note, besides spin magnetism also orbital one
occurs and is typically enhanced compared with bulk one. Also generally the orbital magnetism is dependent on atomic site.

Mie--scattering by spherical small particles is an interesting
phenomena. One expects that magnetism leaves a characteristic
fingerprint on the Mie--scattering profile. The magnetic field of
the incident light couples to the cluster magnetization. This
affects the Mie scattering in particular the backscattering
intensity \cite{B4}.

This effect of magnetism on Mie scattering needs be studied more.
It offers an interesting alternative to deflection experiments by
a magnetic field regarding study of small particle magnetism.

In Fig.3 an ensemble of clusters (or quantum dots) arranged on a
lattice is shown. In general the lattice sites may be occupied by
ferromagnetic (or paramagnetic) clusters or may be empty.
Dependent on the cluster pattern one gets rich physical behavior.
For example, while the single clusters are ferromagnetic, for
larger spacing between the clusters one might get (for higher
temperature) no global magnetization of the whole cluster
ensemble. For dense spacings of the grains, dots and sufficient
strong interaction amongst them, local and also global
magnetization may occur as a function of temperature $T$. Note,
typically $T_{ci}(N)>T_c$, where $i$ refers to the cluster $i$ and
$T_c$ to the Curie--temperature of the ensemble, cluster lattice.
Of course, also antiferromagnetism may occur.

In general interesting phase diagrams are expected for an ensemble
of magnetic grains (quantum dots). One gets for interactions $J>0$
and $J<0$ ferromagnetism or antiferromagnetism for the grain
ensemble. For a mixture of grains, for example a mixture of
superconducting and ferromagnetic ones, one may observe like for
alloys interesting behavior and one may get Josephson currents
between the dots and due to electronic charge transfers odd--even occupation
effects, etc.
\begin{figure}[htbp]
\centerline{\includegraphics[width=0.5\textwidth]{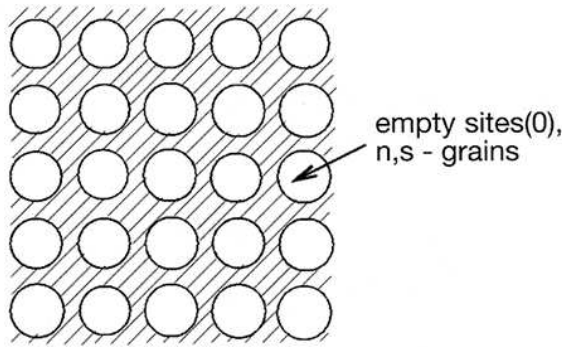}}
\caption{Illustration of an ensemble of small particles (grains) arranged
in a lattice. The lattice sites may be empty or occupied by
ferromagnetic, paramagnetic, or superconducting clusters, for
example. Removing irregularly clusters from the lattice sites
creates all sorts of nanostructures. Electrons move via hopping and tunneling between the lattice sites.}
\end{figure}

An interesting example of an important nanostructure is an
ensemble of quantum dots, an anti-dot lattice for example. In
Fig.4 an anti--dot lattice immersed in a medium is sketched. Here,
the anti--dots repel electrons moving for example between the
quantum dots. An external magnetic field B will change, deform the
electron orbits spin--dependently and this may yield interesting
behavior of the electronic properties of the system. Sensitive
quantum mechanical interferences appear in such nanostructures. As
indicated in Fig.4 a few electronic orbits may determine mostly
the electronic structure, see theory by Stampfli {\it et al.} \cite{B5}.
Quantum mechanical interference cause characteristic oscillations
in the electronic density of states (DOS). These may be spin dependent.
\begin{figure}
\hspace{-.6cm}\centerline{\includegraphics[width=0.36\textwidth]{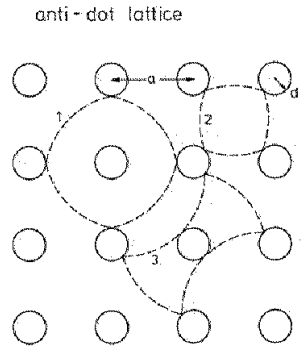}}
\caption{Magnetic field effects on polygonal electron paths in an
anti--dot lattice. Here, a denotes the spacing of the anti--dots
with radius d. The electrons move between the dots and are repelled by the anti--dot
potential and thus selectively the polygonal paths 1, 2, 3 etc.
yield the most important contribution to the spin dependent
D.O.S., magnetoresistance, etc. Details of the theory determining
the electronic structure of such mesoscopic systems are given by
Stampfli {\it et al.}.}
\end{figure}

In summary, these are some cluster like nanostructures. Regarding magnetism transition--metal and rare--earth atoms, metal--oxydes etc. are particularly interesting. The dependence of magnetic spin and orbital moments and spin correlations on particle size and temperature are of interest. Magnetic moments, Curie temperature and magnetization control the magnetic behavior \cite{B1,B2,B6,B7,B8,B9,B10}.

\subsection{Magnetic Films}

For ultrathin films one may get magnetic behavior which could
differ drastically from the one for thick ferromagnetic films. The
latter are approaching as a function of film thickness $d$ the
bulk behavior (b), for the Curie--temperature one finds
$T_c(d)\longrightarrow T_{c,b}$ as film thickness $d$ increases.
The increase of $T_c(N)$ for increasing thickness of the
ferromagnetic film is often described by the scaling law
$\Delta T_c/T_{cb}\propto N^{-\lambda}$, or more empirically by $\Delta T_c / T_c(N) \propto N^{-\lambda^{'}}$ with non--universal exponent $\lambda^{'}$ which fits better experiments, for details see Jensen
{\it et al.} \cite{B2}.

Note, in ultrathin films, one or two atomic layers thick,
ferromagnetism results from magnetic anisotropy (spin--orbit
coupling etc.) suppressing two--dimensional magnetic fluctuations.

Various film structures are shown in Fig.5. Neighboring magnetic films may order parallel or antiparallel.
\begin{figure}
\centerline{\includegraphics[width=0.6\textwidth]{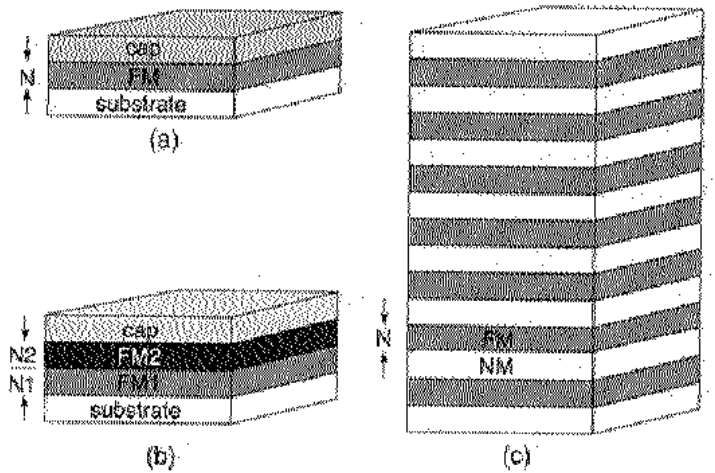}}
\caption{Thin magnetic film structures are shown. Typical
configurations are given. (a) Thin ferromagnetic film of thickness
N on a substrate, (b) two ferromagnetic films (FM1, FM2) on a
substrate and (c) multilayer film structure. The ferromagnetic
films are separated by a nonmagnetic one (NM). At the surface the
magnetic film may be covered by nonmagnetic material (cap). Of
particular importance is the interplay of magnetism of magnetic
thin films of a film multilayer system.}
\end{figure}
Approximately, first for ultrathin films $T_c(d)\sim d$ typically.
As the thickness of the film increases the Curie--temperature
approaches the bulk one. Generally one gets for multilayer
structures a rich variety of magnetic phase diagrams. For example, the change of the magnetization upon increasing the Cu--spacer thickness of a (Co/xCu/Ni) film system reflects characteristically the coupling between the two
ferromagnets Ni and Co. For a
detailed discussion of theoretical and experimental results see
Jensen {\it et al.} \cite{B2,B6,B7,B9,B10,B11}.

Magnetic anisotropy controls the orientation of the magnetization,
$\overrightarrow{M}(T,d,\ldots)$, at the surface of the film.
Dependent on temperature $T$, film thickness and structure and
effective magnetic field $\overrightarrow{B}$ the orientation of
the surface magnetization $\overrightarrow{M}$ may change from
perpendicular $M_{\bot}$ to parallel one $M_{\|}$,
$M_{\bot}\longrightarrow M_{\|}$. This reorientation transition,
important for example for magnetic recording, is illustrated in
Fig.6 \cite{B7}.
\begin{figure}
\centerline{\includegraphics[width=0.8\textwidth]{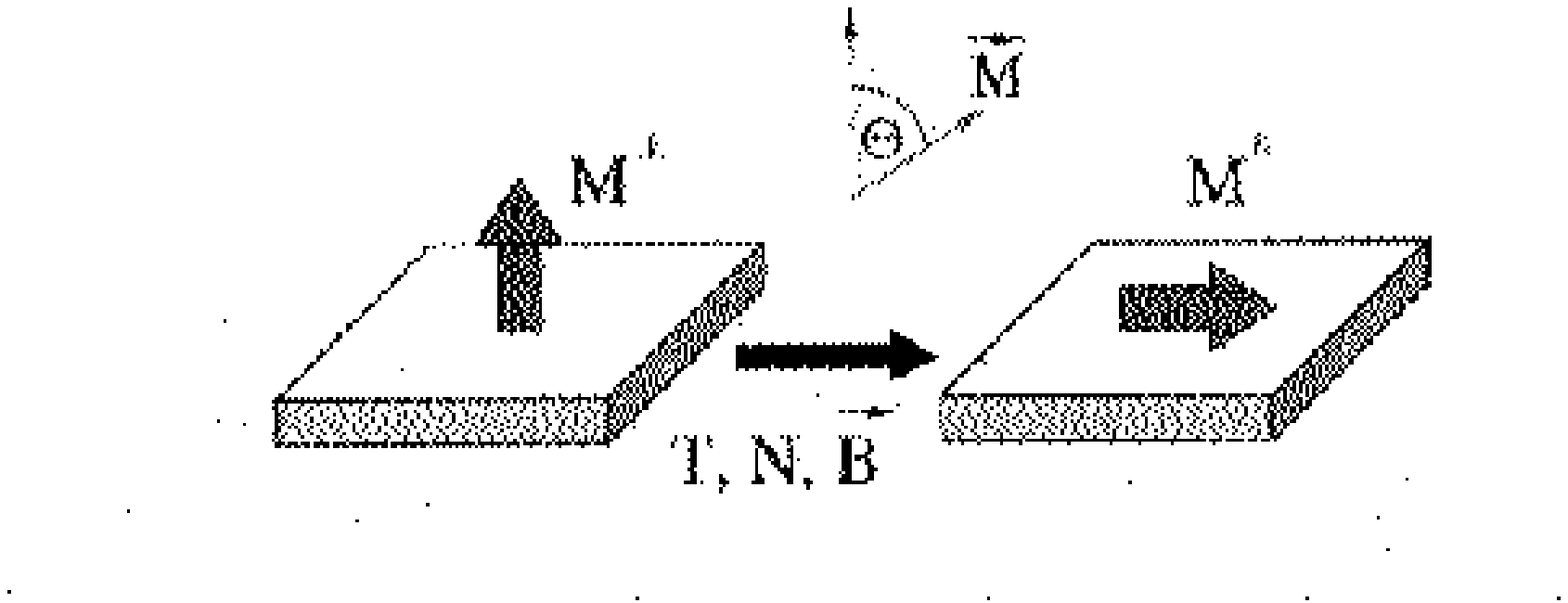}}
\caption{The reorientation transition $M_{\bot}\rightarrow M_{\|}$ of
the magnetization at the surface or interface of films is shown. This
transition may be induced by temperature $T$, increasing film
thickness $N$ and film morphology and external magnetic field $B$.}
\end{figure}

Obviously, the orientation of the magnetization at surfaces is
related to the magnetic domain structure occurring in thin films.
This is also clear from the illustration of domain structure in
Fig.7.
\begin{figure}
\centerline{\includegraphics[width=0.8\textwidth]{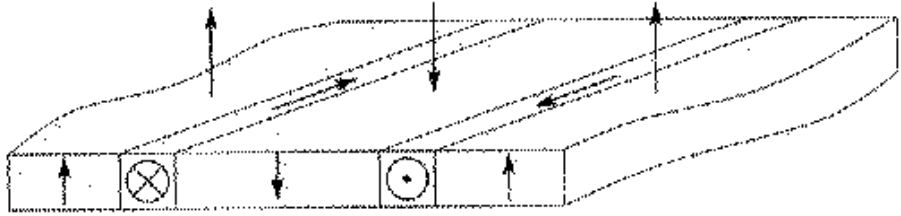}}
\vspace{.3cm}\caption{Illustration of a magnetic stripe domain
phase of a thin film. Neighboring domains are
antiferromagnetically oriented and separated by Bloch type domain
walls. The magnetic structure is generally controlled by the film
morphology and magnetic anisotropy. Of course, dependent on this
other domain structures may result, see Jensen {\it et al.}.}
\end{figure}

As expected on general physical grounds, dependent on film growth
conditions one gets for thin film growth on a substrate a variety
of nanostructures. This is illustrated in Fig.8. For a growing
film the accompanied magnetism may not be at equilibrium, but
changing as the film topology changes. One has a nonequilibrium
situation and magnetic structure changes due to magnetic
relaxation, of domains {\it etc.}. The latter may be relatively slow.
Various pattern of magnetic domains occur in general. Then
reversal of local domain magnetization occur on a ps (picosecond)
to ns (nanosecond) time scale and change the global film
magnetization as function of time, see for example studies by
Brinzanik, Jensen, Bennemann.
\begin{figure}
\centerline{\includegraphics[width=0.7\textwidth]{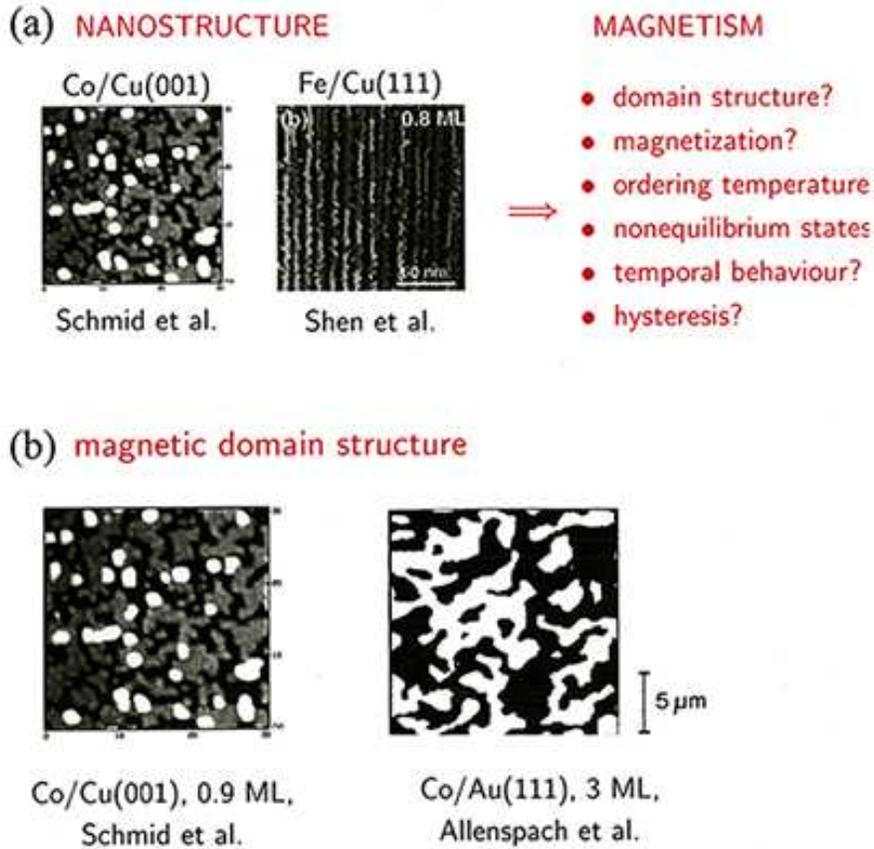}}
\caption{Nanostructures of ultrathin films and accompanying
magnetic domain structures. The upper Fig.(a) shows various
observed atomic nanostructures. Questions are listed regarding
dependence of the resulting magnetic structures on film topology.
Co/Cu(001) refers to 0.9ML of Co (black:substrate, grey:first
layer, white:2.adatom layer), Fe/Cu(111)to 0.8ML of Fe (grey:
chains of Fe atoms which will coalesce for increasing Fe coverage).
Fig.(b) The obtained magnetic domain structures are given. Only domain
structure at surface is shown. Magnetic domains for Co/Cu (white:
2. layer, grey: 1. layer). The irregularly shaped domains (white) for Co/Au
have a perpendicular magnetization and lateral size of about 1um. Note,
such structures are observed in various experiments. Typically
magnetic domain size increases with increasing film thickness. Increasing for Co/Au the Co film thickness to
6ML turns their magnetization to in--plane (reorientation transition). }
\end{figure}

Similarly as in the case of cluster ensembles on a lattice the
magnetic domains resulting in nanostructured films may be dense or
separated by larger distances. Correspondingly the domains are
dominantly coupled via exchange or dipolar interactions, for
example. This then will be reflected in the global magnetization,
size of magnetic domains and in particular magnetic relaxation
during film growth.

In Fig.9 the growth of a thin film resulting from atom by atom
deposition at the surface is shown. For simplicity fast diffusion
of chemisorbed atoms is assumed. Growing of a film occurs, since
deposited atoms prefer to sit next to already deposited atoms thus
gaining optimally cohesion. As time progresses islands of
deposited atoms coalesce and nanostructured film is formed. Of
course, via diffusion of the deposited atoms the film structure
depends somewhat on the film structure of the substrate. Thus, one
may get island formation, striped structures and quasi uniform
growth, see calculations by Jensen {\it et al.}. It is remarkable that already simple growth models yield most of the observed structures during film growth.

\begin{figure}
\centerline{\includegraphics[width=0.56\textwidth]{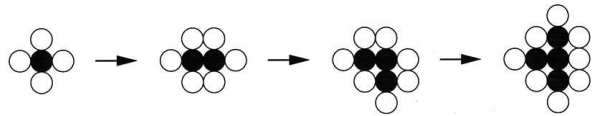}}
\caption{Illustration of simple (Eden type) growth of a thin film
(for a square lattice). One assumes fast surface diffusion and
uses a Monte Carlo simulation of molecular beam epitaxial (MBE)
film growth. Surface atomic sites are indicated. Successively
deposited atoms are given in black and due to cohesive energy
prefer to cluster.}
\end{figure}

Regarding magnetization this reflects the nanostructure and
typically magnetic domains occur before global uniform film
magnetization is present.
\begin{figure}
\centerline{\includegraphics[width=0.2\textwidth]{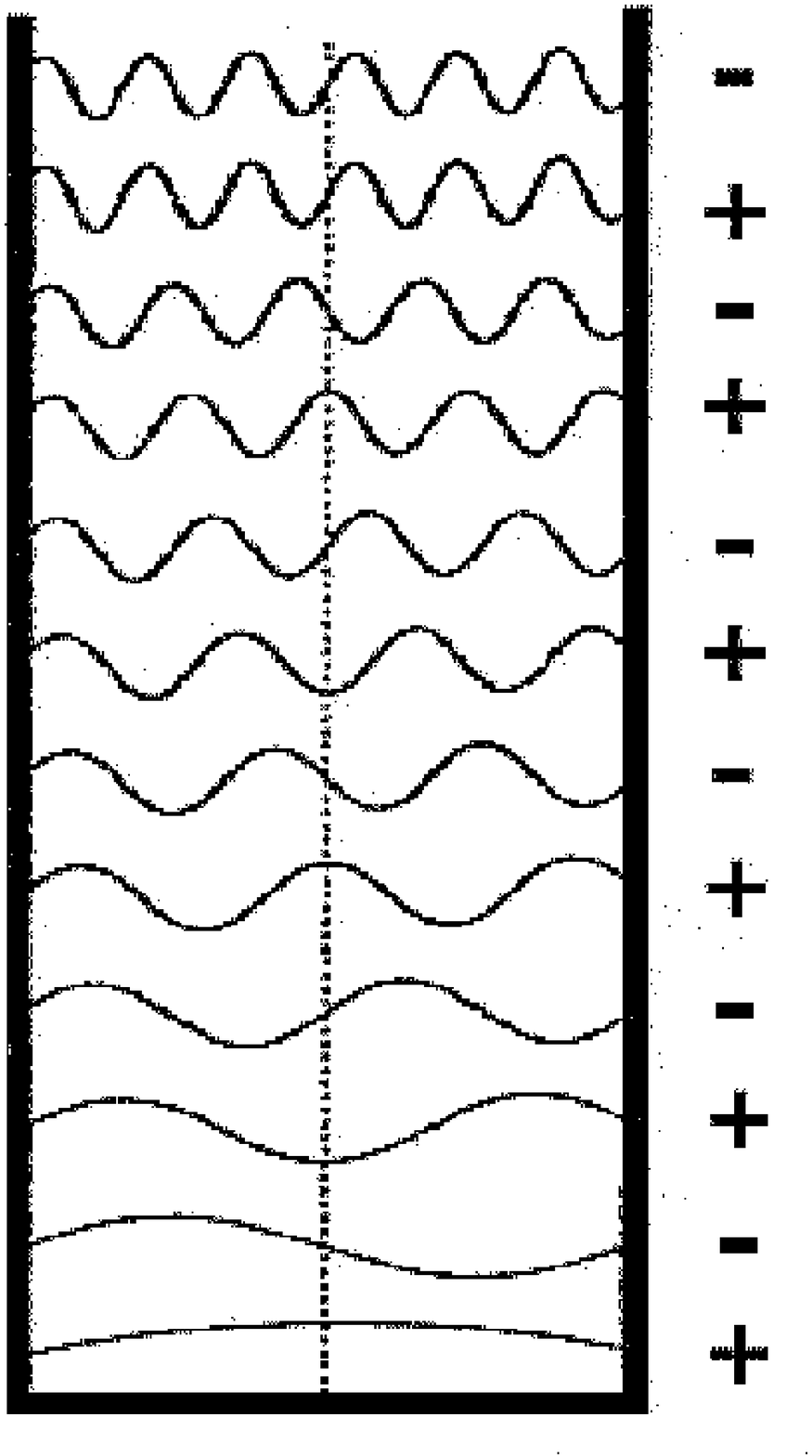}}
\caption{Sketch of electronic Q.W.S. in a thin film and resulting from the confinement of the electrons.
Such QWS occur for example in a Co/xML Cu/Co film system. Describing the film confinement by a square well potential one gets already quite properly
such QWS and which parity is indicated by (+) and (-). Note, parity changes for increasing quantum number.}
\end{figure}

In thin films one expects quantum--well states (Q.W.S.) which will
change characteristically with film thickness. Such states were
calculated for example by H\"{u}bner, Luce {\it et al.}. For details
see Bennemann in Nonlinear Optics in Metals (Oxford University
Press) \cite{B12}. The confinement in thin films causes the Q.W.S. In FM
films these states are spin split reflecting characteristically
magnetism. In Fig.10 QWS are sketched and their parity is indicated by (+) and (-).

Magnetooptics (S.H.G.: Second harmonic light generation) will
reflect magnetism sensitively. Thus in particular for  magnetic nano
film structures with spin split Q.W.S. one will get interesting optical
properties reflecting characteristically magnetism. This is illustrated in Fig.11. Optical interferences
involving QWS cause oscillations in the MSHG as a function of film thickness.
\begin{figure}
\centerline{\includegraphics[width=0.55\textwidth]{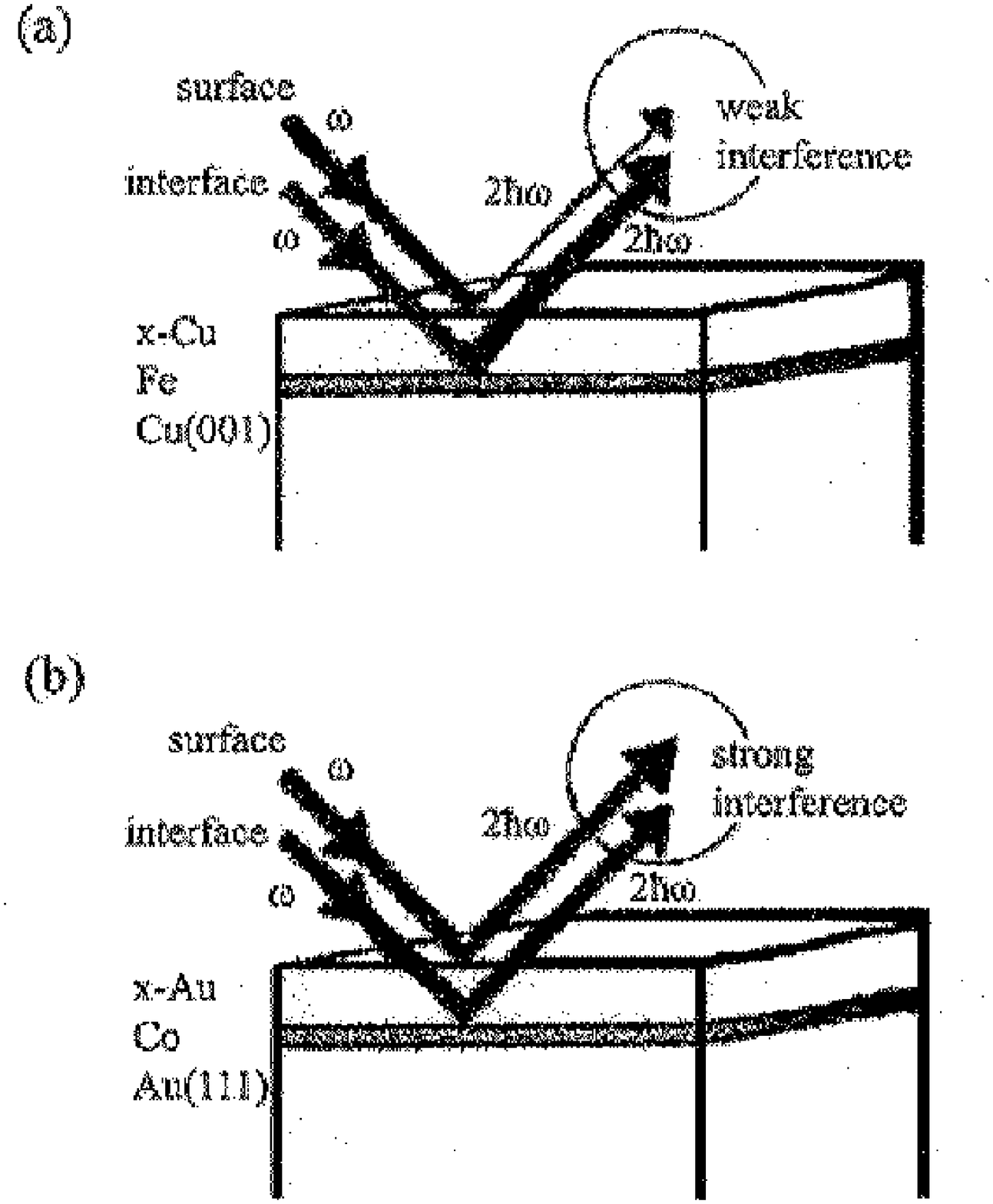}}
\caption{Illustration of magnetic S.H.G.(MSHG) resulting for thin
films with Q.W.S.. Note, in magnetic films the Q.W.S. are
spin polarized and their energies depend on film thickness. Thus, SHG
involving QWS exhibits film thickness dependent effects.
Oscillations in the magnetooptical signals occur. The interference of S.H.G.
from the surface and the interface (film/substrate) yields a
sensitive detection of magnetism. Characteristic differences occur
for (a) weak interference and (b) strong interference. This is
observed for example for xCu/Fe/Cu ( weak interference ) and xAu/Co/Au structures ( strong interference ), respectively. Here,
$x$ refers to the number of layers of the top film.}
\end{figure}

This summarizes then interesting nanostructures consisting of
films. Film multilayers offer particularly interesting magnetic
pattern. Neighboring films interact largely via exchange coupling
and this controls the magnetic behavior of each film and the global one. For example transport properties like the magnetoresistance depend on the relative orientation of the magnetization
of neighboring films. During film growth magnetic relaxation controlled by
anisotropy plays an important role. This is reflected in the
nonequilibrium magnetization of the film and its relaxation
towards the equilibrium one.

\subsection{Tunnel Junctions}

Tunnel junctions involving magnetism are interesting
microstructures, in particular regarding quantum mechanical
behavior, switching devices and charge-- and spin currents and
their interdependence. Coupling of the magnetic order parameter
phases, for both ferromagnets and antiferromagnets, on both sides
of the tunnel medium yields Josephson like (spin) currents, driven
by the phase difference, for details see study by Nogueira {\it et
al.} and others. Obviously, this will depend on the magnetic state
of the medium through which tunnelling occurs, on spin relaxation.
The effective spin coupling between $N_1$ and $N_3$ depends on the
spin susceptibility of $N_2$ and $J_{eff}= J_{eff}(\chi)$. Note,
from the continuity equation follows
\begin{equation}
      \overrightarrow{j_T}\propto J_{eff}\overrightarrow{S_1}\times \overrightarrow{S_3}+\ldots
\end{equation}
The tunneling is sketched in Fig. 12.

Clearly tunnelling allows to study magnetic effects, ferromagnetic
versus antiferromagnetic configurations of the tunnel system
$(\uparrow\mid T\mid\uparrow)$, or $(\uparrow\mid T\mid\downarrow)$ and interplay of magnetism and superconductivity $(T\equiv
S.C.)$, as for example for junctions (FM/SC/FM) or (SC/FM/SC). Such
junctions may serve as detectors for triplett superconductivity
(TSC) or ferromagnetism (FM), antiferromagnetism(AF). Of course
tunnelling is different for parallel or antiparallel magnetization
of $M_1$ and $M_3$.  Also importantly, tunnel junctions may serve
to study Onsager theory on a molecular--atomistic scale.
\begin{figure}
\hspace{.3cm}\centerline{\includegraphics[width=0.4\textwidth]{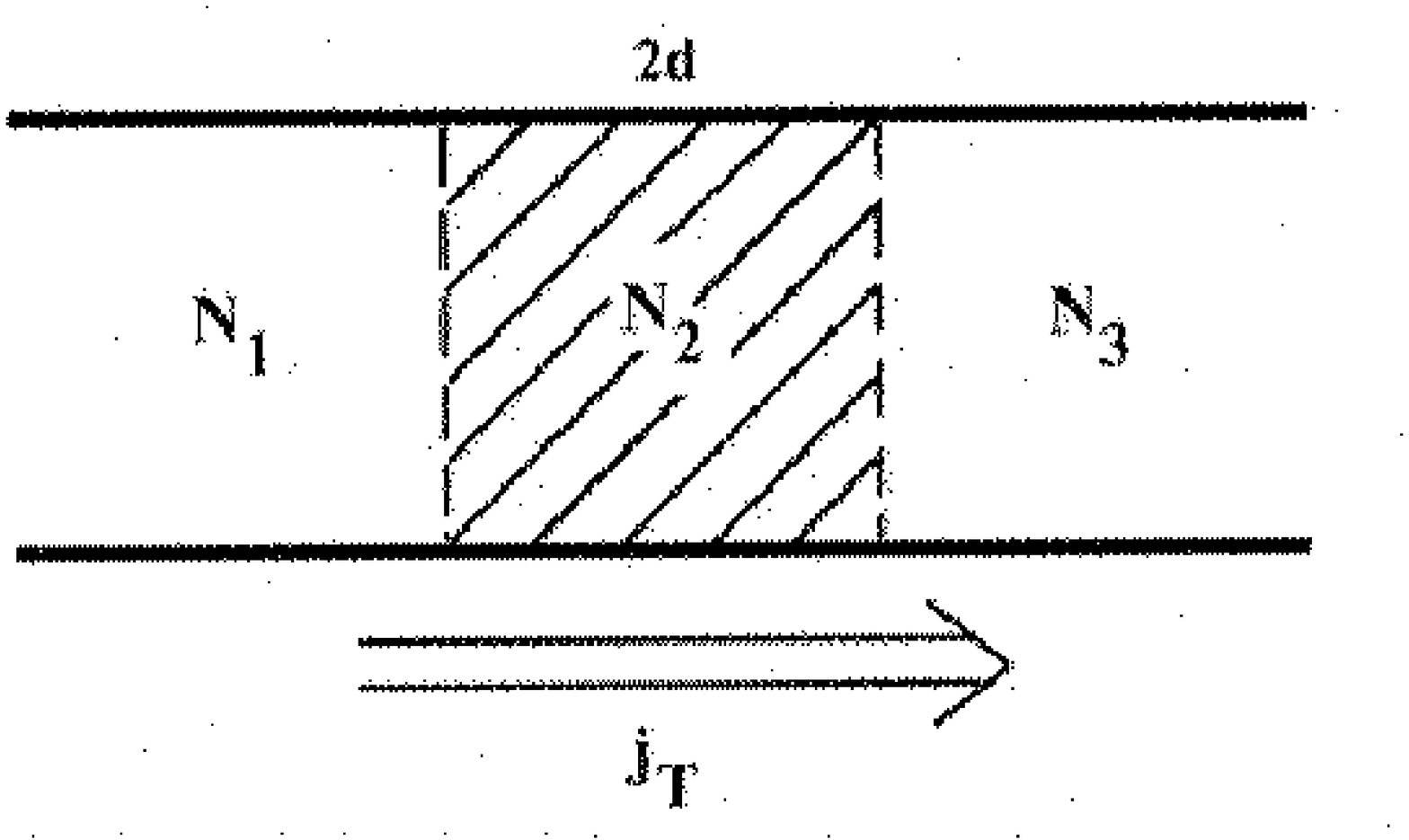}}
\caption{A tunnel junction ($N_1\mid N_2\mid N_3$), where $N_i$ refers
to the state, magnetic, superconducting, normal. The current $j_T$
may refer to spin-- or charge transport and may be driven by the
phase difference $\Delta\varphi=\varphi_3-\varphi_1$, for
example. Both spin and charge current may couple in general. The
continuity equation for the magnetization,
$(\partial_t M_i + \partial_\mu j_{i\mu,s} =0)$,
indicates that magnetic dynamics $M(t)$ may induce spin currents
$j_s$. As a consequence interesting nonequilibrium behavior is
expected. Note, applying a voltage $V$ to the sketched tunnel system
may yield a two--level system (resembling a qubit, etc.)}
\end{figure}

Electron and Cooper pair transfer between two quantum dots
exhibits interesting behavior, for example v.St\"{u}ckelberg
oscillations due to bouncing back and forth of electrons if an
energy barrier is present. Then assistance of photons is needed to
overcome the energy barrier. The current (spin--, charge current) may depend on the pulse shape
and duration, see Garcia, Grigorenko {\it et al.}.

Summary: For illustrating reasons various interesting
nanostructures like clusters, films and tunnel junctions with
important magnetic effects are discussed. In the next chapter some
theoretical methods useful for calculations are presented. Then
results obtained this way are given.

It is important to note that for illustrational purposes the
physics has been simplified. Mostly, the analysis can be improved straightforwardly.
However, likely this will not change the physical insights
obtained from the simplified analysis. For more details see for
clusters studies by Pastor {\it et al.} \cite{B1,B6} and by Stampfli {\it et
al.} \cite{B5}, and for films research by Jensen {\it et al.}\cite{B2,B7,B8}, and for
tunnel junctions studies by Nogueira, Morr {\it et al.} \cite{B3}.

\section{Theory}

In general theory for nanostructures must allow for a local,
atomic like analysis of the electronic structure. Using Hubbard
hamiltonian and tight--binding type theory one may determine

\noindent{\bf (a)} $N_{i\sigma}(\epsilon,\ldots)$, the electron
density of states at an atomic site $i$ and for spin $\sigma$,
dependent on the local atomic configuration surrounding atom $i$,

\noindent({\bf b}) $\mu$, the atomic like magnetic moment, as a function of particle size
(cluster size or film thickness),

\noindent({\bf c}) $M$, the magnetization and the
Curie--temperature ($T_c$).

Note, all properties result from the electronic Green's function
$G_{i\sigma}(\epsilon,\ldots)$. Including spin--orbit coupling ($V_{so}$) yields magnetic
anisotropy. Thus, also

\noindent({\bf d}) orbital magnetism \\is obtained.
Note,
anisotropy and orbital magnetism get typically for nanostructures
more important.

Alternatively to an electronic theory one may use on a more
phenomenological level the Heisenberg hamiltonian including
magnetic anisotropy to analyze magnetism in nanostructures.

\subsection{Magnetism : Electronic theory}

To determine the size and structural dependence of the magnetic
properties of small transition metal clusters the Hubbard
hamiltonian for $d$--electrons which are expected to contribute
dominantly is used (see Pastor {\it et al.} \cite{B1,B6} and Moran--Lopez {\it et al.} \cite{B10})
\begin{equation}
H=\sum_{i=j} t_{ij} c_{i\sigma}^+ c_{i\sigma} + H' \quad ,
\end{equation}
\begin{equation}
H'=\sum_i \epsilon_{i\sigma} n_{i\sigma} - E_{dc} \quad ,
\end{equation}
and effective on--site electron energies
\begin{equation}
\epsilon_{i\sigma}=\epsilon^0 + U\Delta n_i -\sigma\frac{J}{2}\mu_i
\quad .
\end{equation}
Here, $c_{i\sigma}^+$ and $c_{i\sigma}$ are the usual creation and
annihilation operator for electrons on site $i$ and with spin $\sigma$
and $t_{ij}$ denotes the distance dependent hopping integral.
Note, $i,j$ refer to atomic sites and includes orbital character
($d_i: e_g, t_{2g}$ orbitals, $s,p$-orbitals). $H'$ describes
interactions (in the unrestricted Hatree--Fock approximation). The
effective intraatomic Coulomb interaction are denoted by $U$ and
the exchange interaction by $J$
($J=U_{\uparrow\downarrow}-U_{\uparrow\uparrow}$,
$U=(U_{\uparrow\downarrow}+U_{\uparrow\uparrow})/2$). Here,
$U_{\sigma\sigma'}$, refers to electron spins $\sigma,\sigma'$.
$E_{dc}=(1/2)\sum_{i,\sigma}(\epsilon_{i\sigma}-\epsilon^0)<n_{i\sigma}>$
corrects for double counting as usual. The charge transfer $\Delta
n_i$ is given by $\Delta n_i=n_i - n^0$, $n^0=(1/n)\sum_i n_i$.

The quasi local magnetic moment at site $i$ is given by
\begin{equation}
\mu_i\propto  <n_{i\uparrow}- n_{i\downarrow}>\quad ,
\end{equation}
with $\langle n_{i\sigma}\rangle=\int_{-\infty}^{\epsilon_F}d\epsilon N_{i\sigma}(\epsilon-\epsilon_{i\sigma})_t$. Here, $N_{i\sigma}(\epsilon)$ is the
density of states (DOS) at site $i$ for electrons with spin
$\sigma$ (and at time $t$).

This theory applies also to thin films. Then, $i$ may refer to
film layer and one should keep $\mu_i$ and $p_i^{+,-}$.

To calculate the magnetization $M(T)$ one must take into account
the orientation of the magnetic moments. Assuming a preferred
magnetization axis one gets (Ising model, see Moran-Lopez {\it et al.},
Liu) \cite{B10}
\begin{equation}
M(T)\simeq \sum_i\{p_i^+ \mu_i^+ (T) + p_i^- \mu_i^-
(T)\}/\mu^+(0)\quad .
\end{equation}
The probabilities $p_i^{+,-}$ refer to finding moments
$\mu_i^{+,-}$ pointing parallel or antiparallel to the preferred
magnetization axis.  For simplicity one may use the approximation
$p_i^{+,-}=p^{+,-}$. Assuming spherical like clusters, then $i$
refers to the atomic shell of the cluster and $\mu_i^{+,-}$ to the
magnetic moment within shell $i$ pointing in the direction of the
magnetization $(+)$ and in opposite direction $(-)$, respectively.

Then one determines the order parameter ($\propto M$)
\begin{equation}
\mu_i = p_i^+ - p_i^- \simeq p^+ - p^-
\end{equation}
($ \mu_i \approx\mu$ ) from minimizing the free--energy
\begin{equation}
F=\Delta E - TS \quad .
\end{equation}
Here, the entropy $S$ is given by
\begin{equation}
S\simeq - kN\{ p^+ \ln p^+ + p^- \ln p^-\} \quad ,
\end{equation}
and $\Delta E= E(\mu)-E(0)$. The electronic energy is calculated
using a hamiltonian $H$, for example Eq.(4). For calculating the
Green's functions the electronic energies are determined from
\begin{equation}
\epsilon_{i\sigma}^{+,-}\simeq \epsilon_{i\sigma}^0 - \sigma J
\sum_j \mu_i^{+,-}\mu_j^{+,-}\quad .
\end{equation}
Note,
$\epsilon_{i\sigma}^+\simeq\epsilon_{i\sigma}^0-\sigma\mu_i^+
J\sum_j\{p^+\mu_j^+ +p^-\mu_j^-\}$. The Curie--temperature is
given by $M(T_c)=0$.

A similar analysis can be performed using functional--integral
theory as developed by Hubbard {\it et al.}, see Pastor {\it et al.} \cite{B6}.

Note, as already mentioned this theory can also be used for films.
Then $i$ may refer to the film layer, etc..

The magnetization can also be determined using the Bragg--Williams
approximation. Assuming for simplicity Ising type spins one finds
for the magnetization (see Jensen, Dreyss\'{e} {\it et al.})
\begin{equation}
M_i(T)=\tanh \{ \beta J\mu_i (z_0 M_i \mu_i +z_1 M_{i+1} \mu_{i+1}+
z_{-1} M_{i-1} \mu_{i-1})-\Delta h_i\}\quad .
\end{equation}
Here, $\beta =1/kT$, $z_0,z_1,z_{-1}$ are nearest neighbor
coordination numbers and $\Delta h_i$ denotes the Onsager reaction
field. Referring to cluster shells (film layer) $z_0$ gives the
neighboring atoms of $i$ within shell (layer) $i$ and $z_{-1}$ and
$z_1$ the nearest neighbor atoms in the shell (layer) below and
above, respectively. It is
\begin{equation}
\Delta h_i =(\beta J\mu_i )^2 M_i \{ z_0 \mu_i^2 (1-M_i^2) + z_1
\mu_{i+1}^2(1-M_{i+1}^2) + z_{-1} \mu_{i-1}^2 (1-M_{i-1}^2)\}\quad
.
\end{equation}
Applying these expressions to films one has $z_1=0$ (and
$\mu_1$=o) if $i$ refers to the surface plane and $z_{-1}=0,
\mu_1=0$ if $i$ refers to the film layer on a nonmagnetic
substrate.

Note, for $T\leq T_c$ the Eq.(13) can be linearized yielding a
(tridiagonal) matrix equation which largest eigenvalue gives
$T_c(d)$. For the hamiltonian H one may use the Heisenberg one,
for example.

Orbital magnetism, anisotropy: Adding spin--orbit interaction
$V_{so}$ to Eq.(4), $H\rightarrow H +V_{so}$, and using
\begin{equation}
V_{so}=-v\sum_{\alpha,\beta}(\overrightarrow{L_i}\cdot\overrightarrow{S_i})_{\alpha,\beta}
c_\alpha^+ c_\beta \quad ,
\end{equation}
one may also determine magnetic anisotropy and also orbital
magnetic moments ($\langle L_i\rangle$).
$(\overrightarrow{L_i}\cdot\overrightarrow{S_i})_{\alpha,\beta}$
refers to intra--atomic matrix elements between orbitals
$\alpha,\beta$. Of course, the orbital moment
$\overrightarrow{L_i}$ depends on cluster atom $i$ and on
orientation $\delta$ of $\overrightarrow{S}$ with respect to
structural axis: $\overrightarrow{L_i}\rightarrow L_{i,\delta}$
(see Pastor {\it et al.} \cite{B6}). In case of films $i$ refers to the film
layer.

Note, in clusters and thin films and at surfaces the spin--orbit coupling and orbital magnetism
is typically enhanced. Orbital magnetism may play an interesting role for nanostructures.

\subsection{Magnetism : Heisenberg type theory}

One may also calculate the magnetism in nanostructures by using
the Heisenberg type hamiltonian, including magnetic anisotropy
(see Jensen {\it et al.} \cite{B2,B7,B8,B9,B11}). As known magnetic anisotropy controls the magnetic structure like direction of the magnetization, domains, their size, shape, in thin films and at interfaces in particular.
Then one determines the magnetic structure resulting from the quasi local magnetic moments ($\overrightarrow{S_i}$) using the hamiltonian
\begin{eqnarray}
H&=&-\frac{1}{2}J\sum_{i,j} \overrightarrow{S_i}\cdot
\overrightarrow{S_j}+ \frac{1}{2}A
\sum_{i,j}\left(\frac{\overrightarrow{S}_i\overrightarrow{S}_j}{r_{ij}^3}-3
\frac{(\overrightarrow{S}_i\cdot\overrightarrow{r_{ij}})(\overrightarrow{S}_j\cdot\overrightarrow{r_{ij}})}{r_{ij}^5}\right)+ H_{anis.}\quad ,
\end{eqnarray}
with exchange anisotropy hamiltonian
\begin{eqnarray}
H_{anis.}&=&-\frac{1}{4} K\sum_{i,j} S_i^z S_j^z
-\frac{1}{4}D\sum_{i,j}(S_i^x S_j^x +S_i^y S_j^y)\quad .
\end{eqnarray}
Here, $K$ and $D$ are uniaxial and quartic in--plane (exchange)
anisotropy constants, respectively. Note, in $H$ the first term is
the Heisenberg exchange interaction and the second term refers to
the magnetic dipole interaction, and $A=\mu_0(g\mu_B)^2/a_0^3$.
$\overrightarrow{S_i}$ denotes Heisenberg spins ( for Ising model spin
$\frac{1}{2}$) at site $i$. The second term in Eq.(16),
which results as interaction amongst magnetic moments in classical
electrodynamics, reduces in particular at surfaces the
magnetization normal to the surface. it gives the demagnetization
field and is called the shape anisotropy. Although this
long--range magnetic dipole coupling is usually much smaller than
the exchange one, it is on a mesoscopic length scale typically
very important and determines the magnetic domain structure.

Of course, the parameters, in particular $K$ and $D$ may depend on $i$,
$j$ (shell, film layer). Typically anisotropy is larger at surfaces,
interfaces. The interplay of $K$ and $D$, which may depend on film
thickness and temperature ($K\rightarrow K_i$, where $i$ refers to
film layer), determines the normal and in--plane magnetization at
surfaces and the reorientation transition (see Jensen {\it et
al.}). Note, one has also higher order anisotropy constants like
$K_4$ due to noncollinearity of the spins (approximately $K_4
\propto J^{-1}$), see Jensen {\it et al.}, which may play a role.
It is also important to note that exchange and dipolar coupling
have a different distance dependence. The parameters in particular the anisotropy ones depend in general
on temperature. Of course, temperature induces changes of the atomic structure
and this plays a role for thin films and for films during growth on a substrate.

Applying standard methods of statistical mechanism one gets the
free--energy $F$, the magnetization and phase--diagrams in terms
of $J$ and the anisotropy forces for clusters and thin films and
also for nanostructured films with magnetic domain structure. For
a more detailed analysis see in particular P. Jensen {\it et al.}.
The magnetic phase diagram (P.D.) follows from minimizing the
free--energy.

From Eq.(16) follows the magnetic reorientation transition, for
example as driven by temperature:
\begin{equation}
       M_\perp \rightarrow_{T_R} M_\parallel
\end{equation}
occurring at temperature $T_R$. Note, the temperature dependence
of the effective anisotropy parameters and of the dipole coupling
is mainly determined by the magnetization. Minimizing the
free--energy (anisotropy contribution to free--energy) gives the
transition $M_\perp \longrightarrow M_\parallel$ at temperature
$T_R$, see Jensen {\it et al.} \cite{B7}.

Note, the control parameters in Eq.(16) depend on atomic
structure, morphology of the nannanostructures, film thickness for
example. Thus, $T_R$ is affected.

The above Hubbard-tight-binding type electronic theory and the
Heisenberg type plenomenological theory including magnetic
anisotropy permit a calculation of the magnetic properties of
nanostructures. A basic understanding of the dependence of
magnetism in nanostructures on atomic configuration and on more
global morphology is obtained.

Note, for alloys one may
extend above theories using appropriate versions of C.P.A. like
analysis (C.P.A: coherent potential approximation).

\subsection{Electronic Structure of Mesoscopic Systems : Balian--Bloch type theory}

The important electronic structure (shell structure) of
mesoscopic systems like spherical clusters, discs, rings, (quantum) dots can
be determined using a relatively simple theory developed by
Stampfli {\it et al.} extending original work by Balian--Bloch
(Gutzwiller) \cite{B5}. One assumes a square well like generally spin dependent potential ($U_\sigma$). The dominant contribution to the electronic structure (near the Fermi--energy and for electronic wavevector $k_F^{-1}$ larger than atomic distance) results from (interfering) closed electronic orbital paths. Then
the key quantity of the electronic structure of a quantum dot
system, the density of states (DOS), can be calculated from
($n=$number of atoms in nanostructure) from the Green's function
$G$. The Green's function
$G(\overrightarrow{r},\overrightarrow{r}^{'})$ is derived by using
multiple scattering theory, see Stampfli {\it et al.}.
Interference of different electron paths yields oscillations in
the DOS {\it et al.}. This leaves a fingerprint on many properties.

Thus, for example, oscillations in the electronic structure of mesoscopic systems, in the magnetoresistance and other properties of quantum--dot systems can be calculated.

From the electronic Green's function $G$ one determines the (generally spin dependent) density
of states (DOS)
\begin{equation}
N_\sigma(E,n)=\frac{1}{2\pi i}\int_V
d^dr\{G_\sigma(\overrightarrow{r},\overrightarrow{r'},E+i\epsilon)-
G_\sigma(\overrightarrow{r},\overrightarrow{r'},E-i\epsilon)\}_{\overrightarrow{r'}=\overrightarrow{r}}
\quad .
\end{equation}
One gets ($\overline{N}=$ average DOS)
\begin{equation}
N_\sigma (E,n)=\overline{N}_\sigma(E,n) + \Delta N_\sigma (E,n)
\quad ,
\end{equation}
where $\Delta N_\sigma$ refers to the oscillating part of the DOS
due to interference of dominating closed electron paths in
clusters, thin films, and ensemble of repelling anti--dots, see
the theory by Stampfli {\it et al.}. Clearly the scattering
can be spin--dependent (due to a potential $U_\sigma$) and can be manipulated by external
magnetic fields $B$ (s. cyclotron paths, Lorentz--force etc.) \cite{B5}.

Under certain conditions regarding the potential felt by the
electrons in the nanostructures (square--well like potentials
etc., and states with $k^{-1}> a$, interatomic distance) one gets the result (see Stampfli {\it et al.} using
extensions of the  Balian--Bloch type theory)
\begin{equation}
\Delta N_\sigma (E,n)\simeq \sum_l A_{l\sigma}(E,n) \cos
(kL_l+\phi_{l\sigma})\quad .
\end{equation}
Here, $l$ refers to closed orbits (polygons) of length $L$,
$k=\sqrt{\mid E\mid + i\delta}$, and $\phi_{l\sigma}$ denotes the
phase shift characterizing the scattering potential and the
geometry of the system. An external magnetic field affects $\Delta
N_\sigma(E,n)$ via path deformation and phase shifts resulting
from magnetic flux (see Aharonov--Bohm effect).

Clearly, the DOS in particular $\Delta N_\sigma (E,n)$ will be
affected characteristically by the magnetism in various
nanostructures, in quantum dot systems. For details of the
Balian--Bloch like analysis see the theory by Stampfli {\it et
al.} \cite{B5}.

Note, the electronic structure due to the interferences of the
dominant electronic states near the Fermi--energy described
approximately by Eq.(21) results in addition to the one due to the atomic
symmetry of the nanostructure (spherical one for clusters,
2d--symmetry for thin films, etc.). Thus, for example, one may
find for magnetic clusters a phase diagram temperature vs. cluster
size which dependent on magnetization exhibits phases where atomic
(shell) structure and then spin dependent electronic structure
dominates. Of course, for sufficiently large clusters these
structures disappear.

The interference of closed electron orbits in magnetic mesoscopic
systems like spherical clusters, discs, rings, thin films and
quantum dot lattices causes spin
dependently characteristic structure , oscillations in the DOS
$N(E,n)$ (yielding corresponding ones for example in the
cohesive energy, occupation of $d,s$ electron states,
Slater--Pauling curve for magnetism, magnetoresistance, etc.).
\begin{figure}
\centerline{\includegraphics[width=0.5\textwidth]{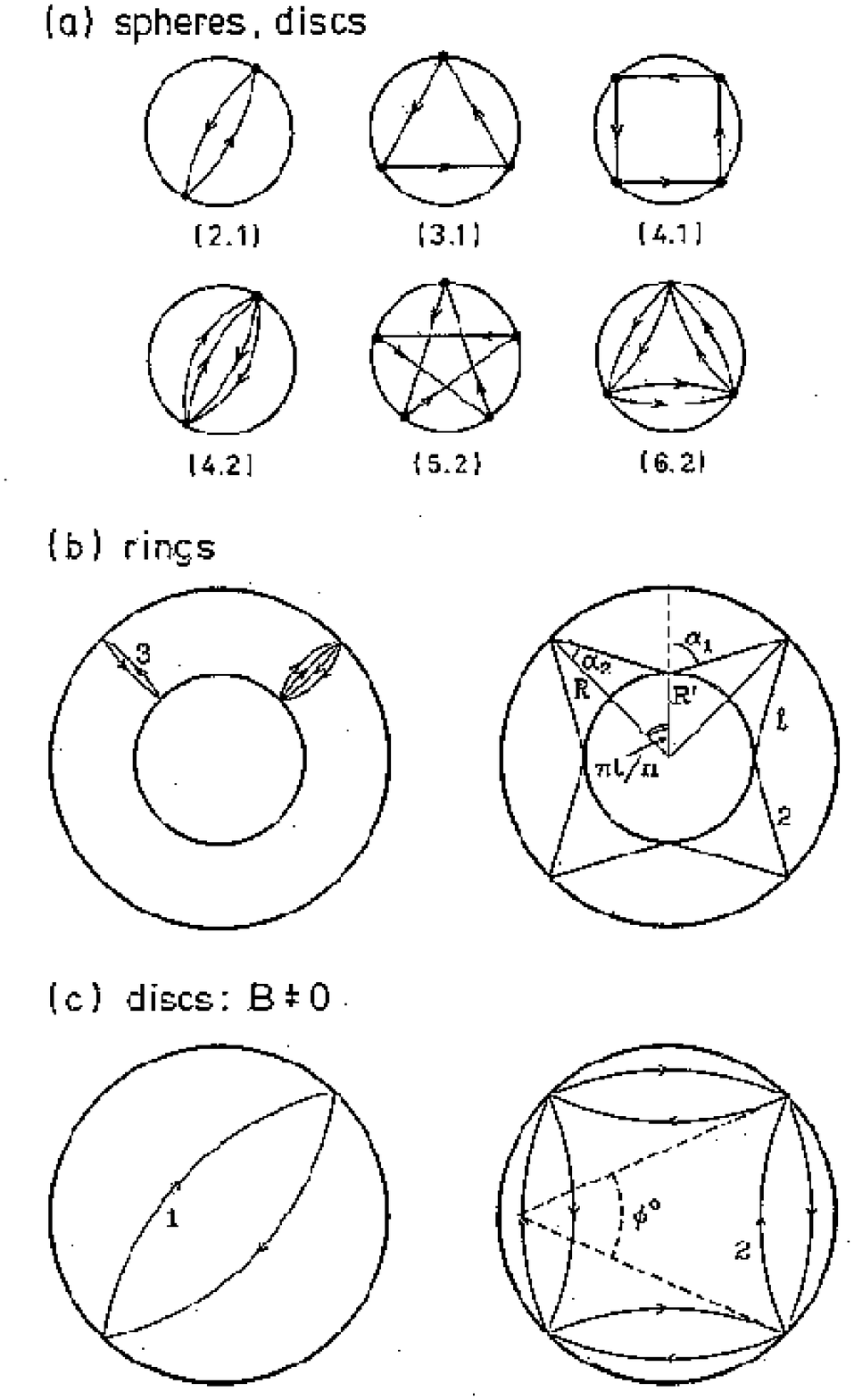}}
\caption{Examples of dominating electron paths characterized by (p,t) in (a) spheres, discs
and (b) rings. Note, $p$ and $t$ refer to the number of corners and t to
the one of circulations around the center, respectively, of the polygonial paths. In (c) the effect of an external magnetic field $B$ is shown. Note, in particular for rings magnetic flux quantization may occur. Also the phase shift due to the Aharanov--Bohm effect may induce currents. For ferromagnets these are spin polarized.}
\end{figure}

(a) Spherical Clusters:
The properties of small spherical clusters
follow from the DOS given approximately by
\begin{equation}
     \Delta N_\sigma = \Delta N_o +
     \Sigma_{t=1}^\infty\Sigma_{p=2t+1}^\infty A_{t,p,\sigma}\sin\phi_{t,p,\sigma},
\end{equation}
where the orbits are characterized by the number of corners $p$ and $t$
describes how many times the center of the sphere is circled.
Here,  $\Delta N_o$ results for $p=2t$ and turns out to be
unimportant. For a detailed derivation of the amplitudes
$A_{t,p,\sigma}$ and phases $\phi_{t,p,\sigma}$ see analysis by
Stampfli {\it et al.} \cite{B5}. One gets
\begin{equation}
       A_{t,p\sigma} = 2R^2a_{t,p\sigma}\sqrt{\sin^3(\pi t/p)/p}\exp-(2pk_2R\sin(\pi t/p)),
\end{equation}
with $a_{t,p}= \sqrt{k_1 R/\pi}(-1)^t \cos(\pi t/p)$. Note, the amplitudes decrease rapidly for
increasing number p of corners. The most important contributions
to the DOS result from triangular, square and lower polygon
orbits. The structure in the DOS affects many properties (conductivity, magnetoresistance, spin susceptibility $\chi$, spin dependent ionization potential, etc.).

(b) Circular Rings: Circular rings with outer radius $R_a$ and
inner radius $R_i$ are particularly interesting, since $R_i/R_a$
controls which orbits are most important. For example, one may
eliminate DOS contributions due to triangular orbits. One gets,
see Fig.13 for illustration,
\begin{equation}
    \Delta N = \Delta N_1 + \Delta N_2 + \Delta N_3  ,
\end{equation}
where $\Delta N_1$ results for orbits inside the ring (with $\pi t/ p < \arccos(R^{'}/R)$)
and which are also present for discs, $\Delta N_2$ due to orbits
scattered by the outer and inner surface of the ring and which
circle around the center, and $\Delta N_3$ results from
ping--pong like orbits between $R_a$ and $R_i$. In all cases one
gets rapid oscillations (due to exponentials) and slower ones
due to $\cos$ and $\sin$ functions.

Note, results for thin films follow for $R_a$, $R_i\longrightarrow\infty$.

(c) Quantum Dots: The DOS of quantum dots is calculated similarly.
Thus, one gets for example oscillations in the magnetoresistance
$r_\sigma$, since $r_\sigma \propto{N_\sigma^2(E,n)}$.

Molecular fields and external magnetic field $B$ will affect the
electronic orbits, see Fig.13 for illustration. For discs, rings and
quantum dots one gets phase shifts $\Delta\phi$ due to path deformation and the flux through the polygons. Note, the Lorentz
force deforms the orbits. Also the magnetic flux due to a vector
potential $\overrightarrow{A}$ perpendicular to the discs plane
for example gives a phase shift
\begin{equation}
     \Delta\phi^{'} = \oint \overrightarrow{A}\cdot d\overrightarrow{r}=
     \overrightarrow{B}\cdot\overrightarrow{S},
\end{equation}
where $S$ is the area enclosed by the orbit. Note, structure in DOS
due to Landau--levels may be included already in
$\overline{N(E,n)}$.

(d) Circular disc and quantum dots: For circular discs and quantum dots one gets (for square well like potentials) in Eq.(22) that $\Delta N_0 \approx 0$ and for the second term that mainly
orbits with p=2,3 are important, since $A_{t,p,\sigma} \sim
1/p^{1/2}$. Note, now $a_{t,p\sigma}= (\alpha_{t,p\sigma}/2)\sqrt{1/(\pi k_1Rp)}$ and
\begin{equation}
      A_{t,p} = \alpha_{t,p}R^2[(1/\pi k_1Rp)\sin^3(\pi t/p)]^{1/2}\exp-[2pk_2R\sin(\pi t/p)],
\end{equation}
with $ \alpha_{t,p}=2$ for $p\neq2t$ and $\alpha_{t,p}=1$ for p=2t and phases
$\Phi_{t,p}=p[2k_1R\sin(pt/p)- \pi/2+ \delta_{t,p}]+ 3\pi/4$ as for spheres. Phases $\delta_{t,p}$ result from (possibly spin dependent) potential scattering at the surface of the system, see books on quantum mechanics \cite{B5}, note $\delta_{t,p\sigma}=-\pi for U\rightarrow \infty$. The phase shift $\delta$ affects level spacing (and vise versa) and appearance of electronic shell structure.

Magnetic field B: The magnetic field is taken to be perpendicular to the disc plane. Then one gets a change of the phase by
\begin{equation}
          \Delta\Phi = \Delta\Phi_1 + \Delta\Phi_2,
\end{equation}
where $\Delta\Phi_1=pkR_c\Phi_c^0$, $\Phi_c^0=2\arcsin(R\sin(\Phi^0)/R_c)$, cyclotron orbit $R_c=k/B$ ($R_c\gg R$), is due to path deformation and
$\Delta\Phi_2=\Delta\Phi^{'}$, see previous Eq., due to the enclosed flux (Aharonov--Bohm).
The angle $\Phi_c^0$ refers to the center of the cyclotron orbit, see Fig.13. Straightforward analysis gives
\begin{equation}
        \Delta\Phi^{'}= \pm BS_{\pm}\quad ,
\end{equation}
with
\begin{equation}
           S_{+,-}= S_0 \pm \Delta\quad , \quad \Delta = (p/2)R_c^2(\phi_c^0 - \sin\phi_c^0)\quad .
\end{equation}
Here, $2\phi^0 = (2\pi/p)t$, $\phi^{0}_c$ is the angle seen from the center of the disc, see Fig., and for the cyclotron orbit one assumes $R\gg R$. $\pm$ refers to clockwise
and counterclockwise motion around the center of the orbit. $\Delta$ denotes
the area resulting from path deformation due to field B. It is then
\begin{equation}
    \Delta N_\sigma(E,B) \propto \Sigma_{t,p} A_{t,p\sigma}\cos(SB)(....)
                          = \Sigma_{t,p} A_{t,p\sigma}\cos(S_{0}B)\sin\Phi_{t,p}\exp{-(pk_2(k_1/B)\Phi_{c}^0)},
\end{equation}
where $\Phi_{t,p}$ denote phase changes due to path deformation, see previous expression. Factor $\cos(S_0B)$ describes Aharonov--Bohm effect plus interference effects and gives oscillations periodic in B. Note,
$\cos BS$ causes oscillations which are periodic in $B$, and
oscillations change with (a-d). For large field $B$,
$2R_c\leq(a-d)$, see Fig.4, path 1 and 3 disappears and only 2 remains (see
Landau level oscillations with periodicity (1/B), since
$S\propto(1/B^2)$). $\Delta N\propto\cos(SB)$, $\phi_2\propto SB$. The DOS oscillations are $\Delta N\propto \cos(SB)$, ($\Phi_2\propto (SB)$).

The ratio (d/a) characterizes strength of antidot potential scattering.
Of course, the oscillations in the DOS cause similar ones in the
magnetoresistance, for example.

Antidot lattice: One assumes a 2d--lattice of antidots scattering via a repulsive potential the electrons. The lattice of antidots is illustrated in Fig.4 and Fig.14. Note, orbits 1, 2, 3 are most important. The scattering by the antidot lattice causes phase shifts $\Phi_{t,p\sigma}$. Note, the area S enclosed by the orbits 1, 2 and 3 is nearly independent of magnetic field
$B$, such orbits occur for $2R_c>(a-d)$ and $R_c\propto 1/B$. One assumes that orbits are dephasing for distances much larger than lattice distance a.
The oscillations in the DOS are given by (performing calculations similarly as for disc, see similar paths in Fig.13 and Fig.14)
\begin{eqnarray}
     \Delta N_\sigma(E,B)&=& ( \sqrt{2}(a-d)/k_1\pi )
     \Sigma_{t=1}^\infty(\sinh\varphi/\sinh(4t+1)\varphi)^{1/2}\cos(BS)\sin\phi_{t,p}\nonumber \\ &\times &\exp(-4tk_2(k_1/B))\varphi_1 + \Delta N_r(E,B),
\end{eqnarray}
with phases $\phi_{t,p}=4t[k_1(k_1/B)\varphi_1 + B/2(k_1/B)^2(\varphi_1-\sin\varphi_1) + \delta_{t,4t}]+ \pi/2,
\cosh\varphi=2a/d -1$, $\varphi_1=2\arcsin[(a-d)/(k_1/B)/\sqrt{2}]$,
$S=2t(a-d)^2$, etc., see Stampfli {\it et al.} \cite{B5}. The first term is due to path 1 and $\Delta N_r(E,B)$ due to path 2, 3 etc..

Note, Eq.32 takes into account that path reflections result from small antidots (rather than from
spherical surfaces of discs with varying curvature). The oscillations of $\Delta N_\sigma$ periodic in $B$ result for orbits 1, 2, 3, 4 and change with (a-d). For very small field $B$, note $R_c\propto(1/B)$, orbits become unstable and the large orbits get irregular due to dephasing. For increasing magnetic field $B$ such that $2R_c\leq(a-d)$ first orbit 3 and then 1 disappears and only 2 survives. This orbit causes then for increasing field $B$ Landau level oscillations with period (1/B), since $S\propto R_c^2\propto (1/B)^2$, flux $\sim SB$ and $\Delta N\propto\cos(SB)$. The oscillation periods are expected to decrease for decreasing lattice constant a and increasing magnetic field $B$. \cite{B5}

The oscillations in DOS $\Delta N_\sigma(E,B)$ affect many properties, in particular the magnetoresistance. A crossover from periodicity proportional in B to one in (1/B) should occur for example for discs.

The theory by Stampfli {\it et al.} can be used also for a system of quantum dots on a lattice, including then
hopping between the dots. Furthermore, an ensemble of magnetic and superconducting quantum dots may exhibit interesting behavior regarding (quantisation effects) magnetic flux etc..

Note, in general for lower dimensions $(3\longrightarrow 2)$
degrees of freedom are reduced and one expects stronger
oscillations of the DOS, for example in a magnetic field. The
interference of the orbits yields generally rapid and slow
oscillations. The spin dependence is calculated from the spin
dependent scattering potential $U_\sigma$ at the surfaces of the
system.
\begin{figure}
\centerline{\includegraphics[width=0.5\textwidth]{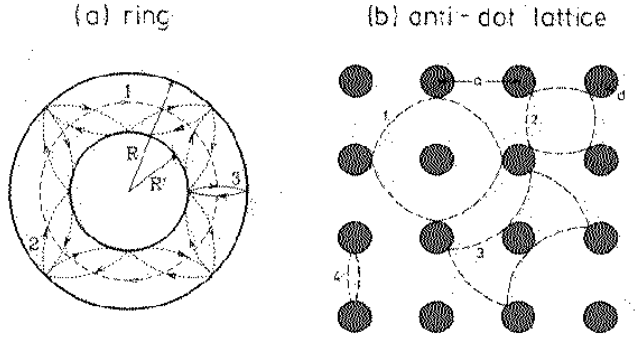}}
\caption{Effect of a magnetic field on various closed electron orbits,
paths: (a) for a ring, (b) for a lattice of anti--dots, which
scatter the electrons by a repulsive potential. Note, in particular for a narrow ring path 1 may dominate and magnetic flux quantization occurs. If the mean free path of the electrons is relatively large, as compared to the circumference of the ring, a magnetic field may induce a current driven by the phase shift due to the Aharonov--Bohm effect. On general grounds this current is expected to be spin polarized for ferromagnets and to decrease proportional to $R^{-2}$ and for increasing temperature.}
\end{figure}

Note, a ring may be viewed as an antidot enclosed by the metallic ring. The electrons are repelled by a strong repulsive potential from the inner core of the ring. If a magnetic field B is present, due to the flux quantization the electronic DOS
reflects this and has fine structure as a result of the flux quantum $\phi_0=2\pi\hbar c/e$. Also a beating pattern of the oscillations may reflect interference of several paths dominating the electronic structure, see Figs.. These oscillations are also present in the electronic energy $E=\int d\varepsilon \varepsilon N(\varepsilon) + ...$. As also discussed later the Aharonov--Bohm effect may induce a current in rings and discs etc. with interesting structure.

For further details of the analysis for rings and films see Stampfli {\it et al.} \cite{B5}. In particular these may exhibit interesting behavior at nonequilibrium ( due to hot electrons, for example ).

It is important to compare the results by Stampfli {\it et al.}
using the Balian-- Bloch type theory \cite{B5} with quantum mechanical
calculations of the DOS. Then
\begin{equation}
     N(E)_\sigma = \Sigma_m \delta(E-E_{m\sigma}),
\end{equation}
where the eigenvalues $E_{m\sigma}$ of the state $\mid m\sigma\rangle$ are
determined by the Schr\"{o}dinger equation.

\subsection{Magnetooptics}
Interesting magnetooptical behavior is exhibited by magnetic
films. Nonlinear optics, SHG is very surface sensitive and
reflects the magnetic properties of the film. SHG is generated at
the surface and at the interface surface/substrate or at the
interface of two films.

In Fig.15 SHG at surfaces is shown (light $\omega\rightarrow 2\omega$). The $2\omega$--light
is characterized by the response function $\chi_{ijl}(\overrightarrow{M},\omega)$. The Fig. illustrates the optical configuration.
Clearly, since the response depends on $\overrightarrow{M}$, note $M_\perp$ and $M_\parallel$
yield different $\chi_{ijl}(\overrightarrow{M})$, the reorientation transition of the magnetization can be studied optically.

\begin{figure}
\centerline{\includegraphics[width=0.65\textwidth]{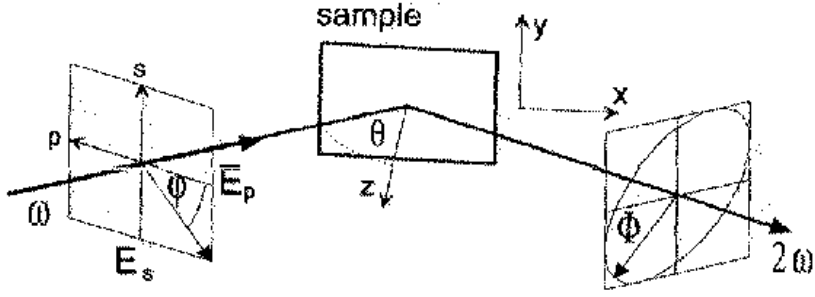}}
\caption{Optical configuration for incoming $\omega$ light and reflected $2\omega$ light. Sketch of polarization of incoming and outgoing light characterized by the angle $\varphi$ and $\Phi$. z is normal to the surface and the crystal axes x, y are in the surface plane. $\theta$ denotes the angle of incidence.}
\end{figure}

For thin films the nonlinear susceptibility, response function  $\chi^{2}$ may be
split into the contribution $\chi^s$ from the surface and $\chi^i$
from the interface. Then owing to the contribution $\chi^s \chi^i$
to the SHG intensity $I\propto |\chi|^2$ the relative phase of $\chi^s$ and $\chi^i$
is important. Furthermore, the magnetic contrast
\begin{equation}
    \Delta I(2\omega, M)\propto I(2\omega, M)- I(2\omega, -M)
\end{equation}
will reflect the film magnetism, since the susceptibility has contributions
which are even and odd in $M$, see Bennemann {\it et al.} in
Nonlinear Optics, Oxford University Press.\cite{B12}

Note, high resolution interference studies are needed to detect
also for example lateral magnetic domain structures of films. Also polarized light
reflects magnetism and in particular the magnetic reorientation
transition, see Fig.16 and following results.

In magnetooptics (MSHG) using different combinations
of polarization of the incoming and outgoing light, see Fig.15, one may analyze
\begin{equation}
     \chi_{ijl}(\overrightarrow{M}).
\end{equation}
Regarding the dependence on magnetization, for the susceptibility odd in magnetization one writes
\begin{equation}
   \chi^o_{ijl} = \chi_{ijlm}M_m + ...
\end{equation}
(Note, $\chi=\chi^e + \chi^o$). Thus, in particular $\chi_{ijl}(M_c)$ changes characteristically for
$c= x$, $y$, or $z$. For
example, for incoming s--polarized light and outgoing $p$--polarized
$2\omega$ light, see Fig. 15 for illustration, the nonlinear
susceptibility $\chi_{ijl}(M_y)$ dominates. Similarly
$\chi_{ijl}(M_z)$ dominates in case of outgoing s--SHG
polarization. For general analysis see H\"{u}bner {\it et al.},
Nonlinear Optics in Metals, Oxford Univ- Press.\cite{B12}

Regarding susceptibility $\chi^e$ and $\chi^o$ note further interesting behavior could result from terms which are higher order in the magnetization M ($\chi^e=\chi_0^e + aM^2+...$, etc.) and $\chi_{ijl}$ reflects also the strength of the spin--orbit interaction. As discussed by H\"{u}bner {\it et al.} in case of

(a) $\overrightarrow{M} \parallel \overrightarrow{x}$ (longitudinal configuration), the tensor $\chi_{ijl}$ involves $\chi_{yxx}$, $\chi_{yyy}$, $\chi_{yzz}$, $\chi_{zxx}$, $\chi_{zyy}$, $\chi_{zzz}$, $\ldots$
and in case of

(b) $\overrightarrow{M} \parallel \overrightarrow{z}$ (polar configuration, optical plane $x,z$) elements
$\chi_{zxx}$, $\chi_{zzz}$, $\chi_{xyz}$ and $\chi_{xzx}$
of susceptibility $\chi_{ijl}$ occur.

For longitudinal configuration and polarization combination $s\rightarrow p$ $\chi_{zyy}$ and for $p\rightarrow s$ $\chi_{yxx}$, $\chi_{yzz}$ are involved. Furthermore, in case of polar configuration and polarization combinations
$s\rightarrow p$  element $\chi_{zxx}$ and for $p\rightarrow s$ the element  $\chi_{xyz}$ occur. This demonstrates clearly that Nolimoke as well as Moke can observe the magnetic reorientation transition and other interesting magnetic properties of nanostructures.

In Fig.16 results by H\"{u}bner {\it et al.} for the polarization dependence of SHG--light are shown. Note, agreement with experiments is very good. Obviously magnetism is clearly reflected.
Similar behavior regarding polarization dependence is expected for the magnetic
transition metals Cr, Fe, and Co. Due to $\langle d\mid z \mid d\rangle \approx  \langle d\mid x\mid d\rangle$ for the dipole transition matrix elements one gets typically same results for s-- and p--polarized light. To understand the behavior of Cu, note $\langle s\mid z\mid s\rangle \approx \langle s\mid x\mid s\rangle \approx 0$. Curves (a) and (b) refer to wavelenghts exciting and not exciting the Cu $d$--electrons, respectively.
Generally as physically expected $I_s(p-SH)_{NM}\longrightarrow_\omega I_s(p-SH)_{TM}$. Again for a detailed discussion
\begin{figure}
\centerline{\includegraphics[width=0.5\textwidth]{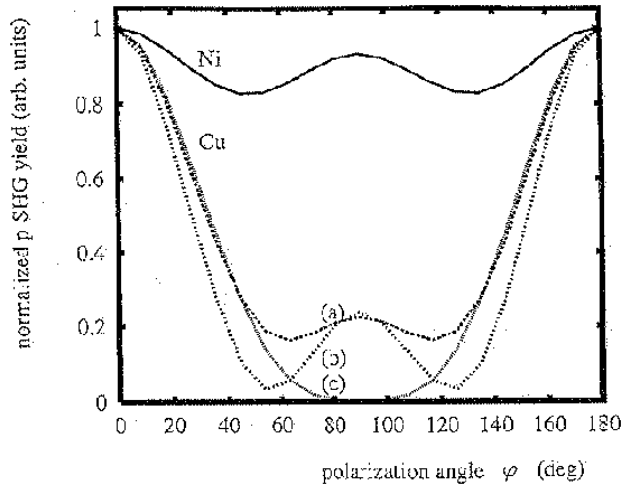}}
\caption{Polarization dependence of SHG. Note, the difference between magnetic Ni and nonmagnetic Cu. The angle $\varphi$ refers to the incoming light polarization. Curves (a), (b), (c) for Cu result from using different input parameter for the diffraction index $n(\omega)$ and for $k(\omega)$, see calculations by H\"{u}bner {\it et al.}.}
\end{figure}
of the interesting polarization dependence see magnetooptics and discussion by H\"{u}bner {\it et al.}.

\begin{figure}
\centerline{\includegraphics[width=0.5\textwidth]{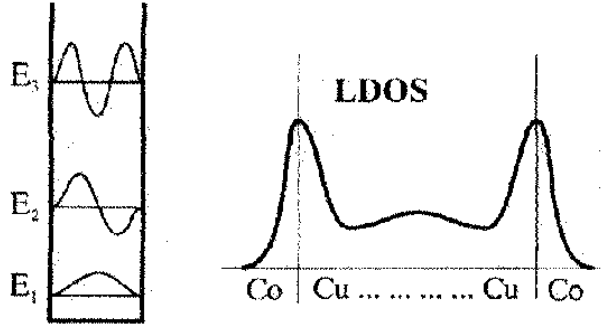}}
\caption{Illustration of quantum well states (QWS) due to confinement in
a thin film of thickness d. Such states occur for example in a Co/10MLCu/Co film
system. Note, the DOS of the QWS is strongly located at the
interface. In magnetic films one gets generally spin split QWS. The parity of the QWS changes for increasing energy.}
\end{figure}

In particular, interesting magnetooptical behavior of SHG and Moke results also
from the spin--dependent
quantum--well states (Q.W.S.) occurring in thin films. These
states result from the square--well potential representing the confinement of the electrons in a
thin film. In magnetic films the resulting electron states are of course
spin--split. Characteristic magnetic properties follow. In
contrast to band like--states for the electrons in the film only Q.W.S. show a strong dependence on
film thickness. Thus, to study film thickness dependent (optical)
behavior SHG involving Q.W.S. needs be studied.

One expects characteristic behavior of the optical response, its
magnitude and dependence on light frequency $\omega$ and
$\overrightarrow{M}$. Characteristic oscillations in the MSHG
signal occur, since QWS involved in resonantly enhanced SHG occur
periodically upon increasing film thickness, see Fig. for
illustration (note, SHG involving transitions $i\rightarrow j
\rightarrow l$ and with one of these states being a QWS).

Clearly, the Q.W.S. energies shift with varying film thickness.
Then  SHG light intensity ($I_{2\omega}$) involving these states
may oscillate as a function of film thickness and this in particular may reflect the magnetism of
the film. Clearly, owing to the periodic appearance of Q.W.S. at
certain energies for increasing film thickness, SHG involving
these states may be resonantly enhanced and then oscillations
occur as a function of film thickness.

A detailed analysis, see calculation of QWS by Luce, Bennemann,
shows that the SHG periods depend on the parity of the Q.W.S., the
position of the Q.W.S. within the Fermi--see or above, and on the
interference of second harmonic light from the surface and
interface film/substrate, etc. If this interference is important,
then SHG response is sensitive to the parity of Q.W.S., light
phase shift at the interface and inversion symmetry of the film.
If this interference is not important, then SHG response and
oscillations are different, see Fig.11 for illustration of the interference.

Thus, for an analysis of SHG involving QWS one may study:

(a) $\chi^i \simeq \chi^s$, when interference is important.
Different behavior of SHG may result then if 1. QWS is involved as final state, 2. as intermediate state, or 3. both QWS as intermediate and final state matter. Note, for $\chi^s \chi^i\rightarrow (-1)$, for example due to inversion symmetric films or a phase shift $\pi$ at the interface, the destructive interference yields no SHG--signal. Also if a final QWS near the Fermi--energy (and which may set the period of SHG--oscillation) has even parity, then in contrast to a QWS with odd parity no SHG signal occurs, since for the latter the product of the three dipole matrix elements is small.

(b) $\chi^i \neq \chi^s$, and interference is unimportant.
Then different oscillation periods may occur. For example if a QWS above the Fermi--energy becomes available at film thickness $d_1$ for SHG then a first peak in SHG appears and then at film thickness $2d_1$ again at which previous situation is repeated, and so on. Of course, strength of signal depends on wavevector k in the BZ, DOS and frequency $\omega$ must fit optical transition. In case of a FM film then the resonantly enhanced SHG transitions are spin dependent and an enhanced magnetic contrast $\Delta I$ may occur.

If occupied QWS below the Fermi--energy are involved, then also oscillations occur, in particular due to DOS, see transition metals. If QWS below and above the Fermi--energy cause oscillations then the period of SHG may result from a superposition, as an example see the behavior of $xCu/Co/Cu(001)$ films.

Characteristic properties of film SHG are listed in table 1.
Various properties are listed for the case of no interference
($\chi^i\gg \chi^s$) and strong interference ($\chi^i\approx
\chi^s$) of light from surface and interface. The first case is
expected for a film system xCu/Fe/Cu(001), for example, since the
interface Cu/Fe dominates due to the QWS of the Cu film and the
large DOS of Fe near Fermi energy $\varepsilon_F$ (see SHG
transitions: $Cu\rightarrow_\omega Fe \rightarrow_\omega QWS
\rightarrow_{2\omega}Cu$). Thus, $\chi^i\gg \chi^s$ and due to
the QWS above $\varepsilon_F$ one gets two and more SHG
oscillation periods.

The case $\chi^i \approx \chi^s$ is expected for a film
xAu/Co(0001)/Au(111), for example, since no Co $d$--states (see
band structure of Co) are available as intermediate states for
SHG transitions and the QWS in Au just below the Fermi--energy
controls the SHG contribution, see Fig.18 for illustration. The
\begin{figure}
\centerline{\includegraphics[width=0.5\textwidth]{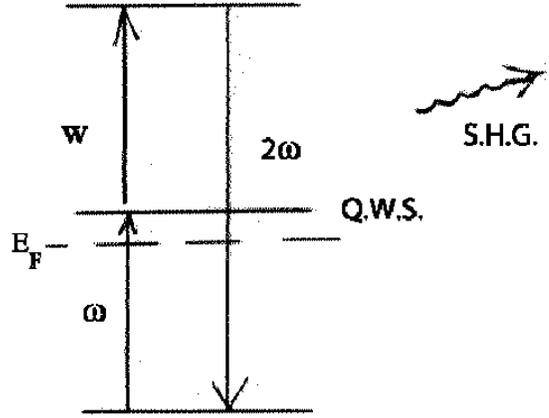}}
\caption{Illustration of electronic SHG transitions involving a QWS determining the
oscillation period. If the QWS has even parity then resulting SHG
is small. Only QWS with odd parity cause due to larger dipole matrix
elements larger SHG and oscillations. In FM--films the QWS is spin split. Note, parity of QWS changes for increasing quantum number.}
\end{figure}
parity of the QWS causes then oscillations $\Lambda = 2\Lambda_M$
($M$ refers to Moke), since clearly the 3 dipole (p) matrix elements
$\langle d\mid p\mid QWS\rangle\langle QWS\mid p\mid d\rangle\langle d\mid p\mid d\rangle$
are much smaller if QWS has even symmetry (note, $p$ is odd and $d$
is odd).

The wave vector $k$ dependence is controlled by the Brillouin zone
structure (BZ). For a detailed discussion see Nonlinear Optics
in Metals, Oxford University Press. \cite{B12}

The interesting S.H.G. (second harmonic light) interference
resulting from surface s and interfaces i and reflecting
sensitively magnetic properties of the film may be analyzed as
follows. The SHG light intensity $I(2\omega)$ is approximately
given by
\begin{equation}
I(2\omega)\sim \mid \chi_{ijl}\mid^2\quad ,
\end{equation}
where $\chi$ denotes the nonlinear susceptibility. Note, $\chi$
may be split as (s:surface, i:interface)
\begin{equation}
\chi_{ijl}(2\omega)=\chi_{ijl}^s + \chi_{ijl}^i \quad .
\end{equation}
One gets
\begin{equation}
I(2\omega) \sim 2\mid \chi_{ijl}^s\mid^2 + 2 \chi_{ijl}^s
\chi_{ijl}^i + \ldots.
\end{equation}
Obviously the intensity $I(2\omega)$ depends on the resultant
phase of the susceptibilities $\chi^s$ and $\chi^i$.

Then assuming, for example, that $\chi^s$ and $\chi^i$ are of
nearly equal weight $(\mid\chi^s\mid\sim\mid\chi^i \mid)$, it may
happen that the 2.term in Eq.(38) cancels the first one, see for
example inversion symmetry in films ($\chi^s_{ijl}\rightarrow
-\chi^i_{ijl}$) or phase shift of the light by $\pi$ at the
interface.

If the interference of light from the surface and interface is
negligible then different oscillations of the outcoming SHG light
as a function of film thickness occur.

The weight of the optical transitions $i\rightarrow j\rightarrow l$ changes as the film thickness increases, see later
discussion. Thus, Moke and Nolimoke oscillations occur.
\begin{table}
\caption{Characteristics of SHG response from thin films. Its dependence on the film thickness involves QWS. The SHG oscillations reflect magnetic properties of the film.}
\begin{center}
    \begin{tabular}{ | l | p{5cm} | p{5cm} |}
    \hline & & \\
       & \quad\quad\quad\quad$\mid \chi^{i}\mid\gg\mid \chi^{\delta}\mid$ &   \quad\quad\quad\quad $\mid\chi^{i}\mid=\mid \chi^{\delta}\mid$ \\
       \hline\hline
    $k$ selectivity & ($x$ Cu/Fe/Cu(001) for example)  &  \\ \hline
     & Strongmanetic signal due to strong (magnetic) interface contributions & Weak SHG signal, from only few $k$ points and without strong interface contributions \\ \hline
     & Sharp SHG peaks due to few contributing $k$ points resulting in strong resonances & Doubled period and additional periods are frequency dependent \\
    \hline
     & Strong frequency dependence of the SHG oscillation of the QWS in the $k_{\perp}$ direction & MOKE period absent; doubled and additional SHG period visible \\
    \hline
     & MOKE period and larger periods visible; no exact doubling of the MOKE period & \\
    \hline\hline
    No $k$ selectivity &  & ($x$ Au/Co(0001)/Au(111) for example \\
    \hline
     & Strong magnetic signal, since strong interface contribution Broad SHG peaks, since contributions come from many $k$ points & Smaller magnetic contribution, since interface and (nonmagnetic) surface contributions are of the same magnitude \\
    \hline
     & Weak frequency dependence of the oscillation period & Broad, smooth peaks, since interference effects do not change magnitude \\
    \hline
     & MOKE period and larger periods present & SHG oscillation periods rather independent of the frequency, since the SHG signal is caused by the QWS near $E_F$ \\
    \hline
     &  & MOKE period absent, doubled period present \\
    \hline
    \end{tabular}
\end{center}
\end{table}


Regarding optical properties, the morphology of the thin film and
its magnetic domain structure should play a role in general.

Film multilayers: For magnetic film multilayers one expects interesting
interferences and magnetic optical behavior. Assuming for example
the structure shown in Fig.19, neglecting for simplicity QWS, one
gets for SHG
\begin{equation}
   I(2\omega) \sim | \chi_s (\overrightarrow{M})+ \chi_{int.1}+
   \chi_{int.2}+ ...|^2.
\end{equation}.
Writing $\chi = \chi^e + \chi^o, \chi^o \sim M $, one has
\begin{equation}
   I(2\omega) \sim | \chi_s^e + \chi_s^o(M) +
   \Sigma_i\chi_{int.i}^e + ... |^2
\end{equation}
for the af structure. In the case of ferromagnetically aligned
films it is
\begin{equation}
   I(2\omega) \sim | \chi_s +
   \Sigma_i\chi_{int.i}(\overrightarrow{M})|^2.
\end{equation}
\begin{figure}
\centerline{\includegraphics[width=0.35\textwidth]{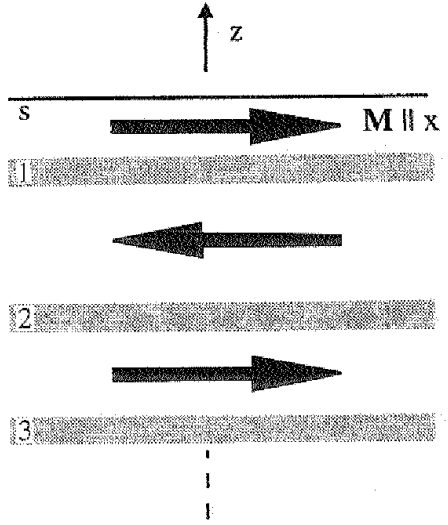}}
\caption{Multifilm system with antiferromagnetically ordered neighboring
films. The shaded areas indicate interface regions important for SHG. Note, SHG results also at the surface. Typically approximately two atomic layers will contribute. The magnetization is assumed to be parallel to the surface which normal is in z--direction.}
\end{figure}

Note, to detect magnetic domain structure experiments must achieve high lateral resolution, via interferences for example, etc. Of course surfaces with a mixture of domains with magnetization $M\bot$ and $M\parallel$ might yield a behavior as observed for the magnetic reorientation transition.

\subsection{Magnetization Dynamics}

The time dependence of the magnetization is of great interest.
Generally due to reduced dimension dynamics in nanostructures
could be faster than in bulk. For example switching of the
magnetization in thin films is expected to be faster. Of course,
this depends on film thickness, magnetic anisotropy and molecular
fields of neighboring magnetic films in a multilayer structure.

To speed up a reversal of the magnetization
$(\overrightarrow{M}\leftrightarrow - \overrightarrow{M})$ within
ps-- times to fs times, one may use thin magnetic films for
example with many excited electrons, hot electrons, resulting from
light irradiation. In heterogeneous magnetic film structures
transfer of angular momentum may be fast so that a fairly fast
switching of the topmost film magnetization may occur.

The magnetic dynamics is described by (see Landau--Lifshitz
equation)
\begin{equation}
d\overrightarrow{M}_i/dt = -
\left(\frac{g\mu_B}{\hbar}\right)\overrightarrow{M_i}\times\overrightarrow{H}_{eff}
+ \frac{G}{M_s^2}
\overrightarrow{M_i}\times(\overrightarrow{M_i}\times
\overrightarrow{H}_{eff}).
\end{equation}
Here, $i$ refers to a film or a magnetic domain in a nano
structured film and $\overrightarrow{H}_{eff}$ to a molecular
field acting on $\overrightarrow{M_i}$ and resulting for example
from neighboring films or magnetic domains. The first term
describes precessional motion and the second one relaxation. $M_s$
is the saturation magnetization (Note, using
$\overrightarrow{M_i}\times\overrightarrow{H}_{eff}\rightarrow
\partial\overrightarrow{M_i}/\partial t$ in the 2. term on the r.h.s. of Eq.(42) yields the Gilbert
equation with damping parameter $G$ describing spin dissipation).

One gets from Eq.(38) (and also from Boltzmann type theory) that the magnetization in thin films changes during times of the order of (controlled by angular--momentum conservation)

\begin{equation}
\frac{1}{\tau_M}\propto A(T_{el})\mid V_{\uparrow\downarrow}\mid^2
\overline{N}\overline{N} + \ldots
\end{equation}
Here, in case of excited electrons $T_{el}$ may refer to the
temperature of the hot electrons and $V_{\uparrow\downarrow}$ to
the spin flip scattering potential causing changes in the
magnetization (for example: spin--orbit scattering or exchange
interactions) and $\overline{N}$ is the average DOS of the
electrons. \cite{B13}

Note, in case of transition metals with many hot electrons Eq.(35)
may yield response times $\tau_M$ of the order of 100 fs. Clearly,
raising the temperature, $T\rightarrow T_{el}$ due to hot
electrons, can speed up the magnetic dynamics.

The Landau--Lifshitz (LL) equation is generally important
for spin dynamics. One may use the spin continuity equation
($\partial_tM_i+\partial_\mu j_{i\mu,\sigma}=0$) to determine the
spin currents $\overrightarrow{j}_\sigma(t)$ (including spin
Josephson one) induced by magnetization dynamics and then
describe the latter by using the $LL$--equation \cite{B3}.

As will be discussed one gets for magnetic tunnel junctions  (presumably best if $\tau_s >\tau_t$) spin current
driven by phase difference between magnets of the form
\begin{equation}
                \overrightarrow{j_s} \propto \partial\overrightarrow{M}/\partial t \sim \sin (\Delta\Phi).
\end{equation}
This follows from the LL--equation and the spin continuity equation (see nonequilibrium magnetism).

\subsection{Theory for Magnetic Films during Growth : Nonequilibrium Magnetic Domain Structure}

The magnetic structure of thin films can be controlled by film
growth conditions, since magnetism depends on the atomic structure
of the film. Obviously this is of great interest for engineering
the magnetic properties of films \cite{B8,B11}.

During growth of the film one has a film structure changing in
time. This changes the magnetization. Thus, also the film
magnetization changes in time (note, magnetic relaxation
processes may occur somewhat time delayed). Then
\begin{equation}
\overrightarrow{M}(\overrightarrow{r},t,\{\mbox{atomic structure
(t)}\})\quad
\end{equation}
describes the nonequilibrium magnetization of growing films. Clearly, the magnetic structure (domain structure) changes in time
$t$ as the atomic structure changes during growth of the film. For
simplicity one may use a kinetic MC simulation (MC: Monte Carlo
method) for molecular beam epitaxial film growth (MBE) to calculate
the film structure and the accompanying magnetic one.. The
nonequilibrium behavior reflects sensitively magnetic properties
of the film, relaxation processes, range of magnetic forces and magnetic domain structure.

The atomic structure of the growing film is determined from a
particular film growth model. For example, in the Eden type growth
model the atoms are randomly deposited (to sites $i$) with
probability
\begin{equation}
p_i=\propto e^{-E_i/kT}\quad ,
\end{equation}
where $E_i$ is the atomic binding energy depending on the local
coordination number $z$ ($E_i\approx -A\sqrt{z_i}$). For illustration see Fig.9.

Different film growth is obtained by using layer dependent parameters $A$.
One may also take into account diffusion of the deposited atoms.
This yields then irregular film structures during growth with
varying clusters of deposited atoms (islands of different sizes
and shapes).

The resulting magnetic structure consists typically of an ensemble of
magnetic domains. The magnetic domains may not be at equilibrium
and will respond to the in time changing atomic structure of the
growing film. This may occur time delayed. Thus, successively
atomic structure and magnetic structure change.

Magnetic relaxation flipping the domain magnetizations is
described by using a Markov equation and Arrhenius ansatz for the
magnetic transition rates.

In summary, for each obtained atomic structure of the growing film
one calculates the corresponding magnetic (domain) structure.
Using the Heisenberg type hamiltonian including anisotropy it is
obvious how by changing the atomic coordination the exchange
interactions change and thus magnetism. For simplicity one uses
first Eq.(16). For details of the analysis
see Brinzanik, Jensen {\it et al.} \cite{B11}.

Spin relaxation assumes coherent rotation of the spins of each
atomic cluster (deposited group of atoms). The magnetization
(one may use for simplicity an Ising model) of such atomic clusters is directed along the easy
axis. Magnetic relaxation is (using Arrhenius type rates for
transitions $\overrightarrow{M_i}\rightarrow
-\overrightarrow{M_i}$) given by
\begin{equation}
\Gamma_i = \Gamma_0 e^{-E_i^b/kT} \quad .
\end{equation}
Here, $E_i^b$ gives the energy barrier against switching of the
domain magnetization. Note, if no energy barriers are present then
one may use the Metropolis algorithm. The magnetic barrier energy
of a cluster controlling the domain relaxation is for example
given by
\begin{equation}\label{eq26}
    E_i = N_i K_i (T) \cos^2\varphi - g\mu_B H_i S_i \cos\varphi + ...
    \quad .
\end{equation}
The first term is due to magnetic anisotropy and the second due to
the field $H_i$ (exchange interaction, external magnetic field,
dipolar field) acting on spin $S_i$.

Generally an energy barrier $E_i^b$ is present for the two
magnetic orientations of magnetic cluster $i$. One applies then as mentioned already a
kinetic $MC$ simulation during film growth to obtain the magnetic
structure corresponding to a given atomic structure at time $t$.
Correlated relaxation of neighboring magnetic domains must be
taken into account. Note, the time for each calculational step is
set by the spin precession frequency $\Gamma_0\approx 10^9 -
10^{12}sec^{-1}$.

A successful analysis has been derived by Brinzanik and Jensen
{\it et al.} \cite{B11}. In their studies first the atomic structure of the
film at time t is calculated using a Eden type growth model. Then
the corresponding magnetic structure is determined performing a
kinetic MC simulation and using Markov equation and for the system
of magnetic domains (interacting via exchange and dipolar
coupling) the energy
\begin{equation}
   E \simeq -(1/4)\Sigma_{i,j}\gamma_{ij}L_{ij}\overrightarrow{S_i}\cdot\overrightarrow{S_j}- \Sigma_i K_iN_i(S_i^z)^2 + \Delta E .
\end{equation}
Here, $\gamma_{ij}$ is the domain wall energy, $L_{ij}$ the
island surface area, and $N_i$ the number of island atoms. The 2.
term describes anisotropy. And
\begin{eqnarray}
    \Delta E = &\Sigma_{i>j}&(\mu_i\mu_j/r_{ij}^3)\;[\overrightarrow{S_i}\cdot\overrightarrow{S_j} - 3(\overrightarrow{r_{ij}}\cdot\overrightarrow{S_j})(\overrightarrow{r_{ij}}\cdot\overrightarrow{S_j})/r_{ij}^2
    ]\nonumber \\ &-& \Sigma_i \mu_i\overrightarrow{B}\cdot\overrightarrow{S_i}\quad .
\end{eqnarray}
Here, for simplicity islands with $N_i$ atoms and aligned magnetic moments are treated as particles with magnetic moments $\mu_i(T)$. The film magnetization results from averaging over the domain magnetization.

Typically one gets first for the given nanostructured film
magnetism which is not at equilibrium. A non--saturated
magnetization is typically obtained. As time progresses magnetic relaxation processes,
magnetization reversals of domains, for example, occur. These
change the magnetization of the film. Thus, finally equilibrium
magnetization is obtained. For growing film thickness one might
get an uniformly magnetized film, for details see analysis by
Jensen {\it et al.} \cite{B8,B11}.

\subsection{Tunnel Junctions: Spin Currents}

Tunnel junctions ($N_1\mid N_2\mid N_3$) are interesting
nanostructures regarding (ultrafast) switching effects, interplay
of magnetism and superconductivity, and in general quantum
mechanical interference effects. Note, $N_i$ may refer to material
which is ferromagnetic or superconducting, for example. The
situation is illustrated in Fig. 20. On general grounds one may
get spin currents driven by phase difference between two magnetic
systems, ferromagnets or antiferromagnets (which order parameter
is also characterized by a phase) \cite{B3}.

Interesting tunnel junctions are shown in Fig.(20). In Fig. 20(a)
is shown a two quantum dot system $n_{i\sigma}$ with energy
$\epsilon_{i\sigma}$ controlling the tunnel current
$j_\sigma^{i\sigma}$. The position of the energies $\epsilon_i$
may depend on occupation (see Hubbard like
hamiltonian: $\varepsilon_{i\sigma}=\varepsilon_i^o +
Un_{i\bar{\sigma}}+\ldots$). Note, $n_{i\sigma}$ and thus the current
$j_\sigma$ can be manipulated optically.

The tunnel system shown in Fig.(20b) can be used as a sensor for
triplet superconductivity. Then the configuration of the angular
momentum $\overrightarrow{d}$ of the triplet Cooper pairs and of
the magnetization $\overrightarrow{M}$ of the tunnel medium control characteristically
the tunnel current. The Josephson current exhibits interesting
behavior, for example upon rotating the magnetization relative to
the angular momentum of the Cooper pairs.

In Fig.(20c) is illustrated how tunneling can be used (with the
help of a bias voltage) to control magnetoresistance and to
determine the magnetization of a ferromagnet (see Takahashi {\it
et al.}) \cite{B3}. Note, upon changing ($\uparrow |N_2| \uparrow$) $\rightarrow$ ($\uparrow |N_2|\downarrow$) the
magnetoresistance increases, see GMR effect, and accumulation of spin polarized electrons occurs in $N_2$.
This has been discussed by Takahashi {\it et al.} \cite{B3}. Furthermore, if $N_2$ becomes superconducting a competition between superconductivity and magnetization occurs for configuration ($\uparrow |N_2|\downarrow$) as a result of accumulating nonequilibrium spin density in $N_2$. In Fig.20(b) and Fig.20(c) we assume $\tau_s> \tau_t$, where $\tau_s$ and $\tau_t$ refer to the spin diffusion and electron tunneling time, respectively. Electrons keep spin while tunneling.

For tunnel configuration $(FM_1|TSC|FM_2)$ one expects similar interesting tunneling behavior, for example regarding dependence on relative orientation of $\overrightarrow{M_i}$ and angular momentum $\overrightarrow{d}$ of the triplett Cooper pairs as for junctions $(TSC/FM/TSC)$, see Fig.20(b).

Furthermore, the current $j_s^J$ expected for $\tau_s>\tau_t$ depends on relative magnetization of the two magnets, and on the state of $N_2$ (normal, superconducting singulet or triplet).

Spin Currents :
Using the continuity equation one can find the
relationship between spin currents and magnetization dynamics for
the magnetic tunnel junction illustrated in Fig.(12). One has
\begin{equation}\label{eq27}
    \partial_t M_i + \partial_\mu j_{i\mu,\sigma}=0.
\end{equation}
This may give under certain assumptions straightforwardly the
connection between magnetization and spin polarized electron
currents (induced by hot electrons, temperature gradients or
external fields). Note, the magnetic dynamics (characterized by
$\frac{dM_i}{dt}$) may be described by the LL--equation. This yields generally a
spin Josephson current between magnets.

Also according to Kirchhoff the emissivity (e) of the junction is
related to its (time dependent) magnetization, magnetic
resistance. It is
\begin{equation}\label{eq28}
    \frac{\Delta e}{e}\simeq a \; (GMR) \quad ,
\end{equation}
where (GMR) denotes the giant magnetoresistance resulting for the
junction if the configuration ($\uparrow\mid N_2\mid\uparrow$)
changes to ($\uparrow \mid N_2\mid \downarrow $). This
changes the emissivity to $\Delta e$.
\begin{figure}[!htbp]
\centerline{\includegraphics[width=0.48\textwidth]{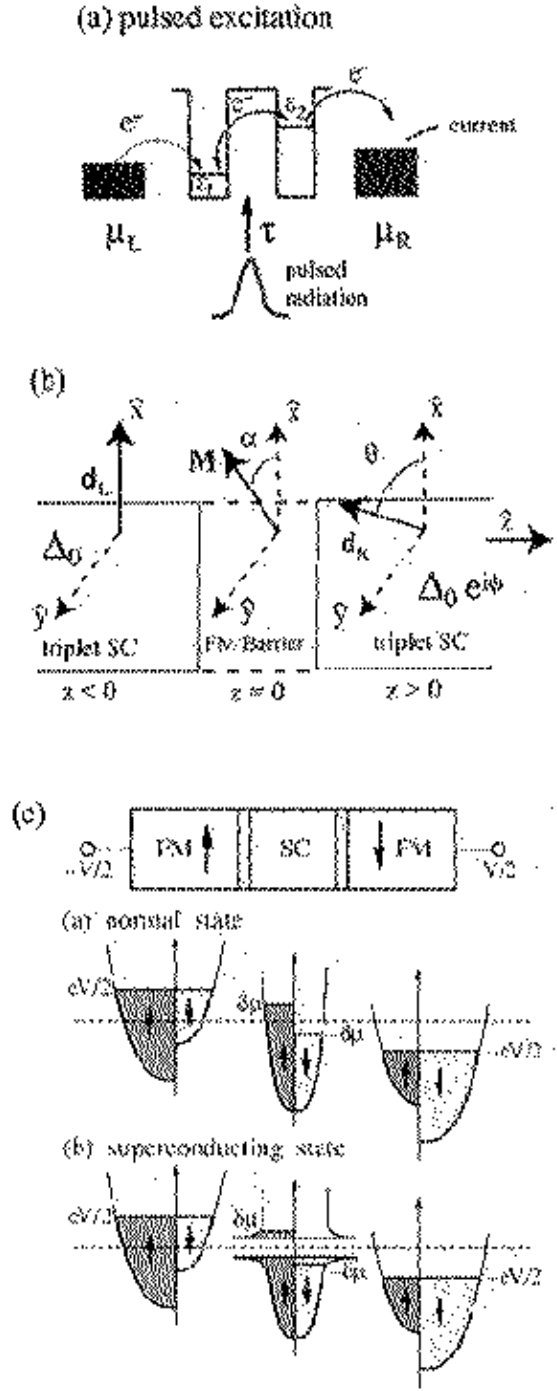}}
\caption{Illustration of various tunnel junctions. (a) A two
quantum dot system with spin--dependent levels
$\epsilon_{i\sigma}(n_{i\sigma},t)$ is shown (for example
$\varepsilon_{i\sigma}=\varepsilon_i^o + U_in_{i\bar{\sigma}}$). Two
levels are separated by an energy barrier which can be overcome
using an external electric field. (b) A triplet superconducting
Josephson junction ($SC_1\mid FM\mid SC_3$) is sketched. Here,
$\protect\overrightarrow{d}$ refers to the angular momentum of the Cooper
pairs which order parameter has the phase $\phi$. (c) A junction
\protect{($FM_1\mid SC\mid FM_2$)} is illustrated. An external electrical
field (potential $V$) shifts the electronic energy levels (bands).
This may control the tunnel current (magnetoresistance), see Takahashi {\it et al.}. 
Of particular interest is the interplay of spin current and superconductivity.}
\end{figure}

Viewing the phase of the magnetic order parameter $\phi_i$
(magnetization $M_i$)
\begin{equation}
\overline{M}=\mid M_i\mid e^{i\phi_i}
\end{equation}
similarly as the phase of the S.C. state, one gets for a junction
$(FM\mid N\mid FM)$ a Josephson like current $j^J$ driven by the
phase difference of the spin polarizations on both sides of the
tunnel junction. Using the continuity equation for the spins,
integrating using Gauß integral, and the Landau--Lifshitz
equation, one gets for the spin current
\begin{equation}
   j_\sigma =  j_\sigma^1(V) + j^J ,    j^J \propto dM/dt \propto \overrightarrow{M}_L\times\overrightarrow{M}_R+\ldots
       \propto |M_L||M_R|\sin( \phi_L - \phi_R +...).
\end{equation}
Here, $L$ and $R$ refer to the left and right side of the tunnel
junction, respectively. $j_\sigma^1(V)$ refers to the spin current
due to the potential $V$ and may result from the spin dependent DOS.
For details of the derivation of the spin Josephson current see
Nogueira {\it et al.} \cite{B3}.

Note, using the general formula $j= \frac{\partial F}{\partial \phi}$,
where $F$ is the free--energy, one gets
\begin{equation}
   j = -(e/\hbar)\Sigma_i \partial E_i/\partial\phi \tanh(E_i/kT).
\end{equation}
Then using
$E_i\propto J_{eff}\overrightarrow{S}_L\cdot\overrightarrow{S}_R +\ldots,$
one gets also
\begin{equation}
    j_\sigma^J \sim J_{eff} \sin(\phi_L - \phi_R)+\ldots.
\end{equation}
Here, $E_i$ gives the energy difference between opposite directions of the
magnetization (molecular field).

Of course, such a Josephson like spin current is expected on
general grounds, since $\phi$ and $S^z$ are canonical conjugate
variables and ( approximately )
\begin{equation}
        [\phi, S^z] = i .
\end{equation}
Note, this holds for the Heisenberg hamiltonian as well as for
itinerant magnetism described by the Hubbard hamiltonian, for
example. The commutator relationship suggests in analogy to the
BCS--theory to derive the spin Josephson current from the
hamiltonian
\begin{equation}\label{eq30}
    H=-E_JS^2\cos(\phi_L-\phi_R)+\frac{\mu_B^2}{2C_s}(S_L^z-S_R^z)^2+\ldots
    \quad ,
\end{equation}
where again $L,R$ refer to the left and right side of the junction
and $C_s$ denotes the spin capacitance. In general spin relaxation
effects should be taken into account (magnetization dissipation,
see $LL$--equation). Using then the (classical) Hamiltonian
equations of motion ($\phi=\partial H/\partial S^z$,
$S^z=-\partial H/\partial \phi$) one gets
\begin{equation}\label{eq31}
    \Delta \dot{\phi}=2\mu_BV_s\ , \quad j_S^J=(2E_JS^2/\mu_B)\sin\Delta
    \phi\quad .
\end{equation}
Here, $\Delta\phi=\phi_1-\varphi_2$ and
$V_s=(\mu_B/C_s)(S_L^z-S_R^z)$. That $\Delta\dot{\phi} = 0$ if $M_L\parallel M_R$ and $\Delta\dot{\phi}\neq 0$ if $M_L$ is antiferromagnetically aligned relative to $M_R$ can be checked by experiment. It is $j_s^J \sim \sin (\Delta\phi_0 + 4Mt)$. Note, details of the analysis for the a.c. like effect are given by Nogueira {\it et al.}

Thus, interestingly the spin current in a $FM\mid FM$ tunnel
junction behaves in the same way as the superconductor Josephson
current. Of course, as already mentioned magnetic relaxation (see
Landau--Lifshitz--Gilbert Eq.) affects $j_S^J$. For further
details see again Nogueira and Bennemann \cite{B3}.

Clearly, one expects that also junctions involving antiferromagnets (af), (AF/F),
(AF/AF) yield such Josephson currents, since using the order
parameter for an AF one gets also that $S_q^z$ and $\phi$ are
conjugate variables. (Treat af as consisting of two fm sublattices). Eq.57 should hold for both $J>0$ and $J<0$, see Heisenberg hamiltonian.

Note, also $J_{eff}=J_{eff}(\chi)$ is a
functional of the spin susceptibility $\chi$, since the effective
exchange coupling between the $L$ and $R$ side of a tunnel junction is
mediated by the spin susceptibility of system $N_2$, see Fig.12.

The analysis may be easily extended if an external magnetic field $\overrightarrow{B}$ is present. Then from the continuity equation one gets $j_s\sim \frac{\partial M}{\partial t}$ and $\frac{\partial\overrightarrow{M}} {\partial t}= a\overrightarrow{M_L}\times\overrightarrow{M_R} - g\mu_B \overrightarrow{B_L}\times\overrightarrow{S_L}+\ldots$ (and similarly $\frac{\partial \overrightarrow{M_R}}{\partial t}= a\overrightarrow{M_R}\times\overrightarrow{M_L} +\ldots$). Alternatively one may use the Hamilton--Jacobi equations with the canonical conjugate variables $S^z$ and $\phi$ $(\dot{S^z}=\frac{\partial H }{\partial\phi}, \dot{\phi}= - \frac{\partial H }{\partial S^z})$ and changing the hamiltonian $H \longrightarrow H - g\mu_B (\overrightarrow{B}\cdot\overrightarrow{S_L}+\overrightarrow{B}\cdot\overrightarrow{S_R})$ to derive the currents $j_s$ and $j_s^J$.

Note, according to Maxwell
Eqs. the spin current $j_S$ should induce an electric field
$E_i$ given by
\begin{equation}\label{eq32}
    \partial_xE_y-\partial_yE_x =-4\pi\mu_B
    \partial_t\langle S^z\rangle =4\pi j_s\quad .
\end{equation}
Here, for simplicity we assume no voltage and $\partial_tB=0$ for
an external magnetic field $B$.

Tunnel Junctions with Spin and Charge Current:
Generally one gets
both a charge current $j_C=-e\dot{N_L}$ and a spin current
$j_s=-\mu_B(\dot{S_L^z}-\dot{S_R^z})$ and these may interfere.
This occurs for example for $SCM\mid SCM$ junctions, where $SCM$
refers to nonuniform superconductors coexisting with magnetic
order (see Larkin--Ovchinikov state), and for a (SC/FM/SC) junction with a ferromagnet between two superconductors. Then one gets after some
algebra for the Josephson currents (see Nogueira {\it et al.})
\begin{equation}\label{eq33}
    j_c^J=(j_1+j_2\cos\Delta\varphi)\sin(\Delta\phi +\frac{2\pi l}{\phi_0}H_y)
\end{equation}
and
\begin{equation}\label{eq34}
    j_s^J=j_s \sin\Delta\varphi\cos(\Delta\phi +\frac{2\pi l H_y}{\phi_0})\quad .
\end{equation}
Here, $\phi_0$ is the elementary flux quantum, $H_y$ an external
magnetic field in y--direction, $l=2\lambda +d$, with $\lambda$
being the penetration thickness, and  $d$ the junction thickness.
$\Delta \varphi$ and $\phi$ refer to the phase difference of
magnetism and superconductivity, respectively. The magnetic field
$H_y$ is perpendicular to the current direction.

(TSC/FM/TSC) Junction:
Regarding switching of tunnel current and
analysis of triplet superconductivity the tunnel junction ($TSC\mid
FM\mid TSC$) is of interest. Here, $TSC$ refers to triplet
superconductivity. The Josephson current flows then through
low--energy Andreev states. Relative orientation of the
magnetization $\overrightarrow{M}$ of the ferromagnet ($FM$) and
$d$--vectors of the triplet superconductors, see Fig.20(b),
control the tunnel current. One gets, see Morr {\it et al.} \cite{B3},
\begin{equation}\label{eq35}
    j_c^J=-\frac{e}{\hbar}\sum_i \frac{\partial E_i}{\partial
    \phi}\tanh(\frac{E_i}{kT})\quad ,
\end{equation}
where $\phi$ is the phase difference between the two
superconductors and $E_i$ are the energies of the Andreev states and
which are calculated using Bogoliubov--de Gennes analysis. Thus
one derives an unusual temperature dependence of the Josephson
current on temperature and even that $j_c^J$ may change sign for
certain directions of $\overrightarrow{M}$, although $\Delta \phi$
did not change.

(SC/FM/SC) Junction:
As discussed recently by Kastening {\it et
al.} \cite{B3} such junctions reflect characteristically magnetism. Again
the tunnel current is carried by Andreev states. No net spin
current flows from left to right, since the spin polarized current
through the Andreev states is compensated by the tunnel current
through continuum states as must be due to basic physics. In case of strong ferromagnetism single
electrons tunnel, while for weak ferromagnetism Cooper pair
tunneling occurs. The Josephson current may change sign for
increasing temperature without a change in the relative phase of
the two singlet superconductors. Of course, dependent on coherence length, temperature and thickness of the FM and strength of FM one may get $j^J=0$
for the Josephson current.

(FM/SC/FM) Junction:
It is already obvious from Fig.20(c) that in
the presence of an applied voltage $V$ junctions ($FM_1\mid
SC_2\mid FM_3$) carry currents which depend sensitively on the
relative orientation of the magnetization of the two ferromagnets.
If these are directed in opposite direction (AF configuration) one
gets a maximal spin accumulation in the superconductor and thus
one may suppress (at a critical voltage) superconductivity in
singlet superconductors. As a consequence the magnetoresistance
changes. Hence, such junctions exhibit currents
\begin{equation}\label{eq36}
    j_{c\sigma}=j_{c\sigma}(\overrightarrow{M_1},\overrightarrow{M_3},SC_2,T)
\end{equation}
which reflect sensitively superconductivity and ferromagnetism.

Of course, the current $j_{c\sigma}$ is affected by temperature
gradients $\Delta T$ between the ferromagnets and resulting for
example at nonequilibrium from hot electrons in one $FM$. This may cause
interesting behaviour.

Note, also a ring structure hollow at the center (and with magnetic flux $\Phi$) and outer ring superconducting and a neighboring inner ring ferromagnetic and which is also next to the hollow center
may exhibit interesting behavior in the presence of an external magnetic field B.

The equation for the current due to $\partial E/\partial\phi$, for the current driven by the phase dependence of the electronic energy, is of general significance. From it one may derive the currents induced in rings, discs etc. due to the Aharonov--Bohm effect if a magnetic field is present.

\subsection{Quantum dot Systems}
Interesting current pattern and interferences are also expected
for the quantum dot grain structure shown in Fig.3. Assuming a
mixture of ferromagnetic and superconducting grains (clusters,
quantum dots) one has the hamiltonian
\begin{equation}\label{eq29}
    H=H_T-(ze^2/C)\sum_{i,j}(N_i-N_j)^2+H' \quad .
\end{equation}
Here, $H_T$ denotes the tunneling hamiltonian. This includes also
spin Josephson currents of the form  $E_{ij}^J\cos\varphi_{ij}$
and also Josephson Cooper pair current contributions resulting
from the phase difference $\varphi_{ij}=\varphi_i-\varphi_j$ of
the phases of the S.C. order parameter of grains $i$ and $j$. The
electrostatic effects due to different charges of grains $i,j$ are
given by the second term. $H'$ denotes remaining affects, for
example due to ferromagnetic grains. One expects characteristic
differences for singlet and triplet superconductivity and
interplay of spin currents $j_\sigma$ between magnetic grains and
Josephson--currents.

In grains the interplay of Cooper pair size and distance between
Cooper pairs manipulated by confinement is expected to exhibit
interesting behavior.

For a lattice like array of quantum dots it is of interest to
study phase ordering of the order parameter of the various dots $i$,
its dependence on distance between dots, etc. (see related
situation for superconductors when $T_c\propto \rho_s$, here
$\rho_s$ is the superfluid density).

As speculated for two--band superconductors and as assumed for $sc$
$q$--bits (strong) phase coupling of two magnetic systems (quantum
dots) may cause a covalent splitting like process (mode--mode
coupling like process). Thus, for example, one expects for the
combined, phase coupled system $N_1$ and $N_3$, see Fig.20, the two
covalently like split states $\mid N_1N_3\rangle_1$ and
$\mid N_1N_3\rangle_2$.

Light irradiation causes interesting responses, since occupation
of electronic states (by single electrons, Cooper pairs) can be
manipulated.

Also note, for a mixture of superconducting and magnetic quantum dots on a lattice (and presence of an external magnetic field $B$) one may expect
interesting quantum mechanical effects.

These examples may suffice already to demonstrate the interesting
behaviour displayed by nanostructures involving tunneling. This
holds also for currents across multilayers of magnetic films.

\section{Results}

Characteristic results are presented for clusters, films and
tunnel junctions.

\subsection{Small Magnetic Particles}

\subsubsection{Single Magnetic Clusters}

In small clusters the local
D.O.S. exhibits the dependence of the electronic structure on the
local atomic environment of a cluster atom. In Fig.21 the local
D.O.S. is compared with the average D.O.S. Results were obtained
by Pastor et al. using the tight--binding Hubbard hamiltonian.

The D.O.S. is the most important property of the electronic
structure and determines the thermodynamical and magnetic behavior
$(\mu_n, M_n(T))$.
\begin{figure}
\centerline{\includegraphics[width=1.\textwidth]{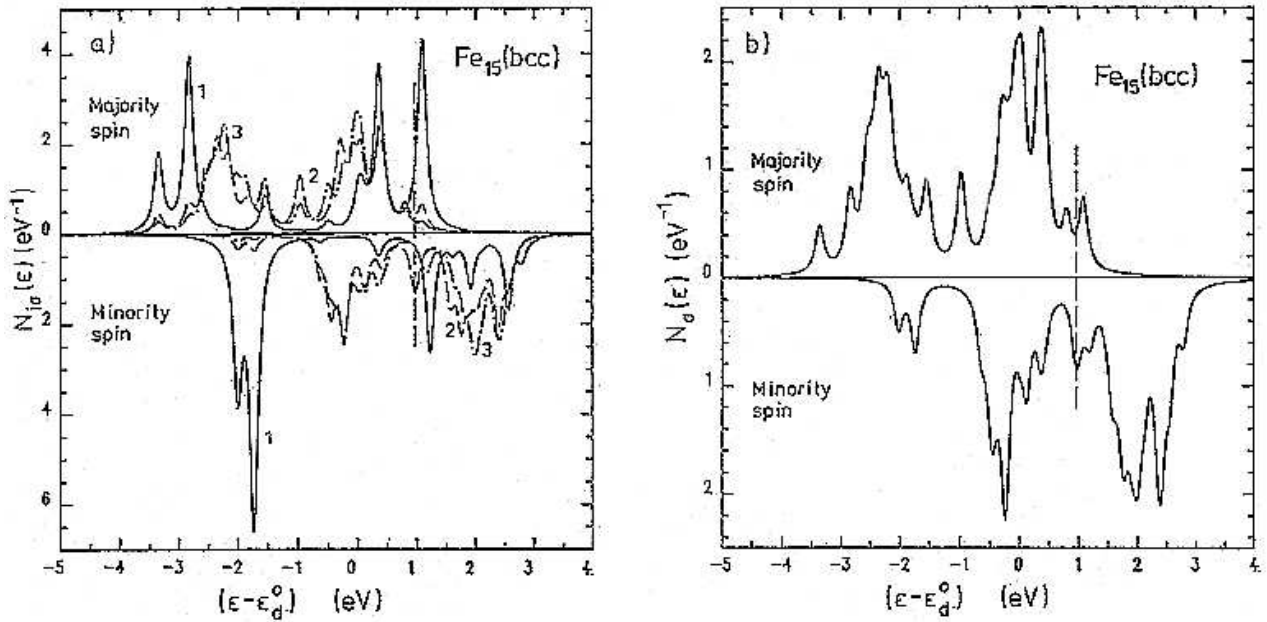}}
\caption{Typical electron density of states (D.O.S.) results for
$Fe_n$ clusters, n=15. The local D.O.S. refers to the atoms 1, 2,
3, see Fig.1. (a) refers to local DOS, (b) to average cluster
DOS.}
\end{figure}

In table 2 results for the bond--length $d_n$ ($d_b$ refers to
bulk), cohesive energy $E_{coh}(n)$ and magnetic moments $\mu_n$
are given for $Ni_n$ clusters.
\begin{figure}
\centerline{\includegraphics[width=1.\textwidth]{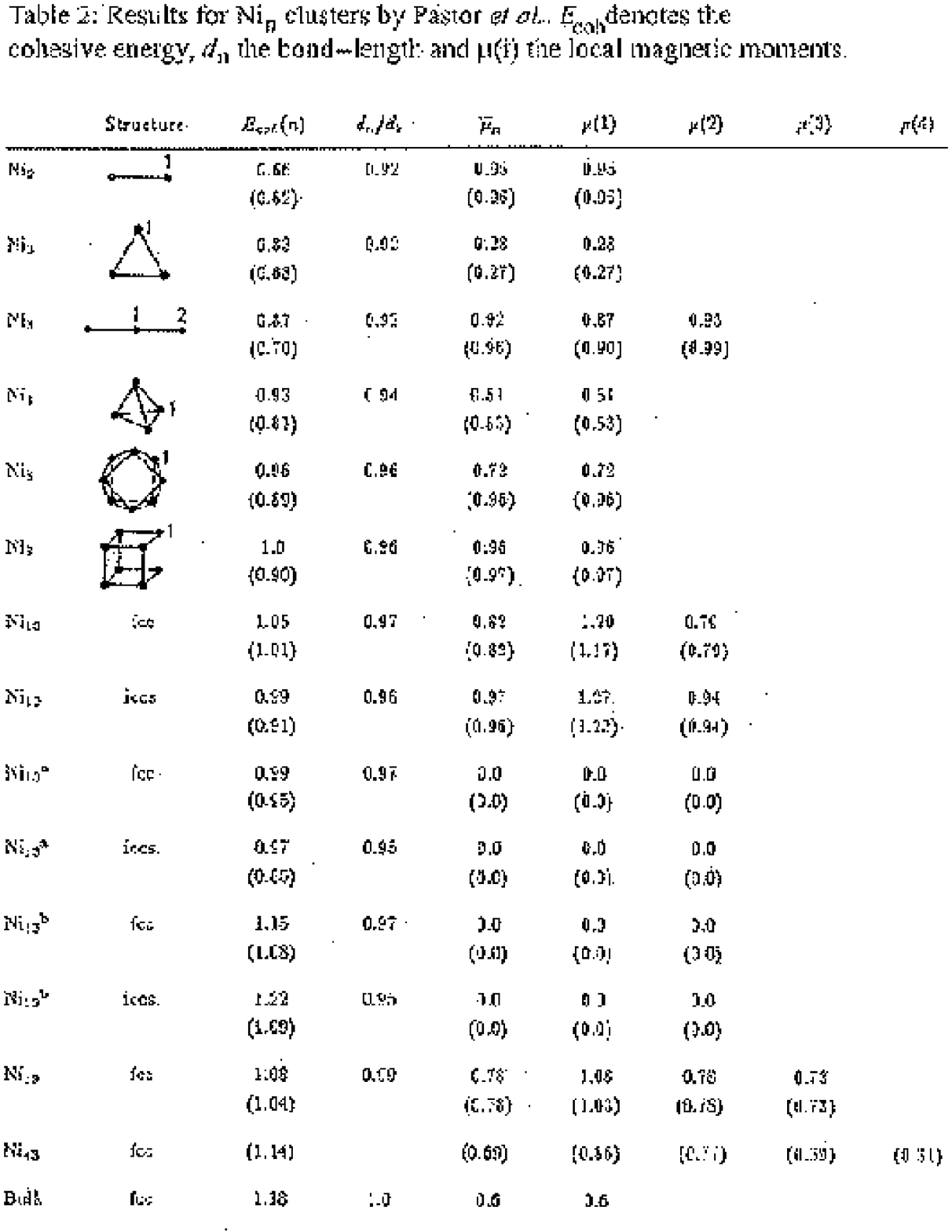}}
\end{figure}
\setcounter{figure}{21}

In Fig.22 the dependence of the magnetic properties on cluster
size is given. The variation of ${\mu_n}$ reflects the
atomic structure (shell structure). These results were obtained by S. Mukherjee
{\it et al.}). In accordance with the electronic
calculations by Pastor et al. a simple model is used assuming that the
moments depend on cluster shell ($i$) and local atomic coordination,
both changing with number of cluster atoms. Such behavior results also for
other clusters, for example $Ni_n$ clusters. Note, the discrepancy between theoretical and experimental results
for larger clusters. Of course, $\mu_i\rightarrow \mu_B$, bulk behavior as clusters get larger.
\begin{figure}[!htbp]
\centerline{\includegraphics[width=0.6\textwidth]{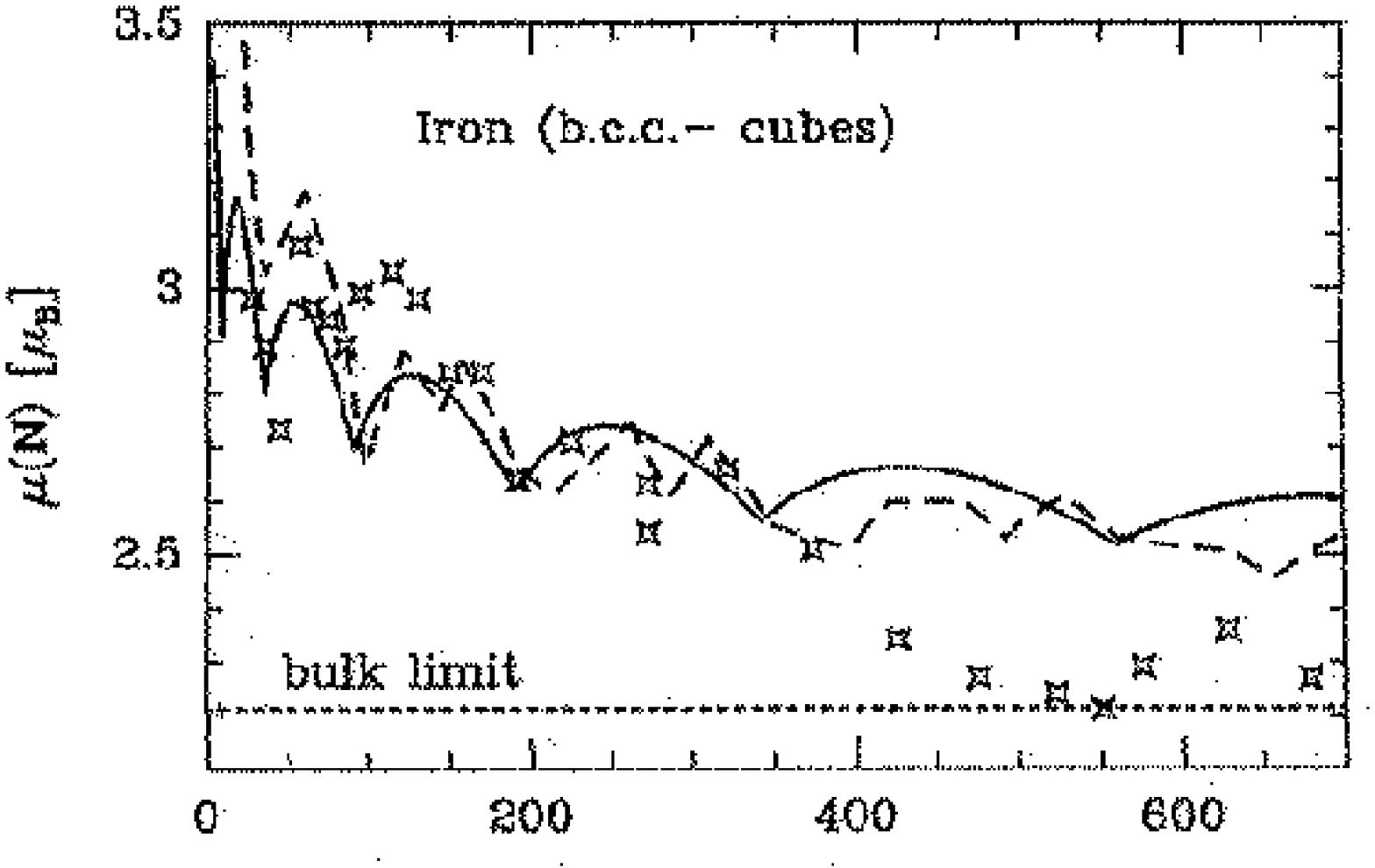}}
\caption{Average magnetic moment $\mu(N)$ of $Fe_n$ clusters as a
function of cluster size. The solid and dashed curves refer to
calculations by Jensen {\it et al.} using different cluster shell
models for the magentic moment $\mu(N)$. $\mu(N)$ is calculated by
averaging over the shell (coordination) dependent moments $\mu_i$
and where these are given by the local atomic coordination
(varying with $N$). Crosses refer to experimental results (de Heer
{\it et al.}). Note, solid and dashed curves refer to calculations
using different models for the dependence of the local magnetic
moments $\mu_i$ on atomic coordination. Of course, for larger clusters $\mu\rightarrow \mu_b$.}
\end{figure}

In Fig.23 results for $\mu_n$ of Fe$_n$ are given. The results were obtained using the Hubbard hamiltonian and a tight--binding type calculation by Pastor {\it et al.}. For Fe one expects a particular strong interdependence of magnetism and structure (see bcc vs fcc Fe). For each cluster one assumes a structure which yields the largest cohesion. The variation of $\overline{\mu_n}$ as a function of n is due to the interplay of changes in coordination number and bond--lengths (note, approximately $W_n\propto d^{-5}$, where W is the band--width and d the bond--length). Interestingly, one gets for the small clusters $\overline{\mu_n}$ larger than $\mu_b$, magnetic moments which are larger than the bulk one as expected. The sensitive dependence of the Fe magnetic moments and magnetism on structure is of general importance for nanostructures and of material involving Fe and Fe under pressure.

\begin{figure}
\centerline{\includegraphics[width=0.5\textwidth]{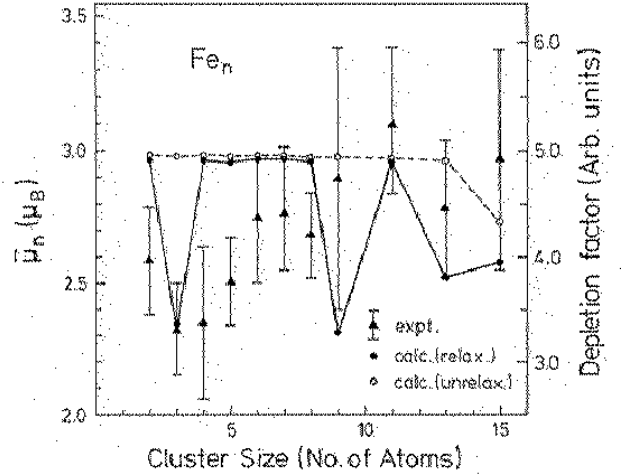}}
\caption{Dependence of average magnetic moment of $Fe_n$ on cluster size.
Note, the dependence of the results on the relaxation of the atomic
structure, see calculations using an electronic theory by Pastor {\it et al.}. For each n a structure with largest cohesion is assumed. For Fe--clusters one expects a sensitive dependence on structure and possibly a particular strong interdependence of magnetism and structure. The vertical bars refer to experimental results for the depletion factor and which is assumed to be proportional to the average magnetic moment, at least approximately.}
\end{figure}
One may also calculate using molecular field type methods the
Curie--temperature $T_c$ as a function of cluster size. Results
are shown in Fig.24. Typically one gets that $T_c$ increases for increasing cluster size. Note, the exchange interaction J, $q_{eff}$ and $\mu_i$ may change as the cluster grows.
\begin{figure}
\hspace{-1cm}\centerline{\includegraphics[width=0.45\textwidth]{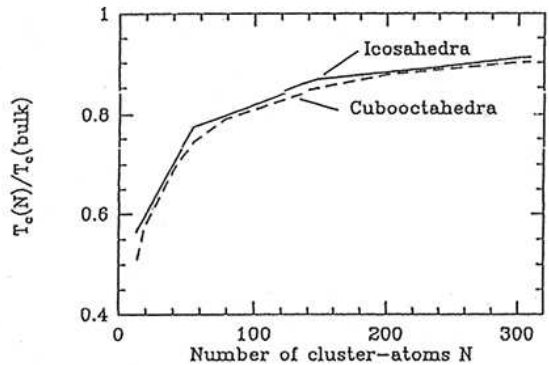}}
\caption{The Curie--temperature $T_C$ $(M(T_c)=0)$ as a function
of number of cluster atoms. The calculations by Jensen {\it et
al.} assume two different cluster structures.}
\end{figure}

Orbital magnetism: Clusters as other low dimensional systems (surfaces, thin films)
exhibit also interesting orbital magnetism. For example, for Ni
clusters a remarkable enhancement of the average orbital moment as
compared to bulk is observed. Note, one gets for the orbital
moment of Ni $\langle L(Ni_7)\rangle \simeq 0.5 \mu_B $ while for bulk $\langle
L(bulk)\rangle \simeq 0.05 \mu_B.$ Using the tight--binding
Hubbard hamiltonian including the spin--orbit interaction
($H_{so}= -a \sum(\overrightarrow{L}\cdot\overrightarrow{S})_{i\alpha,i\beta}a^+ a$,
here $\alpha,\beta$ refer to orbitals), Pastor, Dorantes--Davila {\it et al.} obtain for
the average orbital moment $\langle L_\delta\rangle$ the results
shown in Fig.25. Here, $\delta $ refers to the cluster crystal
axis. The important DOS $N_{im\sigma}^\delta(\varepsilon)$ is determined self consistently
for each orientation $\delta$ of the magnetization with respect to structure. The s--o coupling
connects spin up and spin down states (dependent on relative orientation of $\overrightarrow{M}$ and atomic structure).
$(\overrightarrow{L}\cdot\overrightarrow{S})_{...}$ are intraatomic matrix elements.
\begin{figure}
\centerline{\includegraphics[width=0.5\textwidth]{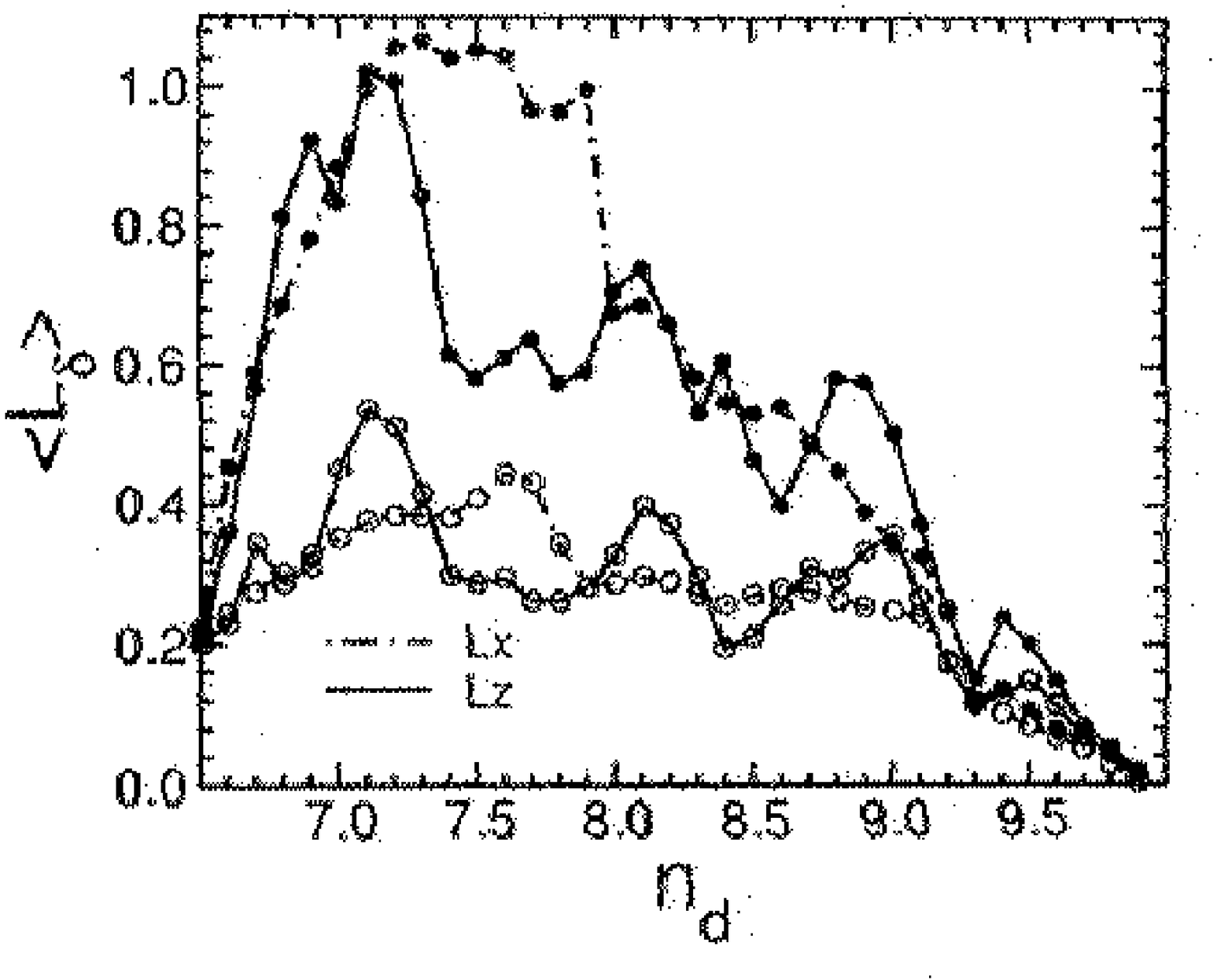}}
\caption{Orbital moment per atom $\langle L_{\delta}\rangle$ for a
pentagonal bipyramid transition metal model cluster of 7 atoms as
a function of d--state filling (band--filling) $n_d$. Bulk n.n.
neighbor atomic distances are taken. $\delta$ refers to the
magnetization direction taken along (principal $C_n$ symmetry)
cluster axis ($\delta=z$: full curves, $\delta=x$: dashed curves,
for details see Pastor, Dorantes--Davila {\it et al.}).Note, full
and open circles refer to calculations using different electron
energies $\epsilon_{i\sigma}$.}
\end{figure}

The results show that the orbital magnetic moment is an important contribution
to the magnetic moment of clusters and quite generally to the one of nanostructures of
transition metals (TM). $\langle\overrightarrow{L}\rangle$ depends sensitively on band filling.
Also note, for example, ($L_z - L_x$), etc. reflect magnetic anisotropy. For Ni the orbital moment aligns parallel
to the spin moment.

Mie scattering: For determining magnetism in small clusters besides Stern--Gerlach
deflection experiments (as performed by W. de Heer {\it et al.})
Mie--scattering could be used. Extending the usual Mie scattering theory to magnetic clusters
one calculates from Maxwell equations with spin orbit coupling, coupling the electromagnetic field and the cluster  magnetization, and with a dielectric function $\varepsilon$ of tensor form the Mie backscattering profile. Characteristic differences are obtained for Ni, Co, Fe. For example, magnetic Mie scattering is important
for spherical clusters with radius approximately larger than 6 nm for Fe and 10 nm for Co.

Generally $\overrightarrow{M}$ has a weak effect on the electromagnetic field. This is strongest when size of cluster is of same order as the wavelength of the electromagnetic field ($k^{-1}$). In view of the weak coupling
one may treat the perturbation due to cluster magnetization $\overrightarrow{M}$ in lowest order (with respect to the off--diagonal element $\varepsilon_{xy}= a\lambda_{so}M+...$ of the dielectric tensor $\varepsilon_{ij}$ which is frequency dependent). The tensor $\varepsilon_{ij}$ may be fitted using magnetooptics (Kerr--effect, ellipticity etc.).

Results obtained by Tarento {\it et al.} are given in Fig.26. Note, the backward scattering and the
angle dependent scattering profile reflect magnetism. Magnetic effects are maximal when magnetization is parallel to y, see Mie scattering configuration, Fig.26(a). Note, for angle $\theta=\pi/2$ the backscattering intensity I(M=0) is very small and then the changes due to the cluster magnetization are relatively large. In Fig.26(b) $\varepsilon_0=-1.569+i5.58$ and $\varepsilon_{xy}=-0.059+i0.122$ is used.
Note, in Fig.26(c) we use for the dielectric function $\varepsilon_o=-4.592+i7.778$ and $\varepsilon_{xy}=-0.1613-i0.0733$.

Mie scattering exhibits also characteristic differences regarding the angular
dependence of the backscattering intensity for Cr, Fe, Co, and Ni.
\begin{figure}
\includegraphics[width=0.45\textwidth]{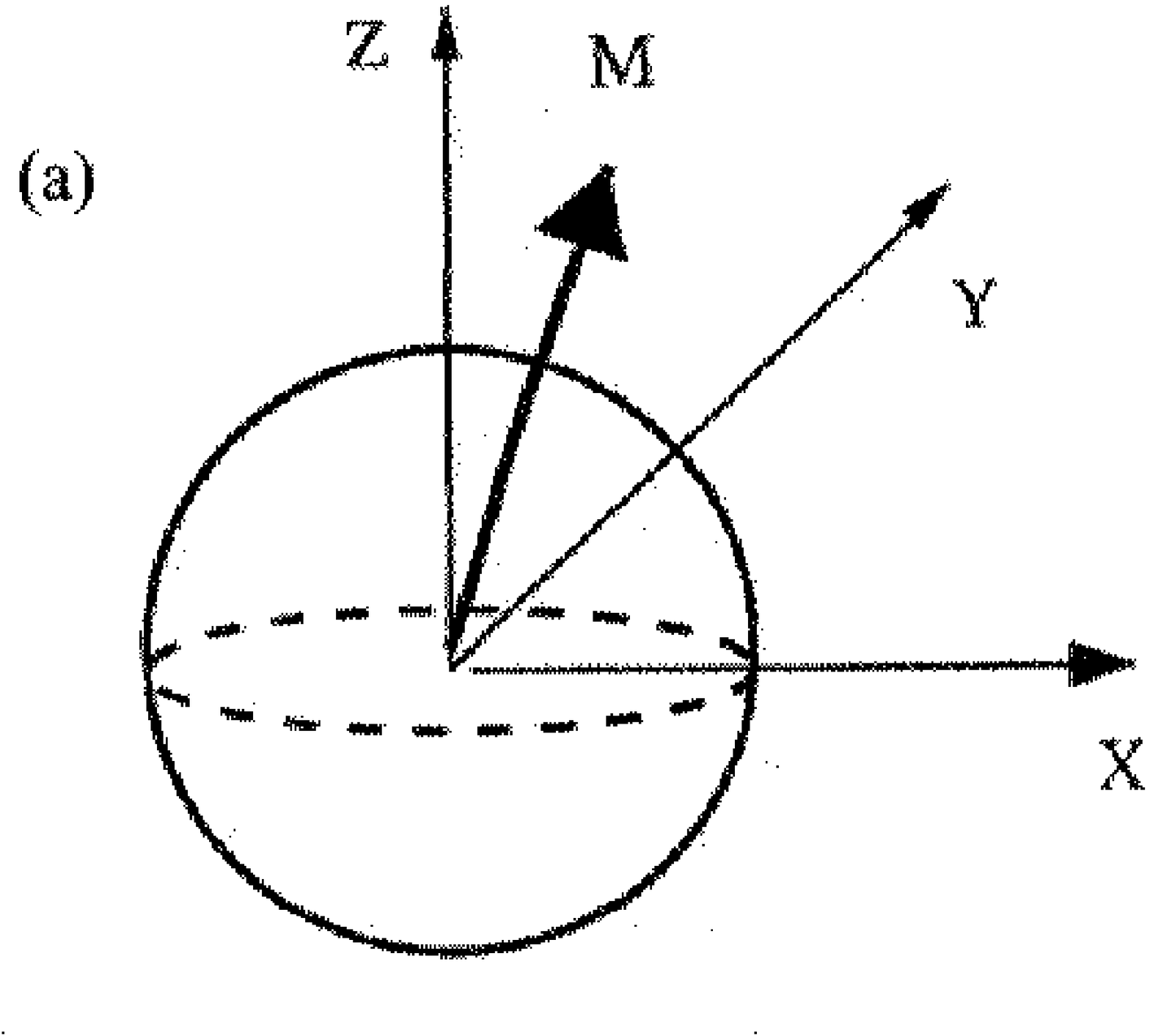}
\hspace{.5cm}\includegraphics[width=0.52\textwidth]{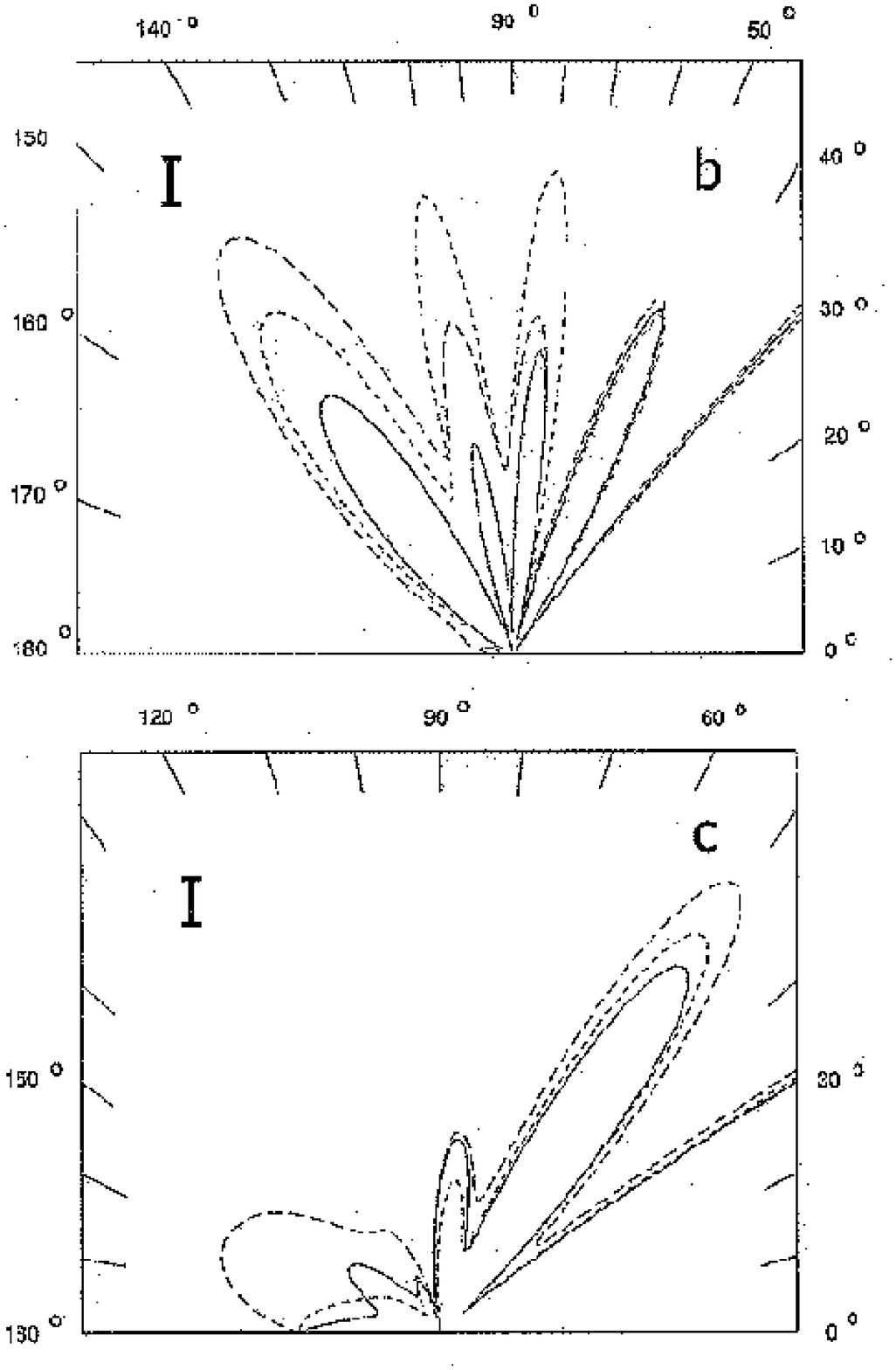}
\caption{Angular dependence of Mie backscattering: (a) Mie--scattering configuration for magnetic clusters
(Co clusters of radius 2600{\AA}). The incident plane wave propagates along z--direction and electric and
magnetic fields are polarized along x-- and y--direction. In Figs. (b),(c) different values for the dielectric tensor $\varepsilon_{ij}$ are used, see text.
Results for the angle--dependent backscattering intensity are given, see
analysis Tarento {\it et al.}. The solid curves refer to M=0, the dashed ones to magnetic clusters with $\protect\overrightarrow{M}$ along y--axis, and
the angle to the polar one in the x--z plane. The axis units are arbitrary. In (c) the two dashed curves refer
to magnetization taken along the x-- and y--axis. (The field frequencies are such that $\hbar\omega$=5eV in Fig.(b), and 3.5eV in Fig.(c))}
\end{figure}

\subsubsection{Ensemble of Clusters}

The behavior of an ensemble of magnetic clusters is expected to
reflect magnetic anisotropy of the cluster and cluster rotation.
An external magnetic field $H$ will align the magnetic clusters
which magnetization points in the direction of the easy cluster
axis. This is set by the anisotropy axis of the cluster. Note, to
align the magnetic clusters the pinning of the cluster
magnetization to the easy axis has to overcome. This should be
reflected then in the dependence of the cluster ensemble
magnetization on an external magnetic field.

$T_{bl}$ characterizes the energy which pins the cluster
magnetization to its easy axis. Therefore, different behavior is
expected for the ensemble magnetization $<\mu_Z>$ for temperature
$T>T_{bl}$ (Langevin behavior) and $T<T_{bl}$.

In Figs. 27(a) and 27(b) results are given for the magnetization
of a system of clusters. Note, in a magnetic field $H$ cluster
anisotropy characterized by the blocking temperature $T_{bl}$ and
$H$ interfere. For temperatures $T>T_{bl}$ the internal anisotropy
needs to be overcome to align the cluster magnetization via the
external field.

Results shown in Fig.27 give the magnetization observed in
Stern--Gerlach experiments for an ensemble of clusters in a
magnetic field $H$. Obviously magnetic anisotropy within the
cluster determines the magnetic behavior. The temperature $T_{bl}$
characterizes the pinning of the magnetization to the easy axis of
each cluster. Thermally activated depinning occurs for $T>T_{bl}$
(see calculations by Jensen {\it et al.}). Clearly, for
temperatures larger than the blocking one Langevin type behavior
is expected. For lower temperatures one needs a certain strength
of the external magnetic field ($\mu H\geq kT$) for unpinning the
cluster magnetization from its easy axis.
\begin{figure}[!htbp]
\hspace{3.cm}
\includegraphics[width=0.6\textwidth]{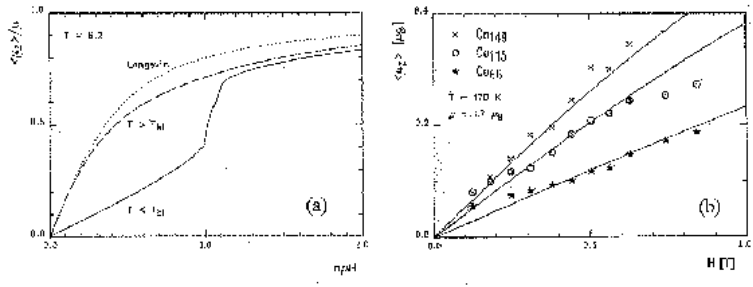}
\caption{Results for cluster ensemble: The component of the
average cluster magnetization in z--direction. (a) Comparison for
$T>T_{bl}$ and $T<T_{bl}$, where $T_{bl}$ refers to the blocking
temperature. (b) Comparison with experimental results for $Co_n$
clusters, see de Heer {\it et al.} and Bloomfield and others.
Note, here $T_{bl}<T$.}
\end{figure}

Quantum dot lattices:
Typical results for a lattice of quantum dots or anti--dots are shown in the following figures.
Of course, one expects that the density of states $N(E)$, generally if magnetism is involved spin dependent, reflects characteristically the nanostructure.

In Fig.28 results for the DOS oscillations are given which demonstrate their dependence on dimension and geometry
of the nanostructure. For a sphere mainly paths with t=1 and p=3 and 4 contribute. Their interference yields
a beating pattern. For circular rings path 3 is most important (see Fig.13) and note larger discrepancy
between classical and quantum--mechanical calculations. For circular discs note $A_i\sim k^{-1/2}$ ( in contrast to spheres with $A_i\sim k^{1/2}$ ).

Approximately the spin up and spin down DOS are split by the molecular field $h_{eff}$ in case of ferromagnetic nanostructures. It is of interest to study the interplay of superconductivity and magnetism and magnetic field B.

In Fig.28 the DOS results obtained from the semiclassical Balian--Bloch--Stampfli theory are compared with quantum--mechanical calculations. The oscillating part of the DOS due to the interference of the dominant electronic paths is shown for various nanostructures. The quantum--mechanical calculations were performed using Schr\"{o}dinger equation and Green's function theory
\begin{equation}
        N(E,..) = (\gamma/\pi)\int dE'\frac{N(E')}{(E-E')^2 + \gamma^2} .
\end{equation}
Of course, the parameter $\gamma$ must be choosen as usual such that it does not affect the structure in the DOS.

Note, in case of magnetism one has a spin dependent DOS $N_\sigma(E)$ and to simplify one may assume that spin up and spin down states are split by the molecular field $h_{eff}$ which may include an external magnetic field $B$.

Also hollow particles (see coated nanoparticles, for example) can be treated similarly as rings.
This is of interest, for example, to study confined and bent 2d electronic systems.
\begin{figure}
\centerline{\includegraphics[width=0.6\textwidth]{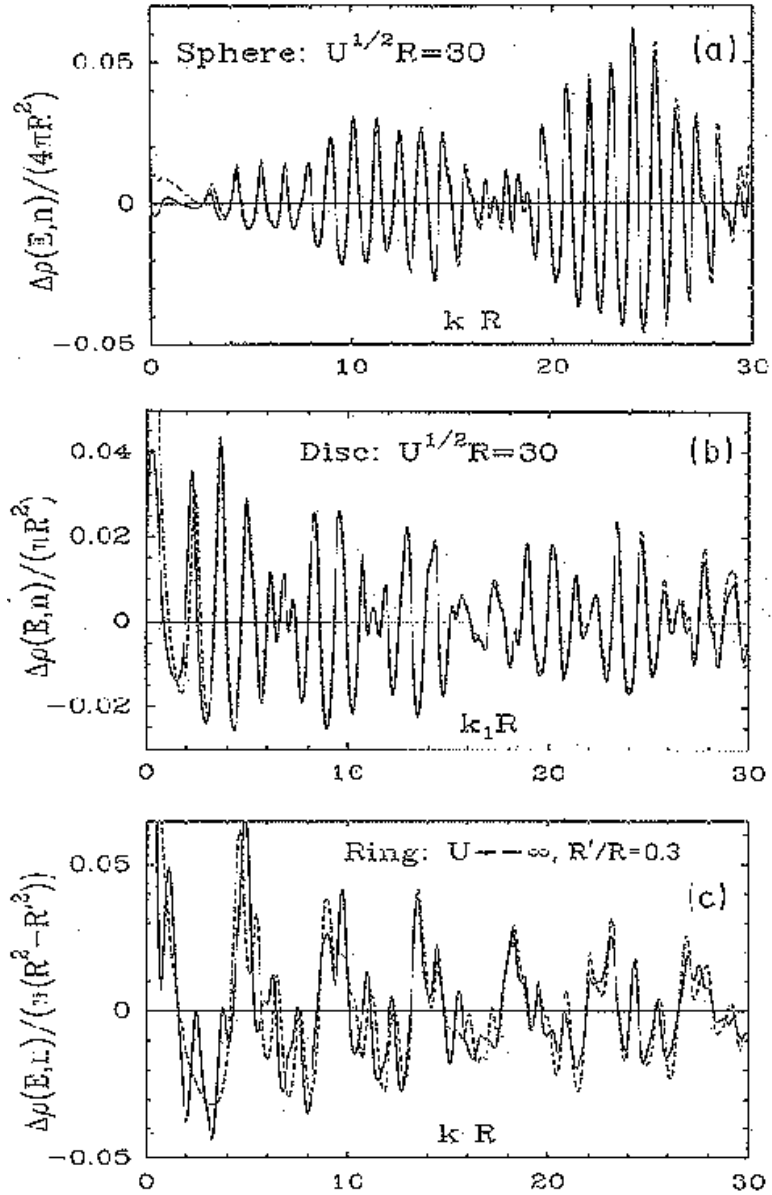}}
\caption{Dependence of the DOS oscillations on dimension and geometry of a
mesoscopic system obtained by Stampfli {\it et al.} using an extension of the Balian--Bloch theory.
Results for (a) sphere, (b) disc and (c) ring are given. The electrons are scattered by the potential U. This potential may scatter spin dependently. The dashed curves refer for comparison to quantum--mechanical calculations.}
\end{figure}

In Fig.29 results are given for the oscillating part of the D.O.S.
as a function of energy and external magnetic field for a lattice
of anti--dots which repel the electrons. The magnetic field $B$
changes the orbits and thus their interference. As a result DOS
oscillations occur. Note, only a few closed electron orbits
(1,2,3,$\ldots$) give the dominant contribution to the electronic
structure, the DOS, see theory by Stampfli {\it et al.} and
Eq.(19) and Eq.(21). The magnetic field causes paths deformation and phase shifts. The flux due to the external field $B$ is $\phi = SB$ and for orbits 1, 2, 3 $S$ is approximately independent of $B$ for small field. The cyclotron orbit radius is $R_c=\frac{k}{B}$ and we assume for simplicity $2R_c > (a-d)$ and $R_c \gg R$. Then, $R_c \sim \frac{1}{B}$ and the DOS depends on (d/a) and $\Delta N \sim \cos SB$. This controls also height of oscillations (note $S\propto R_c^2\propto\frac{1}{B^2}$).

For increasing external magnetic field $B$ the oscillations in the DOS change. Interestingly, the contributions of some orbits may be nearly eliminated. Oscillation period in $B$ decrease for increasing $B$ and decreasing lattice constant $a$.

Note, an internal molecular field due to magnetism is expected
to affect the DOS similarly as the external magnetic field. Note also, the above oscillations in the DOS result from the confinement (closed electronic orbits) and additional structure may result from detailed atomic structure yielding the well known electronic shell structure in clusters etc.

Note, for orbit 4 (see Fig.) and similar ping pong orbits one has $\cos(SB)\approx 1$. Of course, for increasing scattering potential U the amplitude and interference changes ($R_c\sim 1/B$). For decreasing field $B$ dephasing occurs and must be included.

For increasing radius the results for thin rings may be similar to the ones for planar surfaces of thin films. For thin films one may perform the Balian--Bloch type calculations as sketched in the figure below.
\begin{figure}[bp]
\centerline{\includegraphics[width=0.3\textwidth]{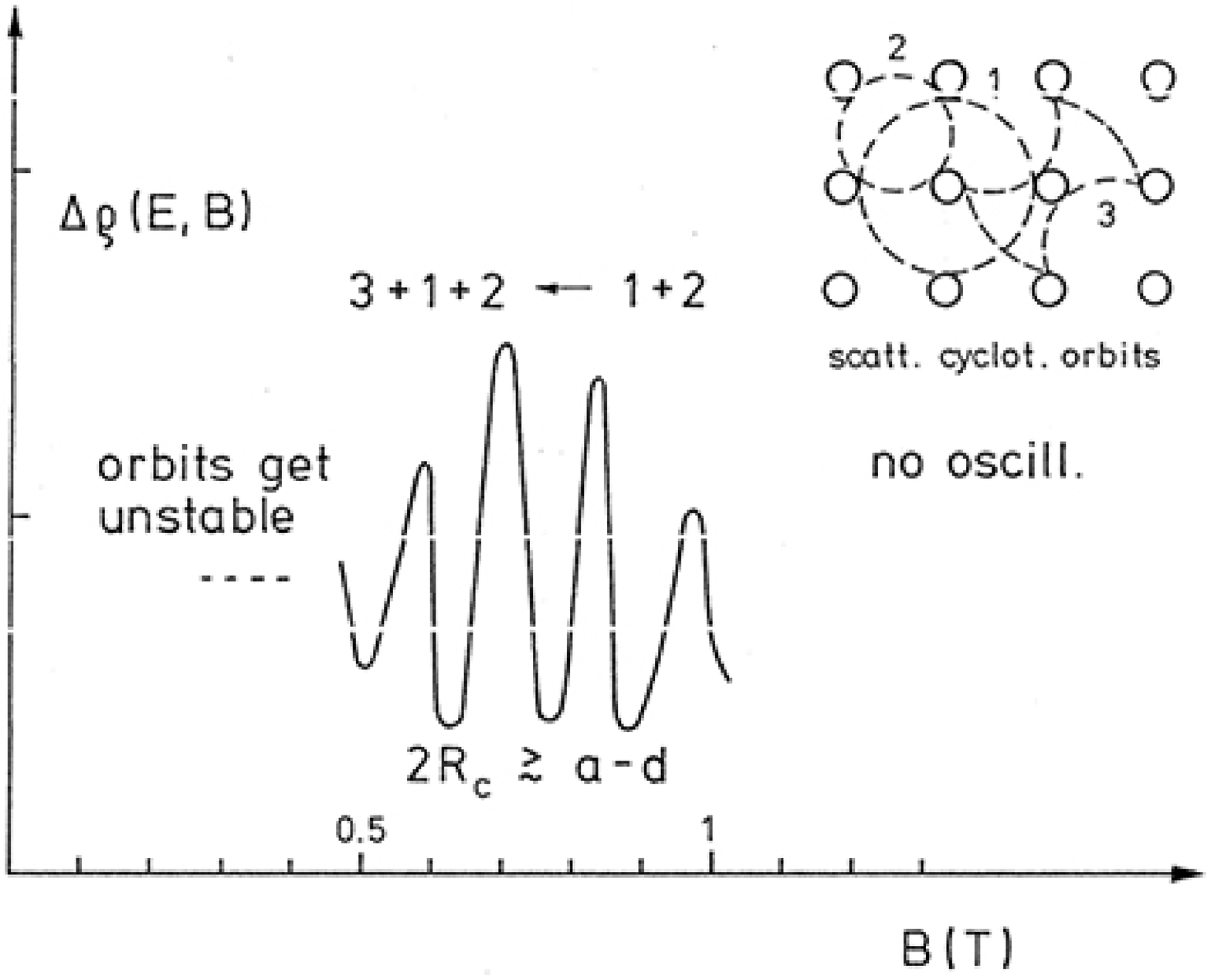}}
\caption{Oscillations of the D.O.S. as a function of magnetic field B for the orbits 1, 2, 3 due to
scattering by repelling anti--dots, see calculations by Stampfli
{\it et al.}. The inset shows the lattice of anti--dots. Note, the
scattering may be spin--dependent and the density of states is spin split. Note $2R_c\geq(a-d)$ is assumed.}
\end{figure}
\begin{figure}
\centerline{\includegraphics[width=0.4\textwidth]{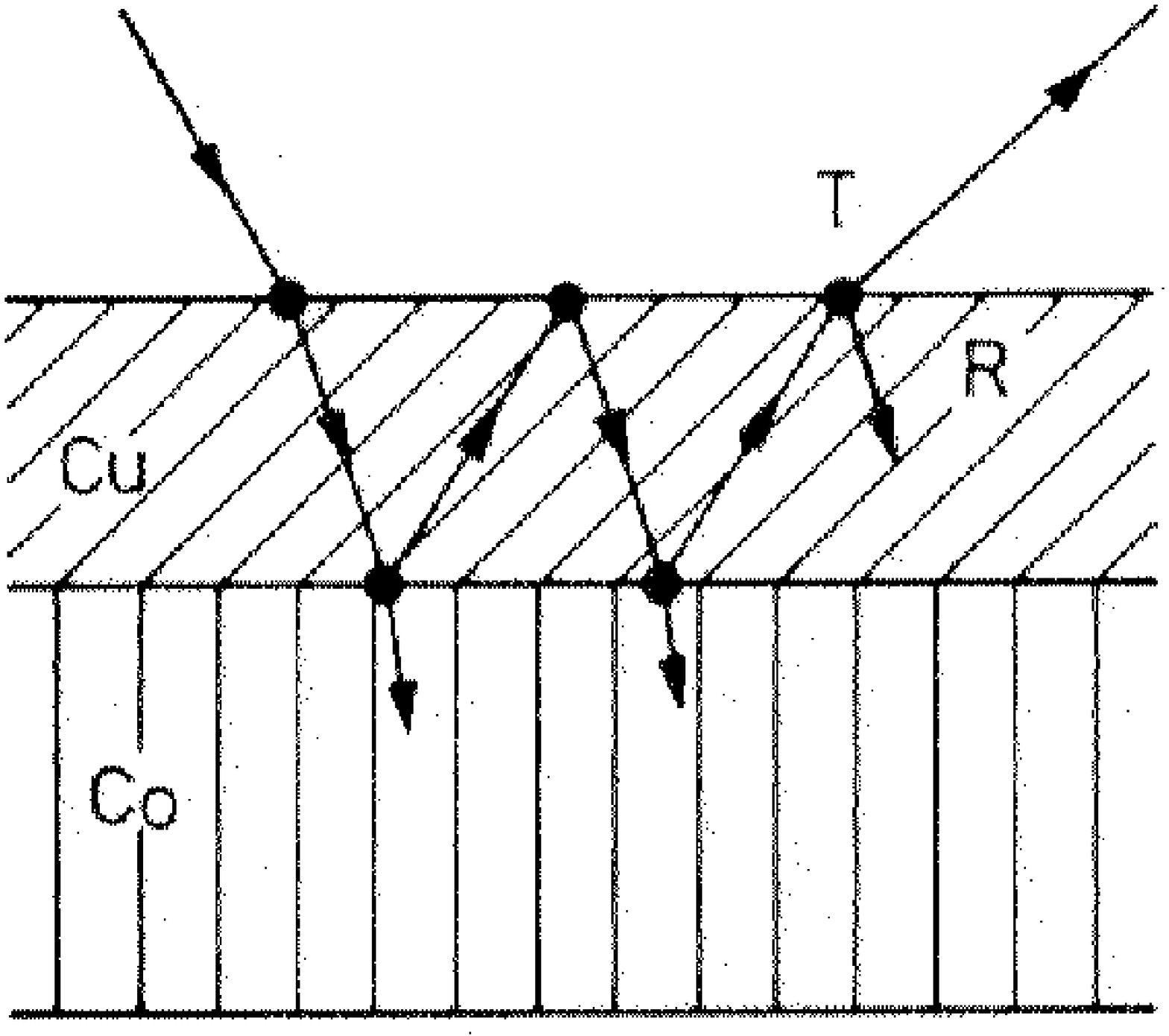}}
\caption{Sketch of the important orbits for thin films. Here, one has to include transmission ($T$)
and reflection ($R$) coefficients at the corners of the paths. Magnetic films yield spin dependent DOS. Note T and R are spin dependent at interface Cu/Co.}
\end{figure}

\subsection{Thin Magnetic Films}

Thin films are of central importance for engineering magnetic
material on a nanoscale. Thin transition metal ( TM ) films, for example, exhibit magnetic properties of
interest for many technical applications like magnetic recording.
In the following typical results are given which show the
dependence of film magnetism on the atomic structure, topology of
the film and on film thickness. Magnetic anisotropy is very
important. Magnetism is characterized by the magnetization as a
function of film thickness, anisotropy, by the orientation of the
magnetization at the film surface and by the Curie--temperature
$T_c$. Nanostructured films exhibit depending on film growth
conditions (controlling the film morphology) characteristic
magnetic domain structures. Then depending on the distances
between the domains various magnetic interactions controlling the
global film magnetization come into play.

\subsubsection{Magnetic Structure}

The magnetic properties of thin ferromagnetic films have been
studied intensively. Due to anisotropy long--range ferromagnetic
order occurs even for ultrathin, quasi 2d--films. Competing
anisotropic forces determine the orientation of the magnetization
relative to the atomic lattice. We use the hamiltonian given in
Eq.(16) with a local anisotropy
\begin{equation}
 H_{anis} = - \Sigma_i (K_2 \cos^2\theta_i + K_4 \cos^4 \theta_i
 + K_s \sin^4 \theta_i )\quad .
\end{equation}
Note, for $T>0$ the in--plane anisotropy $K_s$ gets important.
After some standard algebra (see Jensen {\it et al.}) one finds
the energy per spin. One gets for

\noindent 1. uniform magnetic phases:
\begin{equation}
E_u(\theta)= -(9/2)J - K_2\cos^2\theta - K_4\cos^4\theta -
K_s\sin^4\theta + E_o (\cos^2\theta - 1/3)\quad ,
\end{equation}
where $\theta$ denotes the angle between film normal and
magnetization, and for

\noindent 2. stripe domains with wall width $b$ and domain
periodicity $2L$:
\begin{eqnarray}
  E_{dom}(\theta) = &-&J/2 (q - \pi^2/bL) - K_2
  (\cos^2\theta(1-b/L)+ b/2L ) \nonumber \\ &-& K_4(\cos\theta^4(1-b/L))+
  3b/8L-K_s(\sin^4\theta(1-b/L)+3b/8L) +...
\end{eqnarray}
Here, $\theta$ denotes the canting angle of the domain
magnetization along the $z$--axis ($\theta=0$). $q=4$ for a square
monolayer. The domains have the periodicity 2L along the
y--direction and have an uniform magnetization along the
x--direction, inside the domains the magnetization is along
$\pm z$--direction. This domain configuration is assumed for
simplicity. One gets rich magnetic behavior for thin films. Magnetic anisotropy is important.

Typical results for the magnetic structure are given in Fig.31.
Generally one gets interesting phase diagrams, see Jensen {\it et
al.} \cite{B2,B11}.

Multilayers of magnetic films reflect how the magnetization of neighboring films interfere, affect each other,
act like a molecular field on the magnetization of a particular film.
\begin{figure}[!htbp]
\hspace{-.5cm}\centerline{\includegraphics[width=0.5\textwidth]{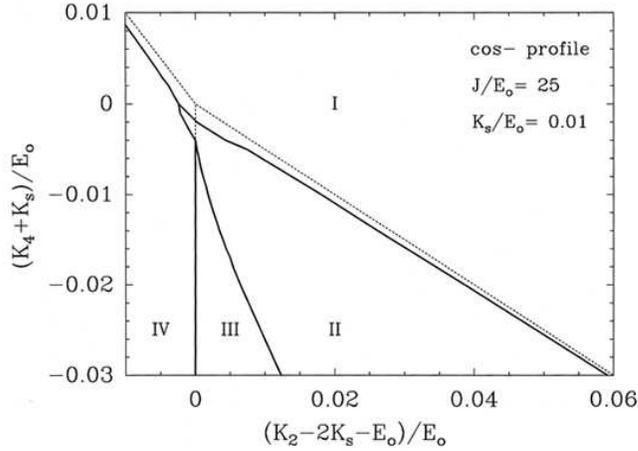}}
\caption{Magnetic structure of thin films obtained at $T=0$ due to
exchange coupling and magnetic anisotropy. Phases $I,II$ refer to
stripe domain structure with perpendicular and canted
magnetization, respectively. Phases $III,IV$ refer to uniform
magnetization with canted and in--plane magnetization,
respectively. ($E_o$ is the demagnetization energy). The
anisotropy constants $K_2$, $K_4$ and $K_s$ are given in Eq.(49), their physical significance is
clear. The dashed curves indicate what happens if only uniform
phases are considered.}
\end{figure}

In Fig.32 results are given for multilayer films Cu|Ni|Cu and
Co|Cu|Ni. Note, one finds for the Curie temperature
$T_c(Co)>T_c(Ni)$. Clearly, the Co film causes that the magnetization
of the Ni film is present at higher temperatures. This results
from interlayer exchange coupling.
\begin{figure}
\centerline{\includegraphics[width=0.6\textwidth]{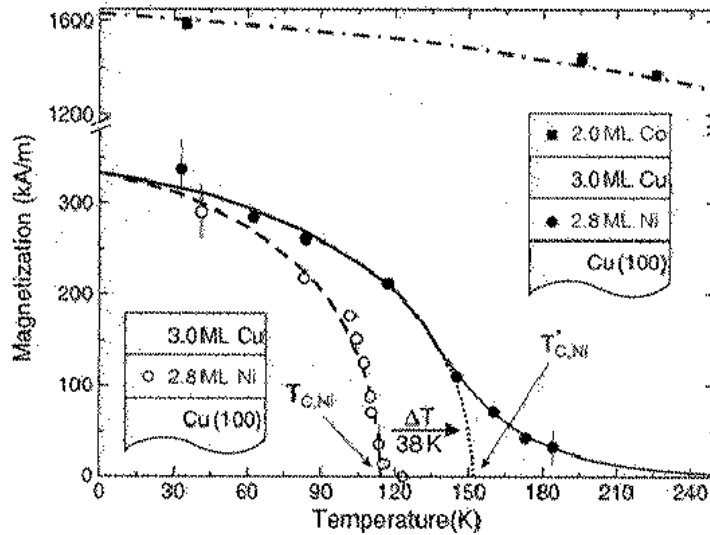}}
\caption{Magnetization of a Co-- and Ni--film separated by 3ML
Cu--film, see inset Fig.. The experimental results (from XMCD) are
indicated by squares and circles. The indicated shift of the Ni
Curie--temperature results from interlayer exchange coupling
between Ni and Co film. }
\end{figure}

The results shown in Fig.33 demonstrate
the dependence of $T_c$ on film structure. Obviously, interesting characterization of magnetic films is also given by
their Curie--temperature.
\begin{figure}
\centerline{\includegraphics[width=0.56\textwidth]{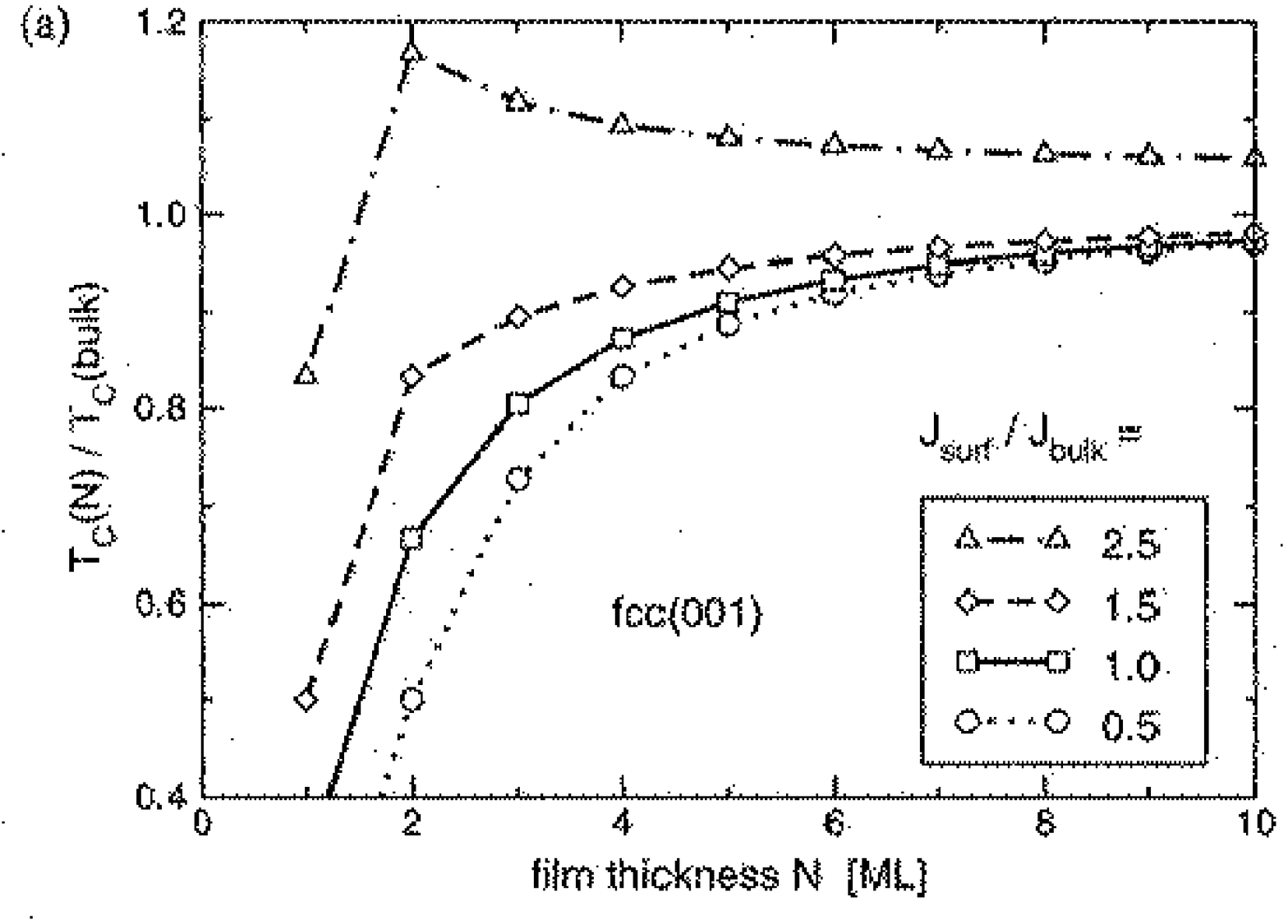}}
\centerline{\includegraphics[width=0.55\textwidth]{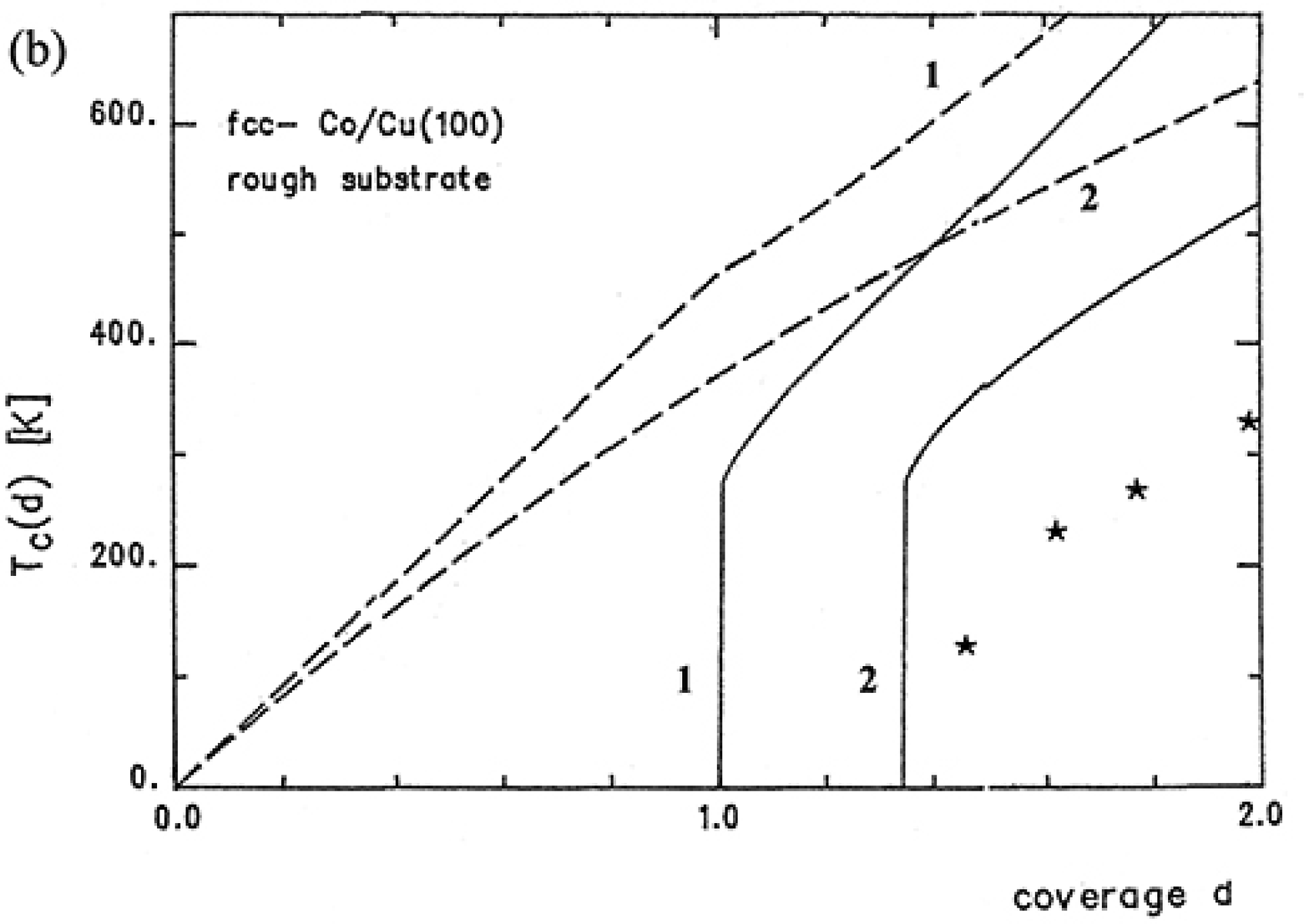}}
\centerline{\includegraphics[width=0.55\textwidth]{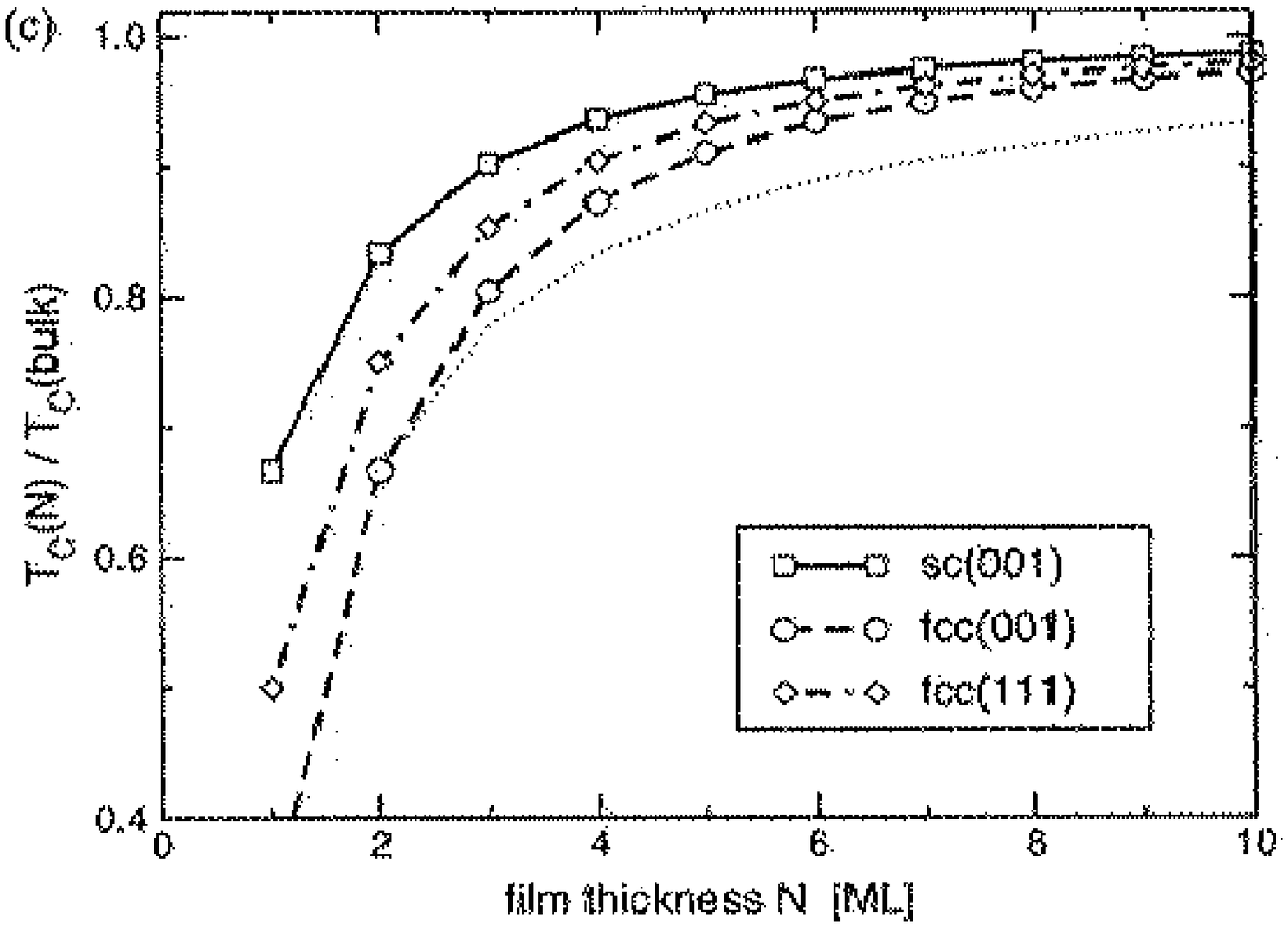}}
\caption{Curie--temperature $T_c$ as a function of film thickness
N and morphology. (a) Results obtained using MFA for f.c.c. films.
(b) $T_c$ dependence on film topology: (1) Layer by layer film
growth. (2) Results refer to rough films and those obtained by
random deposition of atoms. The dashed curves refer to
Bragg--Williams mean field--theory. The stars refer to
experimental results for f.c.c.--Co on Cu(111). (c) Homogeneous
ferromagnetic films assuming different film structures, see Jensen {\it et al.}.}
\end{figure}

\subsubsection{Reorientation Transition of the Magnetization}

The reorientation transition of the magnetization at surfaces (and
interfaces) is very important regarding applications, but also
regarding understanding of magnetic anisotropy. Such a transition may be
driven by temperature, film thickness and film topology (and at
nonequilibrium by hot electrons). Magnetooptically the reorientation transition is characterized by the linear response $\chi_{ij}$ or the nonvanishing elements of the nonlinear susceptibility $\chi_{ijl}$, for example.

In the following typical results for the reorientation transition
of the magnetization are presented. In Fig.34 results are given
(a) for thickness induced reorientation transition in b.c.c. Fe
films,(b) for temperature induced transition at $T_R$, and (c) for
the dependence of the phase diagram (PD) on anisotropy constants,
$K_4$,$K_4^\perp$ and $K_4^\parallel$ are included. The
free--energy $\Delta F=F(T,\theta=0)- F(T,\theta=\pi/2)$ is
calculated with $F=-\Sigma_i K_{2i}(m(T))\cos^2\theta +\ldots$.
Here, $i$ refers to the film layer and $\theta$ to the angle between
film normal and direction of magnetization, see Jensen {\it et
al.} \cite{B2,B7}. Note, $\Delta F<0$ indicates a perpendicular magnetization
and $\Delta F>0$ an in plane one.
\begin{figure}
\centerline{\includegraphics[width=0.62\textwidth]{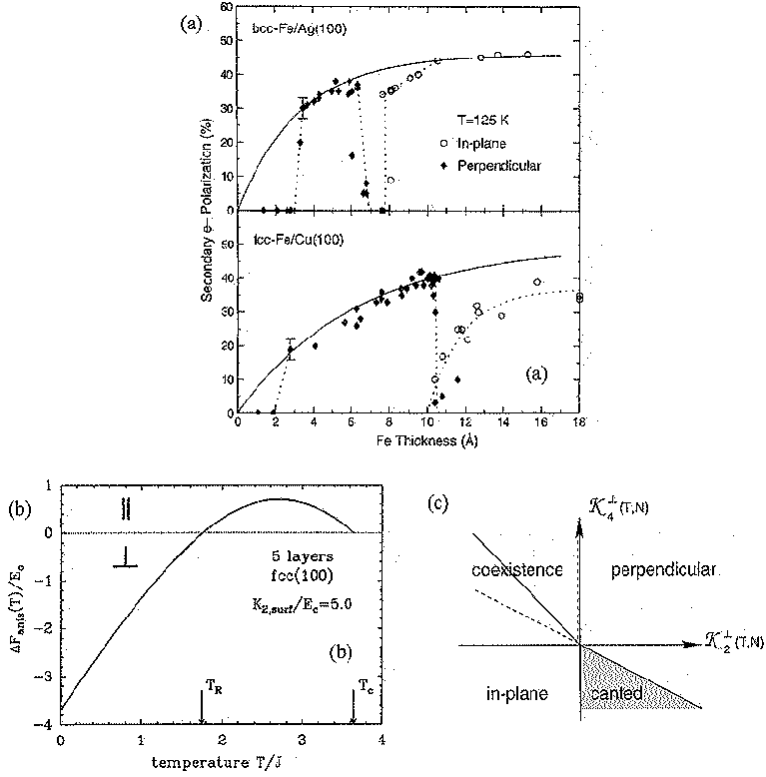}}
\caption{The magnetic reorientation transition. (a) Transition
$M_\perp \longrightarrow M_\parallel$. The solid curve gives the
magnetization of an uniform (single domain) film. occurring for
increasing film thickness and for Fe$\mid$Cu films. (b)
Free--energy change $\Delta F=F(M\bot)-F(M\|)$ per spin and for
increasing temperature. Magnetic anisotropy is temperature
dependent and this causes the reorientation transition at $T_R$. A
fcc(100) film is assumed, see Jensen {\it et al.}. (c) Phase
diagram for the orientation of the magnetization is controlled by
the anisotropy parameter $K_2^\bot(T,N)$ and $K_4^\bot(T,N)$, see
Eq.(49). Note, in the coexistence phase the energy minima of the
states with perpendicular ($\bot$) and parallel ($\|$)
magnetization are separated by an energy barrier. The P.D. results
from minimizing the free--energy ($K_4^{\perp}$ refers to $M_{\perp}$).}
\end{figure}
\begin{figure}
\centerline{\includegraphics[width=0.38\textwidth]{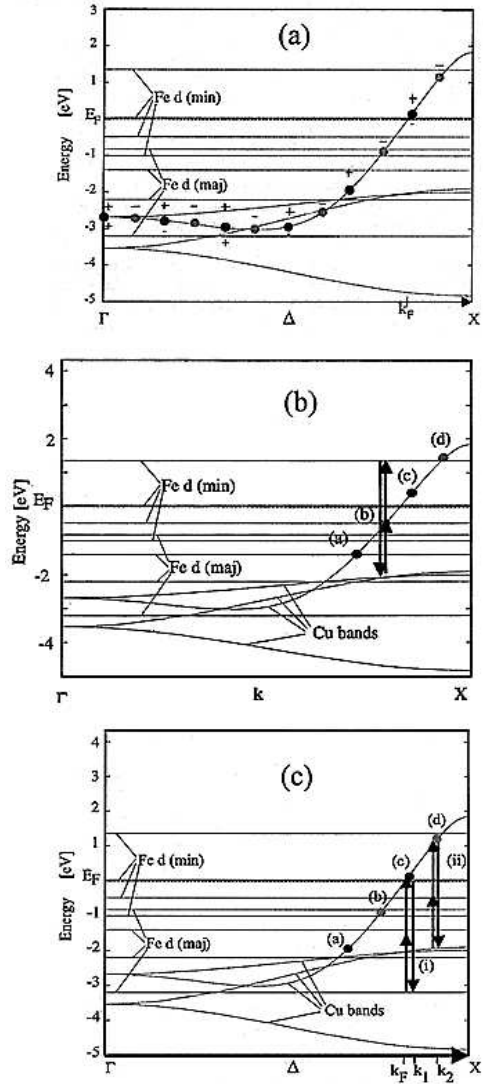}}
\caption{Magnetooptics of thin films (xCu/Fe(001) involving quantum well states
(QWS) due to confinement. As a consequence of  QWS presence the SHG is strongly film thickness dependent.
SHG involving QWS exhibits characteristic SHG oscillations.
(a) QWS in thin films, the parity of the wave
functions is indicated by (+) and (-). Energies of QWS along
$\Gamma\rightarrow X$ direction of BZ are given. Black dots refer to a 6ML film and black
and grey dots to a 12ML film. The parity of the states is
indicated. (b) SHG involving as intermediate state an occupied QWS below the
Fermi--energy, the final state is a Fe d--state. Transitions
contributing dominantly to SHG are shown. One gets corresponding
SHG oscillations as a function of film thickness. (c) Interference
of SHG from surface and interface is negligible ( for example (xCu/Fe(001)). The important SHG
optical transitions involving QWS as final states are indicated. Transitions (i)
are possible for Fe films with 6ML and 12ML, (ii) only for films
with 12 ML Fe. Thus, one gets SHG oscillations having a 6ML
period. For details see discussion by Luce and Bennemann
({\it Nonlinear Optics, Oxford}).}
\end{figure}

\subsubsection{Magnetooptics}

Magnetism of thin films and multilayers thereof is well determined
by their magnetooptical behavior. In particular SHG sensing
sensitively structural symmetry, surfaces and interfaces, is very
useful for studying magnetic properties of nanostructures.
Magnetic thin films cause characteristic SHG reflecting magnetism,
since the nonlinear susceptibility $\chi_{ijl}$ determining the
light intensity $I(2\omega)$ of SHG depends on magnetization
$\overrightarrow{M}$.

As usually one splits the susceptibility in
parts even and odd in $M$ and then the magnetic contrast
\begin{equation}
      \Delta I(2\omega)\propto I(2\omega, M)- I(2\omega, -M)
\end{equation}
yields sensitively the magnetic behavior. Resonantly enhanced SHG
transitions of the ferromagnet which states are spin split are a
sensitive measure of magnetism. Note, the interference of SHG from
the film surface and interface (film/substrate) is of interest, in
particular if magnetism is different at surface and interface, see Fig.11.

In table 1 characteristic features of SHG response from thin films are given. We refer to weak interference of SHG when $\chi^i\gg\chi^s$ and to strong interference when $\chi^i\approx\chi^s$. Note, the wave vector $\overrightarrow{k}$ selectivity is controlled by band--structure. For the film system Cu/Fe/Cu the dipole matrix elements with $d$--states cause $\chi^i\gg\chi^s$. For Au/Co/Au films no Co $d$--states are involved and
therefore $\chi^i\approx \chi^s$. The periodic appearance of QWS for increasing film thickness may cause oscillations in SHG.

In thin films (which may be described with respect to their
electronic structure by potential wells) one gets QWS (quantum
well states) and corresponding contributions to MSHG (magnetic
second harmonic light). As supported by experiments QWS in thin
films may contribute strongly to SHG (quasi resonantly enhance
SHG). Since the occurrence of QWS is dependent on film thickness,
one gets QWS involving characteristic oscillations in SHG as a
function of film thickness. These oscillations depend of course on
light frequency , magnetism, parity of the QWS, inversion symmetry
of the film and interference of SHG from surface and interface.

If interference of nonlinear light from surface and interface is
important, then SHG response depends on QWS parity. As a
consequence only a period twice as large as for the linear Kerr
effect optics appears. If this interference is not so important,
then SHG exhibits typically two oscillation periods. For details
of such behavior see Luce, Bennemann (\cite{B12}) and also discussion given previously, see Fig.18 etc..

SHG involving QWS is shown in Fig.35. In Fig.35(a)
thin film  QWS and their parity are sketched. The Figs. refer to
xCu/Fe(001) system. For other systems the analysis is similar.
Fig.35(a) shows the energies of the QWS. The parity is indicated
by (+) and (-). Note, at the Fermi energy a parity change occurs.
The spin polarized d--states of Fe are also shown. In Fig.35(b)
dominant SHG contribution resulting from an occupied QWS below the
Fermi--energy is illustrated (final state of SHG transition is a
d--state of Fe). In Fig.35(c) SHG involving unoccupied QWS is
shown. Note, states along $\Gamma\rightarrow X$ direction in the Brillouin
zone (BZ) are most important. Interference of surface and
interface SHG is negligible. Dominant SHG transitions are
indicated: (i) are possible for 6ML and for 12ML of Cu. This gives
oscillations with a period of 6ML. (ii) possible only for 12ML.
Thus, one gets two different oscillations (which are frequency
dependent, at $k_1$, $k_2$ in the BZ).

Note, for $\chi^i\chi^s \longrightarrow -1$ one gets a
cancellation of the thickness dependent contribution to SHG due to
QWS. If $\chi^i$ and $\chi^s$ become comparable parity of QWS gets
important. For example, for a dominant QWS close to the
Fermi--energy one gets oscillations only for one with odd parity,
see discussion by Luce {\it et al.} \cite{B12}.

This shows the interesting magnetooptical behavior of thin films.

\subsubsection{Nanostructured Thin Films with Magnetic Domains}

General remarks: In the following the magnetic properties of generally irregular nanostructures at equilibrium and nonequilibrium are discussed. Generally one observes that the magnetic properties of nanostructured films depend on the film growth conditions. During growth irregular film structures with varying island sizes, shapes appear. Competing interactions between islands (exchange, dipole,..) occur. Magnetic anisotropy controls the formation of domains, their size and shape and magnetic relaxation of the domains. The relaxations determine the transition from nonequilibrium to equilibrium. In view of the complex behavior one may use Monte Carlo simulations. To describe the transition from isolated islands to connected ones correlations of magnetic relaxation (of neighboring islands) must be taken into account. For example cluster (island) spin flips will lead to a much faster magnetic relaxation than single spin flips.

Main questions are (1) what is the domain structure, (2) what is the time dependent approach of the film magnetization towards equilibrium, (3) what are the controlling forces. If only short range interactions (exchange ones) are active
then one expects that percolating atomic structures (coverage $\theta\geq\theta_p$) are necessary for long range global magnetic ordering. Ordering for coverages below $\theta_p$ might indicate dipolar interactions.

The time--dependent structural and magnetic behavior of films
during growth reveals many interesting properties and in
particular the important interplay of atomic structure and
formation of magnetic domains and global film magnetization. As
the film growth the size of the domains, its surface roughness and
the orientation of the domain magnetization changes. Magnetic relaxation
occurring typically on a ps to ns time scale causes in thin films
a time dependent magnetization which approaches only at relatively
long times (via thermal activation etc.) the equilibrium
magnetization. Depending on the domain size and distances between
the domains exchange or magnetic dipole coupling may dominate and
cause corresponding, different relaxation. Thus, the approach of
inhomogeneous ultrathin films towards equilibrium magnetization
and its dependence on film topology requires careful studies.

Important parameters controlling film behavior are the blocking
temperature $T_{bl}$ ($T_{bl}\sim KN/k_B\ln(t_{exp}\Gamma_o)\sim 5K$),
for $N\simeq 1000$ atoms resulting from magnetic anisotropy and
the percolation coverage $\theta_p$. While magnetic ordering due
to exchange coupling occurs for $T<T_c\sim J$, long--range
magnetic dipolar coupling causes ordering for $T\leq
T_{dip}\sim(N\mu)^2/R^3$ (which for N=1000 atoms gives about 5K).
$R$ refers to the distance between domains. As mentioned due to its complexity
analysis requires generally Monte Carlo simulations, see Jensen {\it et al.} \cite{B8,B11}.
\begin{figure}
\centerline{\includegraphics[width=0.5\textwidth]{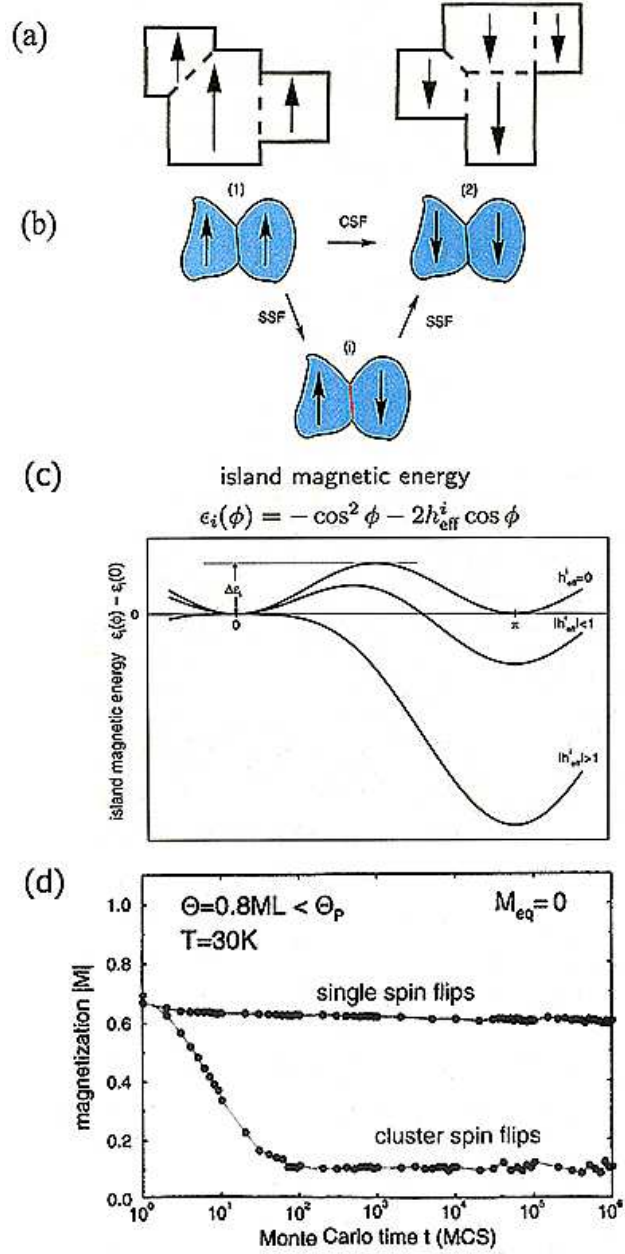}}
\caption{Sketch of (a) coherently relaxing  connecting magnetic
domains. (b) relaxing domains illustrating model assuming coherent
relaxation and successively single domain relaxation used by
Jensen {\it et al.}(CSF: Coherent spin flip of neighboring
domains, SSF: Single domain spin flip). (c) energy barrier of a
domain on which a molecular field $h_{eff}$ acts for magnetization
reversal. The barrier energy ($\epsilon_i=E_i^b/KN$) controls the
reversals of the domain magnetization and results from magnetic
anisotropy ($\sim K$) and (interdomain, interlayer) molecular field
$h_{eff}^i$. (d) Test results for the relaxation of the magnetization using CSF approximation.}
\end{figure}
\begin{figure}
\centerline{\includegraphics[width=0.6\textwidth]{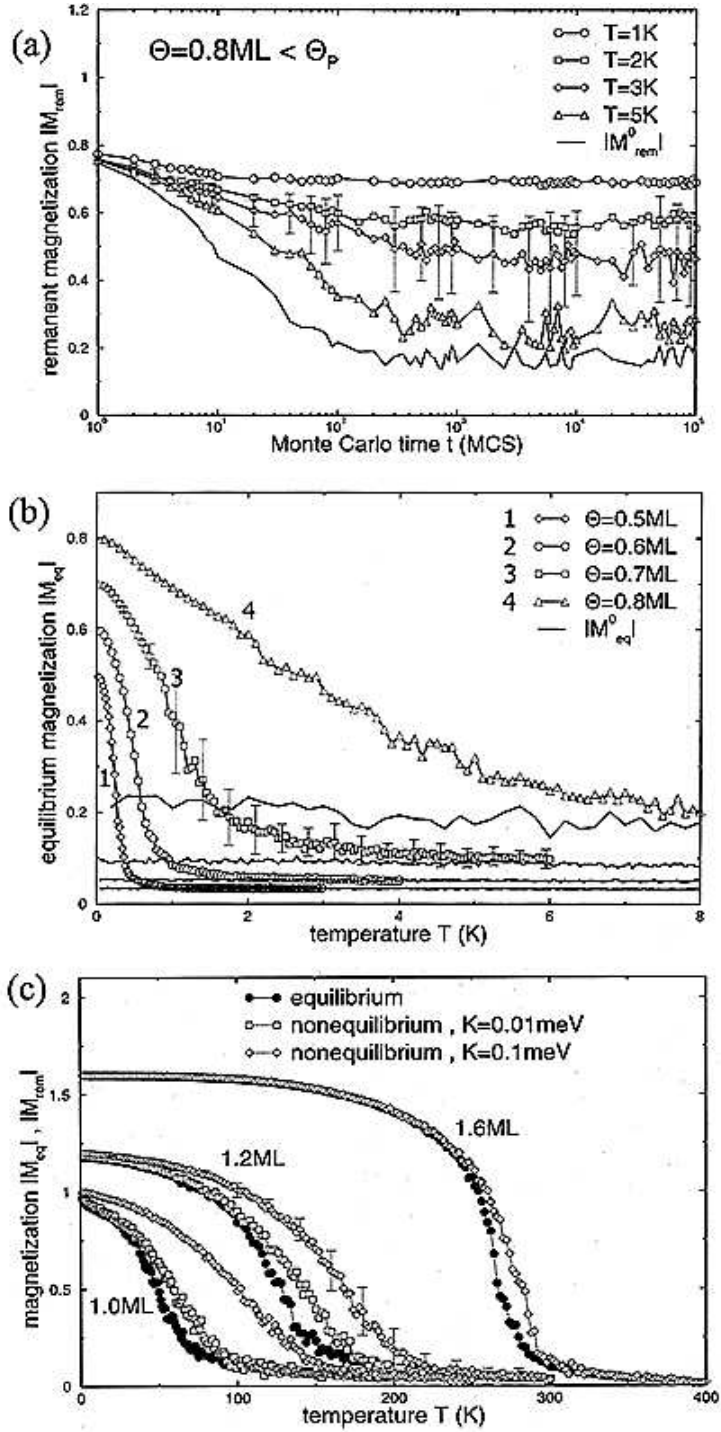}}
\caption{Magnetic relaxation in thin films for different film
thickness and temperature. (a) Time dependent long range ordering
in case of exchange and dipole coupling and for varying
temperature. Note, $\theta <\theta_p$. The relaxation starts from
fully aligned domain magnetization. At low temperatures ($T\sim
1K$) a stable magnetization due to dipole interactions is
obtained. (b) Equilibrium magnetization as a function of
temperature for $\theta < \theta_p$. Here, $\theta_p$ denotes the
percolation coverage ($\theta_p=0.9ML$). (c) Equilibrium magnetization as a function
of temperature for different anisotropy (K) and film thickness.
Note, the difference between nonequilibrium and equilibrium
magnetization.}
\end{figure}
\begin{figure}
\centerline{\includegraphics[width=0.6\textwidth]{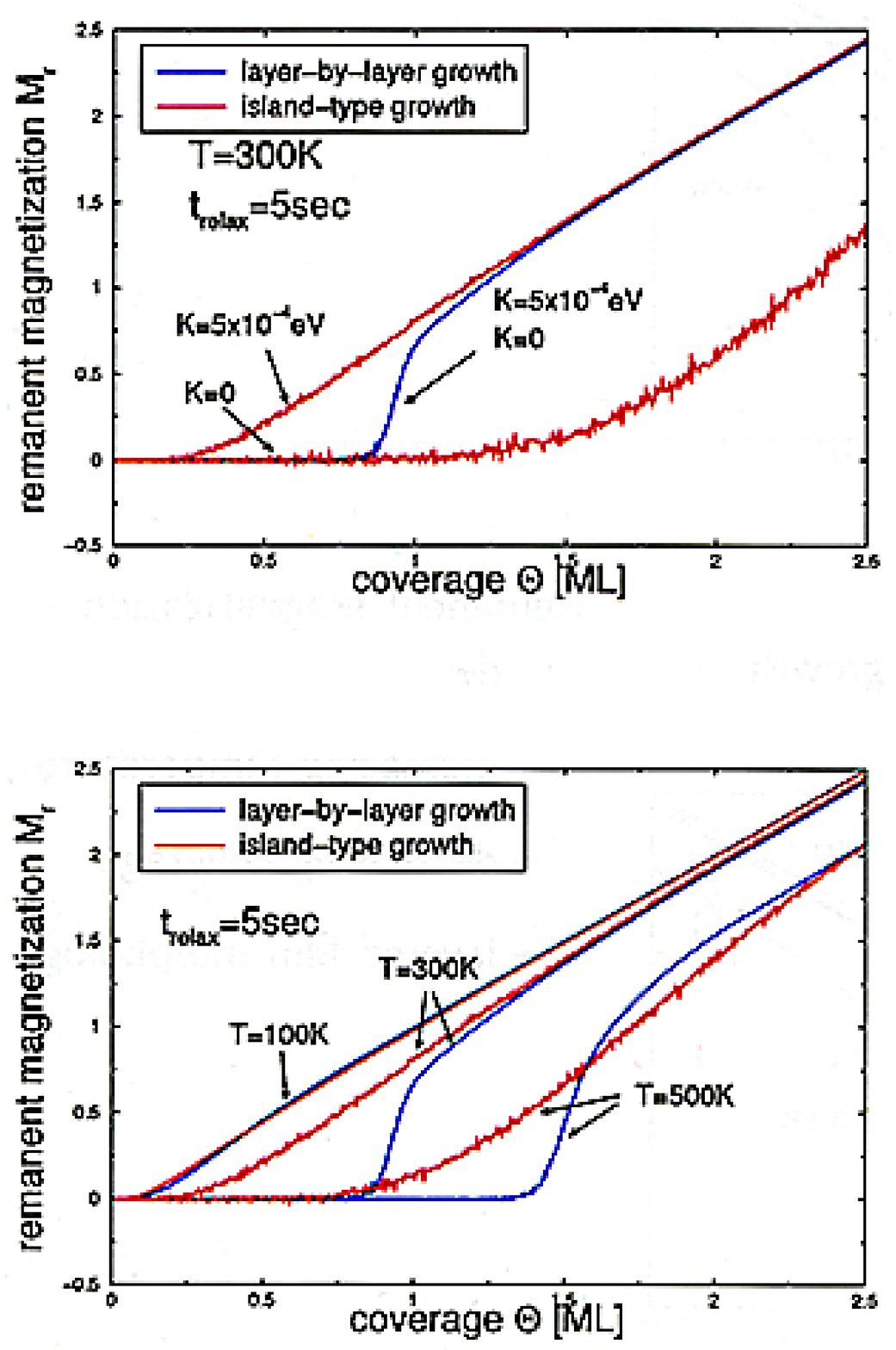}}
\caption{Magnetic relaxation dependence on film structures, on
growth mode, layer by layer vs. island growth. Note, dependence on
anisotropy $K$ and temperature. (upper Fig.) For island growth the
magnetization depends strongly on magnetic anisotropy ($K$). (lower
Fig.) Magnetization in case of layer by layer growth depends
strongly on exchange interactions between islands.}
\end{figure}
In Fig.36 magnetic domain relaxation is sketched and the typical
energy barriers against reversal of the direction of an island
(domain) magnetization are shown.

The magnetic relaxation is calculated using a Monte Carlo method
and a Markov master equation
\begin{equation}
dP(x,t)/dt = -\Sigma\Gamma(x\rightarrow x^{'})P(x,t) +
\Sigma\Gamma(x^{'}\rightarrow x)P(x^{'},t)\quad , \nonumber
\end{equation}
where $P$ gives the probability of a spin state $x={S_1,S_2,\ldots}$ at time $t$. $\Gamma(x\rightarrow x^{'})$ denotes the transition rate.
$\langle M(t)\rangle \propto\Sigma_i M_i(t)$, $i$= time steps
taking into account possibly coherent spin flips of connected
domains, see Brinzanik, Jensen {\it et al.} \cite{B11}.
The Markov equation in combination with the M.C. statistical methods yields the time
dependent non-equilibrium
magnetization and its relaxation towards the equilibrium one.
Depending on the density of the domains, their distance, correlated relaxation of neighboring domains occurs. This is demonstrated in Fig.36(c) where results obtained assuming cluster spin relaxation vs. single spin relaxation are shown.

For a domain structured film the film magnetization is given by
\begin{equation}
               \overrightarrow{M} \propto \Sigma_i m_iN_i \langle\overrightarrow{S_i}\rangle,
\end{equation}
where $\overrightarrow{M}$ points along the easy axis and $i$ refers to the $i-th$ island with $m_i$ aligned magnetic moments. Here, one averages over the magnetization directions of the domains. (The kinetic Monte Carlo method gives after equilibration for the average magnetization $\langle M\rangle \approx (1/R)\Sigma_{r=1}^{R} M_r(t))$.

\begin{figure}[tp]
\centerline{\includegraphics[width=0.5\textwidth]{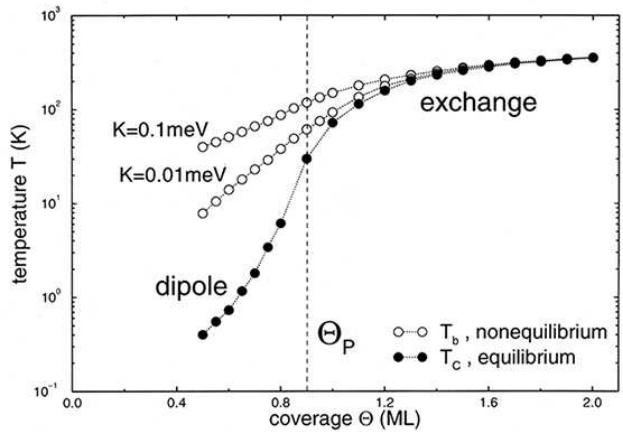}}
\caption{Dependence of Curie--temperature $T_c$ and blocking
temperature $T_b$ on film thickness (coverage $\theta$).
$\theta_p$ denotes the percolation coverage. Note, a
semilogarithmic plot is used. The curve for the ordering
temperature due to dipole coupling neglects for simplicity domain
wall energy. As $\theta\longrightarrow\theta_p$, the exchange
interactions begin to dominate, for $\theta < \theta_p$ dipole
coupling plays a role and yields a small $T_c$. The nonequilibrium
behavior is mainly due to magnetic anisotropy.}
\end{figure}
In Fig.37 magnetic relaxation of thin films is shown (for details
see Jensen {\it et al.}) \cite{B11}. Results obtained below and above percolation
coverage $\theta_p$ are compared. Note, (for $\theta>\theta_p$) the strong influence of single island anisotropy $K$ on relaxation. For increasing connectivity of the islands faster relaxation results from exchange interactions between islands. Note, the temperature and film thickness dependence of the relaxation.

The MC results demonstrate that faster relaxation occurs for correlated spin flips (CSF), as expected of course. As also expected different behavior occurs below and above percolation coverage ($\theta_p$).

In Fig.38 it is shown how the relaxation of the magnetization depends on the atomic film structure. The dependence of the remanent magnetization of a film on its growth mode, layer by layer or island type one is given.
Note, the dependence of the results by Jensen {\it et al.} on
anisotropy and temperature.
\begin{figure}
\centerline{\includegraphics[width=0.75\textwidth]{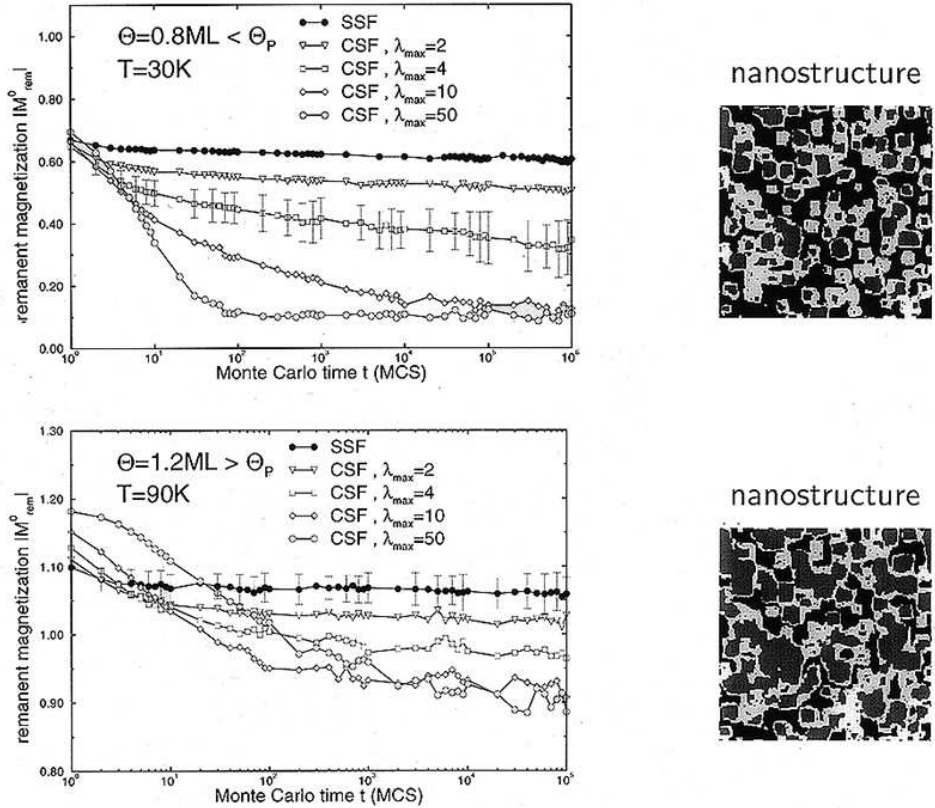}}
\caption{Magnetic relaxation in domain structured thin films using Monte Carlo type calculations performed by Jensen {\it et al.} \cite{B11}. $\lambda$ refers to different spin clusters and $\theta_p$ to the percolation coverage. CSF refers to correlated cluster (island) spin flips. For comparison also results are given assuming single spin flips (SSF). The atomic nanostructures corresponding to the magnetic domain structures are also shown and were obtained using an Eden growth model, for details see Jensen {\it et al.}. Regarding nanostructures white and grey areas refer to magnetic domains and black areas to uncovered surface of the thin film.}
\end{figure}

In Fig.39 ordering in thin films is characterized. Depending on
the density of domains exchange and dipolar coupling act
differently. Clearly the percolation coverage is an important
control--parameter for the magnetic behavior and range of magnetic
interactions. At low coverage dipole coupling may dominate and
then ordering occurs at low temperatures.
\begin{figure}[tp]
\centerline{\includegraphics[width=0.68\textwidth]{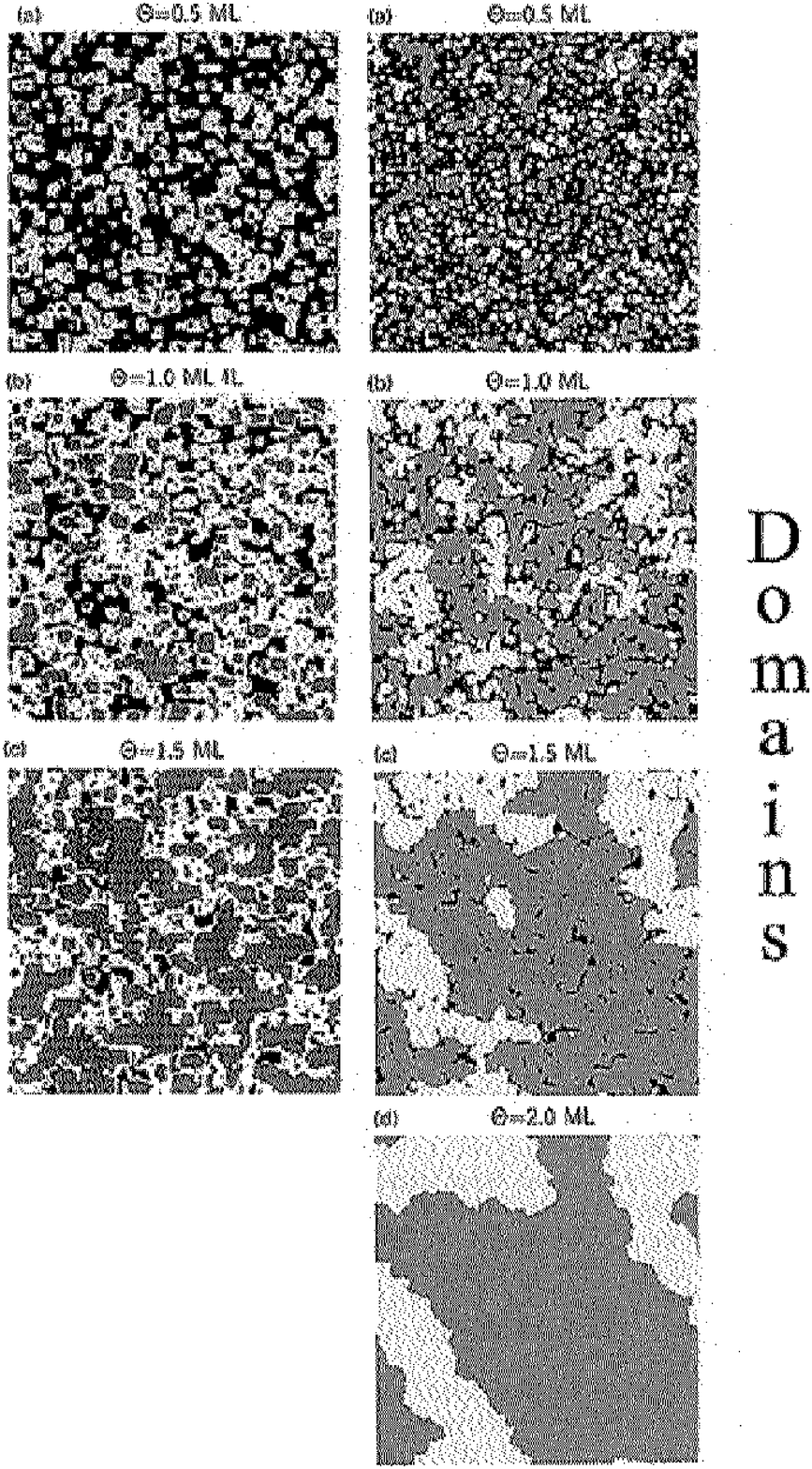}}
\caption{(left) Film nanostructures using Eden type growth model, (black: uncovered surface, grey (light) to first layer atoms, dark to first layer atoms);
(right) corresponding resulting magnetic domains and their dependence on film
thickness. (black: uncovered surf., grey: two opposite magnetic directions). Only domains at the surface are marked, see results by Brinzanik {\it et al.}.}
\end{figure}

In Fig.40 the dependence of magnetic behavior, of domain structure on film morphology is shown. Results were obtained by Jensen {\it et al.}
using a kinetic M.C. simulation and Markov equation for the
relaxation of the magnetic domains are given. Film structures
resulting from an Eden growth model and resulting magnetic domains
are shown \cite{B8,B11}. Regarding calculations an area of 500$\times$500 lattice constants is taken
which contains about 1250 islands. $\lambda$ refers to different spin cluster sizes.\cite{B2,B8,B11}

Of course, magnetic ordering in thin films depends sensitively on
topology and film thickness and the dominant magnetic coupling
between domains. Results presented demonstrate this.

The results presented in Fig.41 for the domain size (and results
on their roughness and surface) are very important regarding
understanding of magnetic behavior of irregular thin films and
applications.

In Fig.41 results are given for the magnetic domain structure in thin films having different thickness.
Of course, the magnetic domain structure changes and the size of
the domains increases as the film thickness increases.

In Fig.42 results for magnetic domains and their size dependence on film thickness and
temperature are given. These were obtained using the previously discussed analysis
by Jensen {\it et al.}. Note, as a function of temperature one may
get a maximum in the domain size. The reason is that for low
temperatures domain growth is hindered by energy barriers and at
higher temperatures thermal activation disintegrates domains.

Note, generally for increasing film thickness the domains which are first
isolated begin to merge and form larger connected areas, still
markedly affected by the atomic nanostructure of the film. The
roughness of the domains is affected by the nanostructure and
decreases for thicker films and for increasing temperature. Of
course, this is expected on general physical grounds.
\begin{figure}[tp]
\centerline{\includegraphics[width=0.95\textwidth]{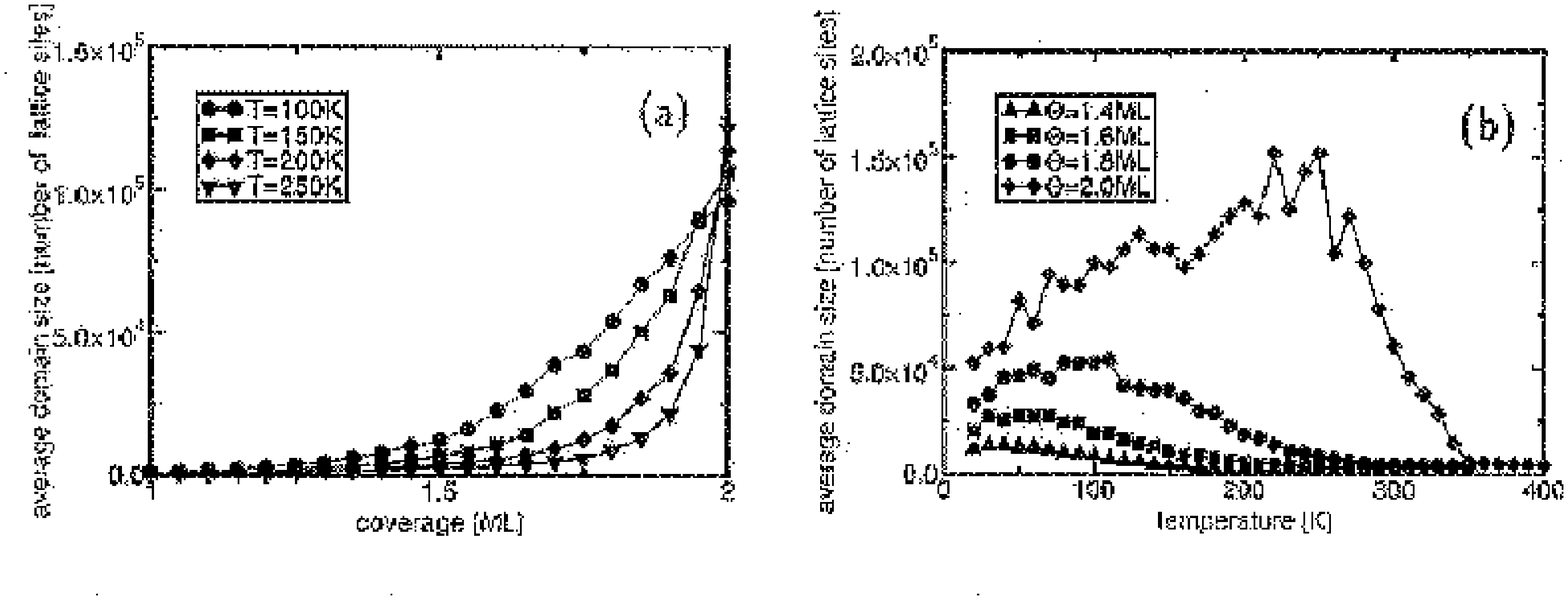}}
\caption{Average magnetic domain size as a function (a) of coverage (film
thickness) and (b) temperature. The film growth model is described
in the text (see Jensen {\it et al.}). The percolation threshold
is $\theta_p\sim 0.9ML.$}
\end{figure}

\subsection{Tunnel Junctions: Magnetic Effects}

Nanoscaled tunnel junctions offer interesting physics in
particular regarding quantum mechanical effects and spin dependent
currents, separation of charge and spin currents and their
interdependence. Also one may use such nanostructures ($FM\mid
M\mid FM$), $M$= normal state, superconducting state, etc.) as
interesting fast switches and magnetoresistance devices. Note, the
Josephson tunnel current is carried by Andreev--states, see
Kastening, Morr {\it et al.}. In the following some
\begin{figure}[tp]
\includegraphics[width=0.49\textwidth]{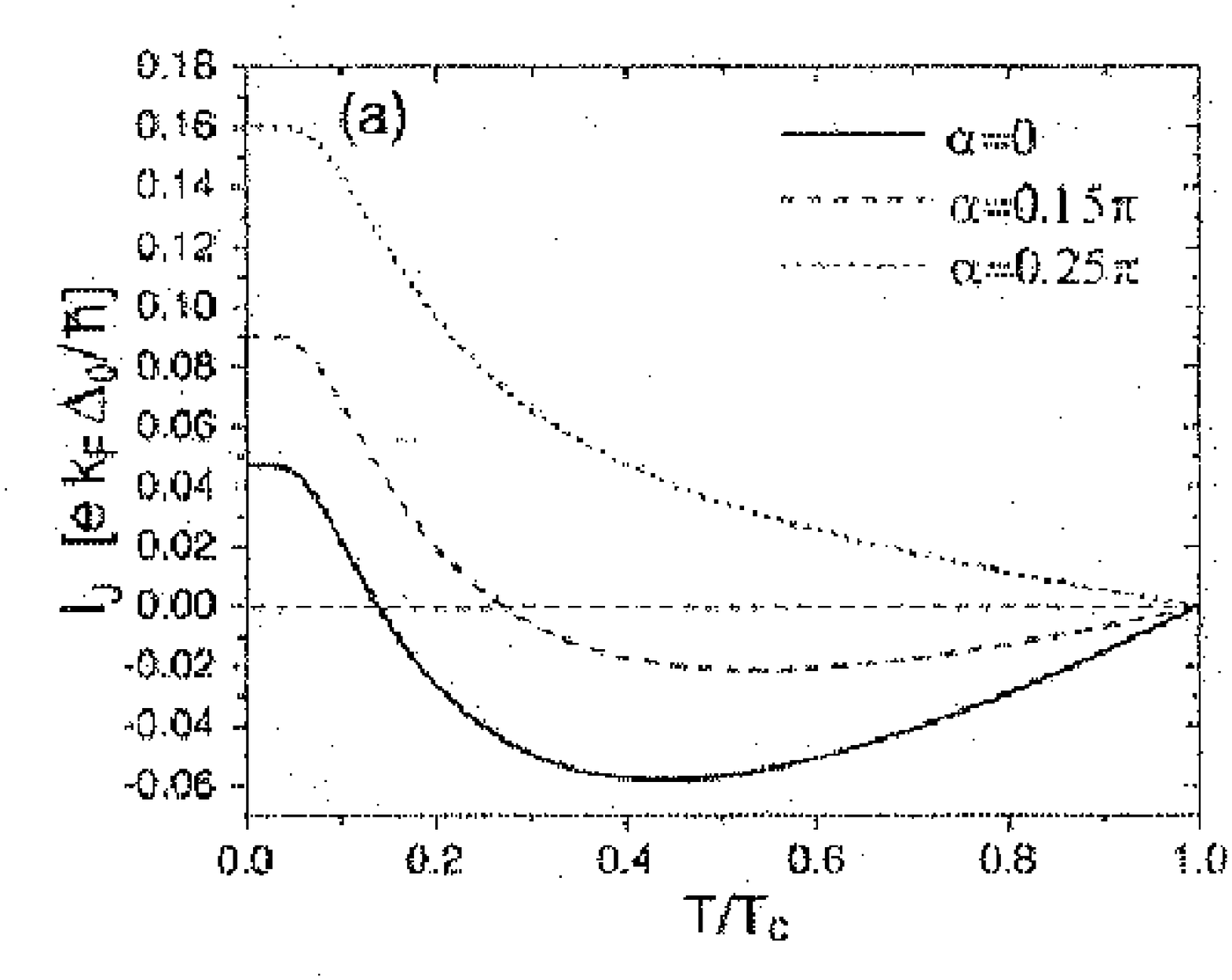}
\includegraphics[width=0.49\textwidth]{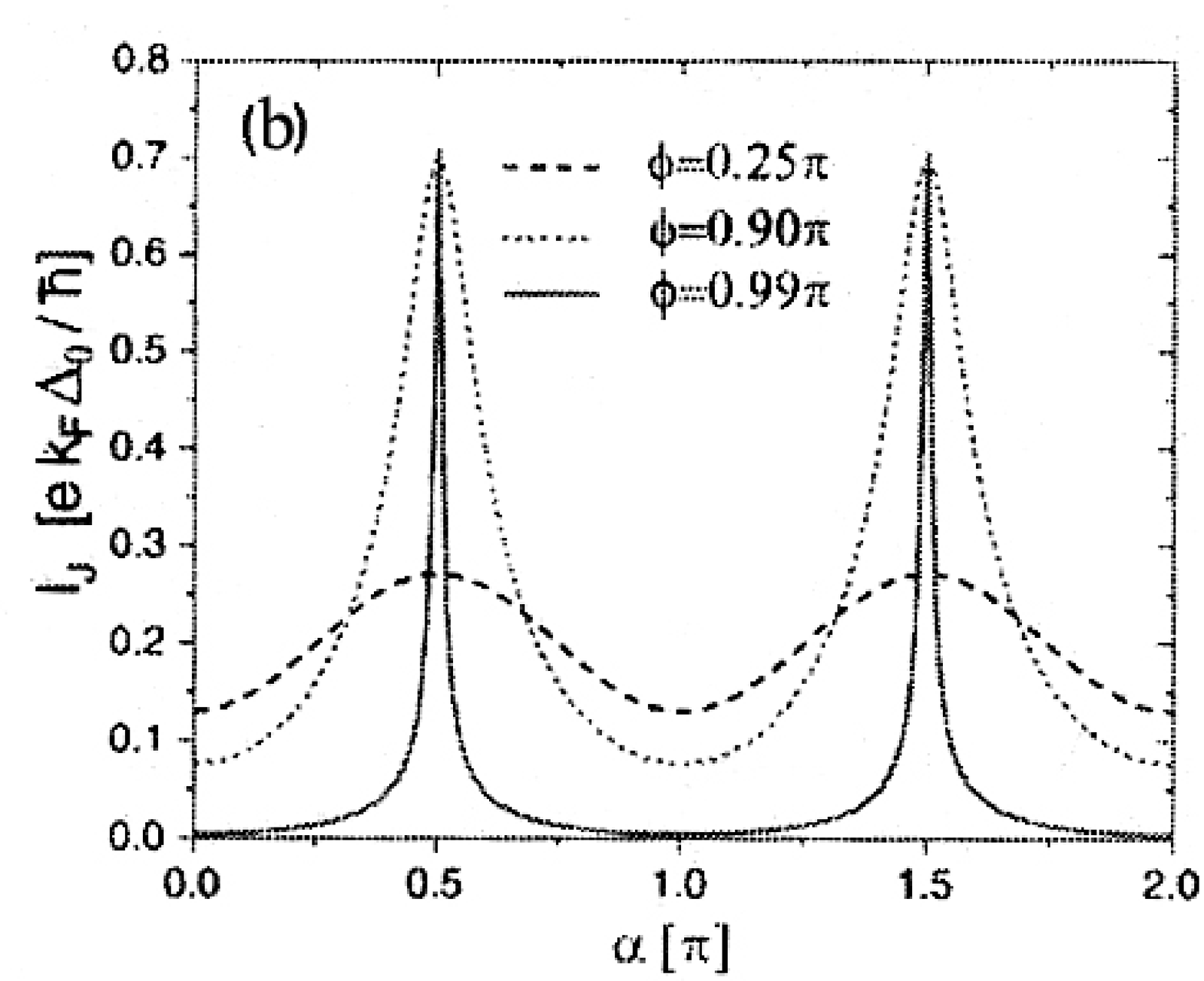}
\caption{Tunnel junction ($TSC\mid FM\mid TSC$), where FM and TSC refer to a ferromagnet and triplett superconductor, respectively, $\phi$ to the phase of the superconductor: (a) Josephson
current $I_J$ as a function of temperature (changing sign), (b) $I_J$ as
a function of $\alpha$, the angle between magnetization
\protect{$\protect\overrightarrow{M}$} and direction normal to the
current. Note, $I_J$ may be switched between zero and a finite
value.}
\end{figure}
examples are discussed. Note, different behavior may result dependent on times $\tau_{CP}, \tau_s,$ and $\tau_t$ referring to Cooper pairs, spin diffusion and tunneling transport time, respectively.

For a tunnel junction
\begin{equation}
    (TSC\mid FM\mid TS)
\end{equation}
where TSC denotes a triplet superconductor and FM a ferromagnet,
one gets interesting behavior of the Josephson current $j_J$ as a
function of temperature, $\alpha$ and $\theta$. The angles
$\alpha$ and $\theta$ determine the direction of the magnetization
of the ferromagnet and the direction of the TSC--order parameter,
see Fig.20(b) for illustration. The Andreev states are determined from $H \psi_j = E \psi_j$ using Bogoliubov--de Gennes method.

As shown by Fig.43 the Josephson current may change sign as a function of temperature and may be switched between
zero and finite value upon varying the angles $\alpha$, $\theta$.
The unconventional change of sign of $I_J$ (without change of phase) for increasing temperature results from the changing occupation of the Andreev states and from the factor $\frac{\partial E_i}{\partial\phi}$ in the expression for the tunnel current. Note, this sign change is different than the one observed for singulet superconductors with a ferromagnetic barrier in between. The behavior of $I_J$ for rotating magnetization, as a function of $\alpha$, suggests to use such junctions as switching devices. Note, for increasing $\phi$ the current changes more drastically.

Of particular interest is the behavior of $I_J$ as a function of phase $\phi$ (for T=0) shown in Fig.43(b). Then only the states $E_i<0$ are occupied. For (1) $M\parallel d_{L,R}$, $\alpha\neq\pi/2$, the ferromagnet (FM) couples Andreev states, and (2) for $ \overrightarrow{M}\perp \overrightarrow{d}_{L,R}$ the Andreev states are not coupled by the FM and spins are well defined.

Clearly the above junction would be a sensitive probe to detect triplet
superconductivity, see results by Morr {\it et al.} \cite{B3}.
\begin{figure}
\centerline{\includegraphics[width=0.6\textwidth]{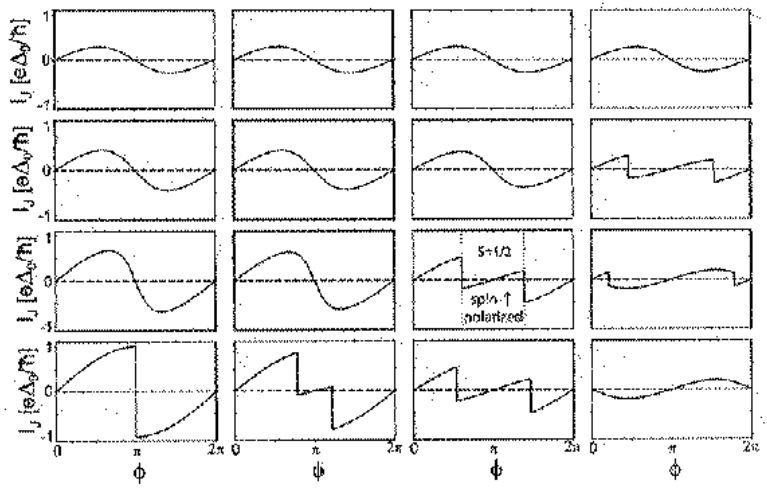}}
\caption{Josephson current $I_J$ at $T=0$ for a junction (SC/FM/SC) as a function of phase
$\phi$ of the superconductor for several values of the scattering strength $g$ and $z$ of the magnetic and nonmagnetic potential, respectively
($g=0,1/3,2/3,1$ from left to right, $z=0,1/3,2/3,1$ from bottom
to top), see results by Kastening, Morr {\it et al.}. Note, the ferromagnet is approximately represented by a potential barrier.}
\end{figure}

Another tunnel system is one involving singlet superconductors,
\begin{equation}
    SC\mid FM\mid SC \quad .
\end{equation}

The FM is approximately represented by a barrier potential scattering spin dependently the electrons ($g$ refers to the magnetic scattering strength, $z$ to the non--magnetic one). Again, depending on nonmagnetic and magnetic scattering of the
tunnelling electrons the Josephson current may change sign as a
function of temperature $T$ and furthermore as a function of the
relative phase of the two superconductors, see Fig.43 and Fig.44.
(Note,$I_J=I_\uparrow+I_\downarrow, I_s=I_\uparrow-I_\downarrow$,
the Josephson current $I_J$ is solely carried by the Andreev states,
the total spin current $I_s$ = 0, since spin current through Andreev states is canceled by the one
through the continuum states).

For $z \gg g$ one gets $I_J=(e\Delta_0/\hbar)\frac{\sin\phi}{2Z^2}$
and non magnetic current is dominated by Cooper pairs. For $g\gg z$ one has
$I_J=-(e\Delta_0/\hbar)\frac{\sin\phi}{2g^2}$ and current is carried by single electrons.
Note, then the phase shift by $\pi$. Results assume for the thickness of the tunnel junction $d$ to be smaller
than the coherence length, otherwise the situation is more complicated.
\begin{figure}
\centerline{\includegraphics[width=0.5\textwidth]{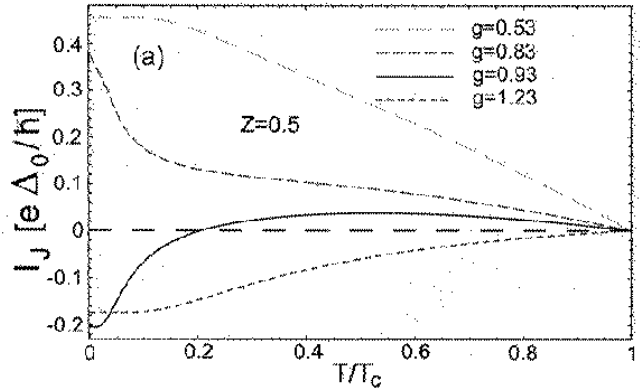}}
\caption{Josephson current $I_J$ of a tunnel junction (SC/FM/SC) as a function of $\frac{T}{T_c}$ for $\phi=\pi/2 $ and nonmagnetic scattering strength $z=0.5$ and several values of $g$, the magnetic scattering strength.}
\end{figure}

Note, the sign change of the Josephson current as a function of temperature shown in Fig.45 results not from a transition of the junction from a $\pi$--state at low temperature to a 0--phase at high temperature, but from a change in the population of the Andreev states (while the relative phase between the superconductors remains unchanged).

Also interestingly, the total spin
polarization of the two superconductors ground state changes from
$\langle S_z\rangle=0$ to $\langle S_z\rangle=1/2$.

Of course, as mentioned already interesting results are also expected
for ($FM_1/SC/FM_2$) junctions. The tunnel current depends on the superconducting state,
singulet vs. triplet. Also a spin current is expected for $\tau_s>\tau_t$, times refer as before to times without spin flip and tunneling time. Singulet superconductivity may block tunneling. For a triplet superconductor the angle between Cooper pair angular momentum $\overrightarrow{d}$ and $\overrightarrow{M}$ may control the tunneling.

Transport between quantum dots: An interesting case of electron transport (electron pump model)
between two quantum dots involving possibly Coulomb-- and
spin--blockade (see Hubbard hamiltonian with spin dependent
on--site interaction) is illustrated in Fig.20(a). Driving the
current with a pulsed (polarized) external field to overcome an
energy barrier one gets spin--dependent charge transport with v.
St\"{u}ckelberg oscillations due to bouncing forth and back of the
electrons between the two quantum dots.

In Fig.46 results are given for the photon assisted tunneling between two quantum dots. The applied electromagnetic field is given by $V(t)=V_0\cos(\omega t)$. Note the dependence of the charge transfer on the duration of the applied light pulse and the Rabi oscillations due to the bouncing back and forth of the electrons. The Rabi oscillation frequency is $\Omega=2\omega J_N(E=V_0/\hbar\omega)$. Here $J_N$ is a Bessel function of order N, where N refers to number of photons absorbed to fulfill resonant condition $N\hbar\omega=\sqrt{\Delta\varepsilon^2+4\omega^2}$.
Different frequencies are used which cause resonant absorption of one, two and three and possibly more photons. Of course this affects the charge transfer. The time resolved analysis of the occupation of the electronic states shows that system if connected to reservoirs acts as electron pump. Before action of the external electromagnetic field the initial state is $n_1= 1$ and $n_2= 0$. After the pulse is over oscillations disappear and one gets again the initial state via transferring one electron to the right quantum dot and the left reservoir donating one electron to the left quantum dot, see Garcia {\it et al.} \cite{B3}. Spin dependent quantum dot electron states cause spin currents.

Of interest is also the coherent control of photon assisted tunneling between quantum dots and its dependence on the shape of the light pulse, see Grigorenko, Garcia {\it et al.} \cite{B3}. The shape of the external electric (or magnetic pulse) may be optimized to get a maximal charge or spin current between the quantum dots connected to two metallic contacts.

Note, these results are also of interest for Fermion or Boson systems on optical lattices, its dynamics and interaction with external fields. Optical manipulation of molecular binding, in particular Boson formation induced by an electromagnetic field is an option. In intense fields nonlinear behavior may be particular interesting. Furthermore, on optical lattices one may study in particular the interplay of magnetism and lattice structure, the transition from local, Heisenberg like to itinerant behavior of magnetism. Important parameter for the occurrence of ferromagnetism (antiferromagnetism) are Coulomb interactions and particle hopping (kinetic energy) between lattice sites, and possibly also spin--orbit coupling. Note, as indicated by Hund's rules ferromagnetism could occur already for a Fermi--liquid with strong enough repulsive Coulomb interactions, quasi irrespective of the lattice structure, since spin polarization (together with the Pauli principle) may minimize the repulsive Coulomb interactions. In case where a.f. dimerization of neighboring lattice sites yields Boson formation and resulting BEC condensation optical influence, also of the magnetic excitations, is an important option.

In Fig.47 the possibility of manufactoring optically an ultrafast
switching device is sketched. Note, in general changing in
nanostructures the magnetization generates in accordance with
Maxwell--equations light.

Via hot electrons fast changing temperature gradients may yield interesting and novel behavior of tunnel junctions. Possibly in such way Onsager theory can be tested for nanostructures.

\begin{figure}
\centerline{\includegraphics[width=1.\textwidth]{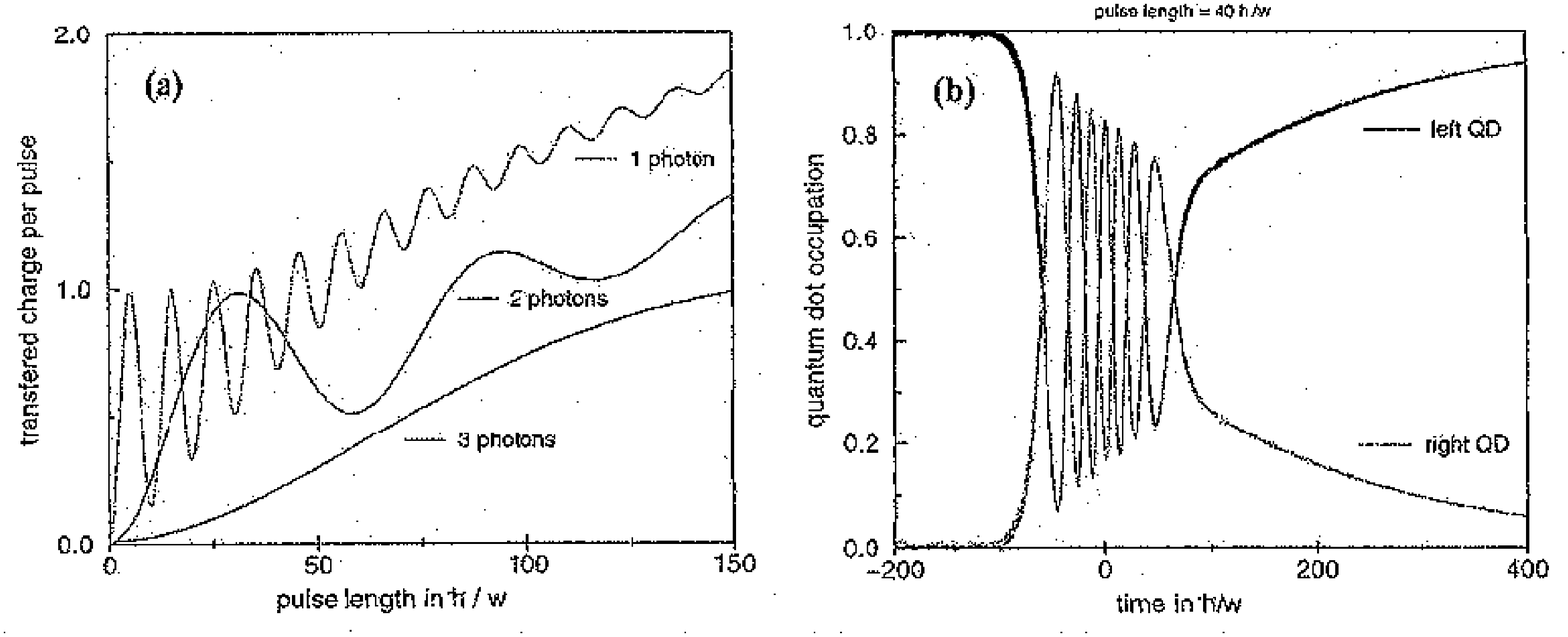}}
\caption{Charge transfer between two quantum dots exhibiting Rabi
(v.St\"{u}ckelberg) oscillations due to bouncing electrons between
the quantum dots. (a) Dependence of transferred charge on pulse
duration of external field and for different frequencies $\omega_1,\omega_2, \omega_3$ which cause resonant tunneling by absorption of one, two and three photons, respectively; (b) Time dependent occupation of
quantum dot states, see results by Garcia {\it et al.}. Note, for spin dependent quantum dot states (magnetic quantum dot), see Fig.20(a), the results depend on the light polarization and may involve spin dependent currents.}
\end{figure}

For further discussion of the interesting physics realized by
tunnel systems, the interplay of spin and charge currents (see continuity equation for the spin density) and of
accompanying light (see Maxwell equations), one should study the literature, see Nogueira, Bennemann {\it et al.} \cite{B3} and Bennemann \cite{B13}.

Currents in Mesoscopic Structures: As mentioned already for mesoscopic structures like discs, quantum dots and rings, for example,
one might get interesting topological effects and persistent currents driven by phases resulting from the Aharonov--Bohm effect \cite{B14}.
One may derive such currents induced by a magnetic field B from $j=\frac{\partial \Delta F}{\partial \phi}$, where F refers to the free--energy. From symmetry
considerations one gets $\Delta F \propto \cos BS$, where BS is the flux enclosed by a ring for example, and for the current
\begin{equation}
                   j \propto \sin BS.
\end{equation}
Here, the phase shift $\Delta\phi = \pm BS$ results from encircling the center
( of the ring, disc, etc.) clockwise or counterclockwise. Note, as in the case of superconductors flux quantization may occur. Dynamics due to B(t) or M(t) is of course reflected in the induced current. Also electron scattering causes dephasing etc. and affects the current.

The induced currents due to the Aharonov--Bohm effect with details of their oscillations can be calculated from
\begin{equation}
       j = \partial\triangle F/\partial\phi = - e/\hbar \sum \partial E_i/\partial\phi \tanh(E_i/kT)
\end{equation}
and using the Balian--Bloch type theory developed by Stampfli {\it et al.} for calculating $E_i(\phi)$ and taking into account clockwise and anticlockwise encircling the center.
For narrow rings one expects that mainly orbits 1, 2, 3 are important.
A magnetic field B causes the phase shift $\Delta\phi = \Delta\phi_1 + \Delta\phi_2 $ due to path
deformation and flux, respectively ($\Delta\phi_2 = BS_{\pm}$, where S refers to the area enclosed by the orbit and $\pm$
to clockwise or conterclockwise encircling of B ), see Aharonov--Bohm effect. Thus, one gets for the current
induced for example by the field B into a narrow ring the expression
\begin{equation}
      j  \propto \Sigma_t\Sigma_p a_{tp}\sin\phi_n \sin\phi_{tp}(\ldots) + \ldots
\end{equation}
Here, $\phi_n = B S_0$ denotes the quantized flux in units of $hc/e$ of the orbit characterized by $t$ and $p$
(describes Aharonov--Bohm effect) and $\phi_{tp}$ results from the path deformation due to the field $B$. Note, the factor
(...) causes rapid oscillations described by exponentials, see Stamfli {it et al.}. Interference of contributions from different paths may yield a beating pattern for the current oscillations. Generally the oscillations resemble those discussed for the DOS, see Figs. Since the coupling of the induced currents is given by terms $j_\mu A_\mu$, where $A_\mu$ refers to the vector potential of $B$, one gets generally that the induced current decreases quadratically with the ring radius R ($j\propto R^{-2}$). \cite{B14} The fine structure of the current oscillations reflects the quantization of the enclosed flux. In accordance with le Chatelier principle the magnetic field associated with the induced current should counteract the external field $B$.
\begin{figure}[tp]
\centerline{\includegraphics[width=.5\textwidth]{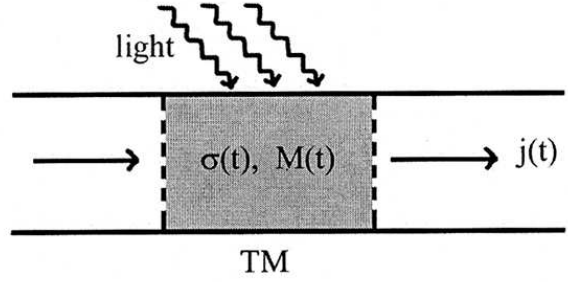}}
\caption{Illustration of a tunnel junction which may act as an
ultrafast switching device due to optical excitation of (hot)
electrons. Magnetization changes as the electronic temperature changes.}
\end{figure}

Of course, dephasing expected for strong scattering of the electrons and large rings (and discs) weakens and possibly destroys the current. Also for ferromagnets (Ni, Fe, ...) one expects spin currents (due to DOS--effects etc.). For spin currents also spin orbit coupling may play a role, see Aharonov--Casher effect. \cite{B14}

As noted a ring shows similar behavior as a (repulsive) antidot covered by a thin film. Of interest would be also to study
a ring of graphene, due to the remarkable properties of graphene, or conducting nanotubes. To study interplay of electric and magnetic fields (Maxwell equations at action in nanostructures) double--rings are of interest. Generally the discussion and results for superconducting cylinders, rings, squids, tunnel junctions, can be frequently applied somewhat to the topologically closed mesoscopic structures in case of relatively long free path of the electrons (see dominant paths in Balian--Bloch--Stampfli theory). This
demonstrates the interesting properties of nanostructures.

\section{Summary}

Magnetic effects in nanostructures have been discussed. Typical
properties of clusters, cluster ensembles, single films and film
multilayers and microscopic tunnel systems are presented. Due to
the reduced dimension and system size external fields may change
sensitively the behavior. Spin dependent transport on a nanoscale,
involving optical manipulation, temperature gradients etc., see
corresponding Onsager theory, offers interesting possibilities.
Also (strong) nonequilibrium behavior of nanostructure needs be
studied.\cite{B13}
\bigskip

{\bf Acknowledgements}

\noindent Essential general help in particular regarding
organisation of the review, technical assistence and critical
reading was given by Christof Bennemann. Furthermore, discussions
and many results by P. Jensen, M. Garcia, P. Stampfli, G. Pastor,
W. H\"{u}bner, K. Baberschke, M. Kulic, and many others are gratefully acknowledged. I am particular grateful to F.Nogueira for ideas, discussions and help.
This review was supported by D.F.G through S.F.B.
\end{ISItext1}

\end{document}